\documentclass[12pt]{article}
\usepackage[dvips]{graphicx}

\usepackage{sectsty}
\makeatletter
\def\@seccntformat#1{\csname the#1\endcsname.~}
\makeatother

\allsectionsfont{\centering}
\subsectionfont{\mdseries\itshape\centering}

\usepackage{natbib}

\usepackage{setspace}
\usepackage[table,xcdraw]{xcolor}
\usepackage[american]{babel}
\usepackage[utf8x]{inputenc}
\usepackage{csquotes}
\usepackage{multirow}
\usepackage{algorithm}
\usepackage{algorithmic}
\usepackage{makecell}
\usepackage{amsmath}
\usepackage{amssymb}
\usepackage{verbatim}
\usepackage{mathtools} 
\usepackage{bm}

\usepackage[colorinlistoftodos]{todonotes}
\usepackage{tikz}
\usepackage{multirow}
\usepackage{xcolor}
\usepackage{pstool}
\usepackage{psfrag}

\usepackage{booktabs,siunitx,caption}
\usetikzlibrary{circuits.logic.US,circuits.logic.IEC,fit}

\usepackage{centernot}

\usepackage{hyperref}
\usepackage{comment}

\newcommand{\cn}[1]{\texttt{#1}}
\newcommand{\lab}[1]{\textbf{#1}}
\newcommand{\vn}[1]{\emph{#1}} 

\usepackage{tikz}
\usetikzlibrary{bayesnet}
\usepackage[dvips]{graphicx}

\usepackage[top=1in, bottom=1in, left=1in, right=1in]{geometry}

\makeatletter
\@ifundefined{MPFourScale}{\def\MPFourScale{1.00}}{}
\DeclareFontFamily{U}{mp4}{}%
\DeclareFontShape{U}{mp4}{m}{n}{<->s * [\MPFourScale]cmb10}{}
\DeclareSymbolFont{boldgreekuc}{U}{mp4}{m}{n}
\DeclareMathSymbol{\bfAlpha}{\mathord}{boldgreekuc}{"41}
\DeclareMathSymbol{\bfBeta}{\mathord}{boldgreekuc}{"42}
\DeclareMathSymbol{\bfPsi}{\mathord}{boldgreekuc}{"09}
\DeclareMathSymbol{\bfDelta}{\mathord}{boldgreekuc}{"01}
\DeclareMathSymbol{\bfEpsilon}{\mathord}{boldgreekuc}{"45}
\DeclareMathSymbol{\bfPhi}{\mathord}{boldgreekuc}{"08}
\DeclareMathSymbol{\bfGamma}{\mathord}{boldgreekuc}{"00}
\DeclareMathSymbol{\bfEta}{\mathord}{boldgreekuc}{"48}
\DeclareMathSymbol{\bfIota}{\mathord}{boldgreekuc}{"49}
\DeclareMathSymbol{\bfXi}{\mathord}{boldgreekuc}{"04}
\DeclareMathSymbol{\bfKappa}{\mathord}{boldgreekuc}{"4B}
\DeclareMathSymbol{\bfLambda}{\mathord}{boldgreekuc}{"03}
\DeclareMathSymbol{\bfMu}{\mathord}{boldgreekuc}{"4D}
\DeclareMathSymbol{\bfNu}{\mathord}{boldgreekuc}{"4E}
\DeclareMathSymbol{\bfPi}{\mathord}{boldgreekuc}{"05}
\DeclareMathSymbol{\bfTheta}{\mathord}{boldgreekuc}{"02}
\DeclareMathSymbol{\bfRho}{\mathord}{boldgreekuc}{"52}
\DeclareMathSymbol{\bfSigma}{\mathord}{boldgreekuc}{"06}
\DeclareMathSymbol{\bfTau}{\mathord}{boldgreekuc}{"54}
\DeclareMathSymbol{\bfVartheta}{\mathord}{boldgreekuc}{"02} 
\DeclareMathSymbol{\bfOmega}{\mathord}{boldgreekuc}{"0A}
\DeclareMathSymbol{\bfVarphi}{\mathord}{boldgreekuc}{"08} 
\DeclareMathSymbol{\bfUpsilon}{\mathord}{boldgreekuc}{"07}
\DeclareMathSymbol{\bfZeta}{\mathord}{boldgreekuc}{"5A}

\DeclareFontFamily{U}{mp4sl}{}%
\DeclareFontShape{U}{mp4sl}{m}{n}{<->s * [\MPFourScale]cmmib10}{}
\DeclareSymbolFont{boldgreek}{U}{mp4sl}{m}{n}
\DeclareMathSymbol{\bfalpha}{\mathord}{boldgreek}{"0B}
\DeclareMathSymbol{\bfbeta}{\mathord}{boldgreek}{"0C}
\DeclareMathSymbol{\bfpsi}{\mathord}{boldgreek}{"20}
\DeclareMathSymbol{\bfdelta}{\mathord}{boldgreek}{"0E}
\DeclareMathSymbol{\bfepsilon}{\mathord}{boldgreek}{"0F}
\DeclareMathSymbol{\bfphi}{\mathord}{boldgreek}{"1E}
\DeclareMathSymbol{\bfgamma}{\mathord}{boldgreek}{"0D}
\DeclareMathSymbol{\bfeta}{\mathord}{boldgreek}{"11}
\DeclareMathSymbol{\bfiota}{\mathord}{boldgreek}{"13}
\DeclareMathSymbol{\bfxi}{\mathord}{boldgreek}{"18}
\DeclareMathSymbol{\bfkappa}{\mathord}{boldgreek}{"14}
\DeclareMathSymbol{\bflambda}{\mathord}{boldgreek}{"15}
\DeclareMathSymbol{\bfmu}{\mathord}{boldgreek}{"16}
\DeclareMathSymbol{\bfnu}{\mathord}{boldgreek}{"17}
\DeclareMathSymbol{\bfpi}{\mathord}{boldgreek}{"19}
\DeclareMathSymbol{\bfvartheta}{\mathord}{boldgreek}{"23}
\DeclareMathSymbol{\bfrho}{\mathord}{boldgreek}{"1A}
\DeclareMathSymbol{\bfsigma}{\mathord}{boldgreek}{"1B}
\DeclareMathSymbol{\bftau}{\mathord}{boldgreek}{"1C}
\DeclareMathSymbol{\bftheta}{\mathord}{boldgreek}{"12}
\DeclareMathSymbol{\bfomega}{\mathord}{boldgreek}{"21}
\DeclareMathSymbol{\bfvarphi}{\mathord}{boldgreek}{"27}
\DeclareMathSymbol{\bfchi}{\mathord}{boldgreek}{"1F}
\DeclareMathSymbol{\bfupsilon}{\mathord}{boldgreek}{"1D}
\DeclareMathSymbol{\bfzeta}{\mathord}{boldgreek}{"10}

\makeatother

\title{Valid standard errors for Bayesian quantile \\ regression with clustered and independent data}

\author{
  Feng Ji\textsuperscript{1}\thanks{Corresponding author: Feng Ji, \texttt{f.ji@utoronto.ca}}, 
  JoonHo Lee\textsuperscript{2}, 
  Sophia Rabe-Hesketh\textsuperscript{3}\\
  \textsuperscript{1} University of Toronto\\
  \textsuperscript{2} University of Alabama\\
  \textsuperscript{3} University of California, Berkeley
}

\date{December 2024}

\begin{document}
\setlength{\baselineskip}{2em}
{\pagestyle{empty}
\maketitle}
\setcounter{page}{1}



\begin{abstract}

\setlength{\baselineskip}{2em}
Bayesian quantile regression typically uses the asymmetric Laplace (AL) distribution as working likelihood, not because it is a plausible data-generating distribution but because the corresponding maximum likelihood estimator is identical to the classical estimator by Koenker and Bassett (1978). While point estimation is consistent, credible intervals tend to have poor frequentist coverage. We propose using Infinitesimal Jackknife (IJ) standard errors (Giordano and Broderick, 2024), which require no resampling and can be obtained from a single MCMC run. Simulations and applications to real data show that IJ standard errors have good frequentist properties for both independent and clustered data. We provide an R package, IJSE, that computes IJ standard errors after estimation with the brms wrapper for Stan.

\end{abstract}
\\

\vspace{\fill}\newpage

\section{Introduction}
Quantile regression is popular across a wide range of fields, including education and psychology \citep[e,g.,][]{betebenner2009norm, castellano2013quantile, furno2011quantile,
konstantopoulos2019using,zhu2024relations}. In frequentist quantile regression, point estimates can be obtained by minimizing an appropriate loss function through linear programming algorithms \citep{koenker1978regression}. Unlike this classic approach, Bayesian quantile regression requires a likelihood. There are several proposals including nonparametric likelihoods \citep[e.g.,][]{kottas2009semiparam, reich2010flexible} and empirical likelihoods \citep[e.g.,][]{otsu2008empir, yang2012bayesian}. However, by far the most commonly applied likelihood is the asymmetric Laplace (AL) likelihood first employed by Yu and Moyeed (2001)\nocite{yu2001bayesian}. The motivation for using an AL likelihood is that maximizing this likelihood corresponds to minimizing the loss function. The general approach of constructing a posterior through a loss function is discussed by, for instance, Syring and Martin (2019)\nocite{syring2019scale}.

The AL likelihood can be thought of as a working likelihood designed to produce consistent point estimates \citep[e.g.,][]{sriram2013ALDconsistent}. There is no particular reason to believe that the AL distribution is the data-generating distribution or that posterior intervals based on a misspecified AL distribution have correct Bayesian or frequentist properties. Unfortunately, this problem does not appear to be well-known, and AL-based Bayesian quantile regression with model-based posterior credible intervals, as implemented in the popular R package \cn{brms} \citep{burkner2017brms} and Stata's Bayesian quantile regression command \citep{stata}, is still widely used without any adjustment \citep[e.g.,][]{he2022distracted, forthmann2022quantity, mcisaac2023days, taheri2021differential, betz2022aldex}.

The AL likelihood has a scale parameter that effectively ``weights the information in the data relative to that in the prior'' \citep{syring2019scale} and therefore determines the spread of the posterior and the width of credible intervals \citep{sriram2015sandwich, yang2016posterior, syring2019scale}. This scale parameter is often set to an arbitrary constant in AL-based Bayesian quantile regression. For this approach, \citet{yang2016posterior} propose an adjustment to the posterior covariance matrix designed to produce correct asymptotic coverage of normal-approximation-based credible intervals \citep[see also][for a similar approach]{sriram2015sandwich}. From now on, we will refer to posterior standard deviations as standard errors to emphasize the requirement for correct frequentist properties. \citet{yang2016posterior} show that their adjusted standard errors are asymptotically invariant to the scale parameter. However, we show that the adjustment is sensitive to the scale parameter when the sample size is small to moderate. While finite-sample frequentist performance of the adjusted standard errors seems to be satisfactory when the scale parameter is set to its maximum likelihood (ML) estimate at the median, as suggested by \citet{yang2016posterior}, we prefer an approach that does not rely on ML estimation before proceeding with Bayesian estimation.

In frequentist inference for quantile regression, it is common practice to use bootstrap or jackknife procedures for standard errors \citep[e.g.,][]{bose2003boot, portnoy2014jack, quantreg}, but such resampling approaches would be computationally intensive in Bayesian estimation. We therefore propose using Infinitesimal Jackknife (IJ) standard errors \citep{giordano2023bayesian} that can be viewed as approximating resampling standard errors. IJ standard errors can be obtained from a single MCMC run if log-likelihood contributions from the units are computed for each post-warmup iteration. We also propose a version of IJ standard errors for clustered data. Simulations and applications to real data show that the IJ standard errors have good frequentist properties, both for independent and clustered data, and that posterior intervals have good coverage, especially when the AL scale parameter is estimated instead of fixed to an arbitrary constant. Although Giordano provided an R package \cn{StanSensitivity} that can calculate IJ standard errors, its usage is mainly in the context of sensitivity analysis. Hence, we provide an R package, \cn{IJSE}, focused on computing IJ standard errors for clustered or independent data after estimation with the \cn{brms} wrapper \citep{burkner2017brms} for \cn{Stan} \citep{carpenter2017stan}.

The structure of this paper is as follows. In Section~\ref{bayes_ALD}, we introduce AL-based Bayesian quantile regression and the adjusted standard errors proposed by \citet{yang2016posterior}. In Section~\ref{bayes_loo_methods}, we briefly motivate and define the IJ standard errors by \citet{giordano2020} and Giordano and Broderick (2024)\nocite{giordano2023bayesian} and show how they can be modified to handle clustered data. For independent (non-clustered) data, Section~\ref{sec:sim1} presents simulation results for IJ standard errors, model-based standard errors, and adjusted standard errors by \citet{yang2016posterior} when the scale parameter is fixed at a range of different values, as well as Bayesian posterior standard deviations and IJ standard errors when the scale parameter is estimated. In Section~\ref{bayes_ALD_cluster}, we evaluate the IJ standard errors for the clustered case. The different kinds of standard errors are compared for real data in Section~\ref{sec:realdata}, and we end with a brief discussion in Section~\ref{bayes_loo_conclusion}.

\section{AL-based Bayesian quantile regression and adjusted standard errors}
\label{bayes_ALD}

\subsection{The AL likelihood}

In quantile regression \citep{koenker1978regression}, models are specified directly for the conditional quantiles $Q_\tau(y|x)$ of the response variable $y$ given a set of covariates $x$, where $\tau$ is the quantile level. Although the methods discussed here apply more generally, we will consider a linear quantile regression model,
\[
Q_\tau(y|x) = x^{\prime}\beta(\tau),
\]
where $\beta(\tau)$ is a vector of regression coefficients for quantile level $\tau$.

In the original frequentist setup, the point estimates are obtained as numerical solutions to the minimization problem
\[
\hat{\beta}(\tau) = \operatorname{argmin}_{\beta(\tau)} \sum_{i = 1}^n \rho_\tau(y_i - x_i^{\prime}\beta(\tau)),
\]
where $\rho_{\tau}(\cdot)$, often called the ``check function'' for its shape, is defined as
\[
\rho_{\tau}(u) = u\{\tau - I(u < 0)\} =
\begin{cases}
u \tau & \text{if } u \geq 0 \\
-u(1 - \tau) & \text{if } u < 0.
\end{cases}
\]

Bayesian inference requires a likelihood, and Yu and Moyeed (2001)\nocite{yu2001bayesian} suggested specifying the asymmetric Laplace (AL) distribution,
\[
f_{\text{ALD}}(y_i|\mu_i, \sigma, \tau) = \frac{\tau(1 - \tau)}{\sigma} \exp\left\{-\rho_{\tau}\left(\frac{y_i - \mu_i}{\sigma}\right)\right\},
\]
where $\mu_i = x_i^{\prime}\beta(\tau)$ is the location parameter and $\sigma$ is the scale parameter. The motivation for the AL distribution is that maximizing this likelihood for a given value of $\sigma$ corresponds directly to minimizing the loss function $\sum_{i=1}^n \rho_{\tau}(y_i - \mu_i)$ used in classical quantile regression. Implementation of MCMC algorithms is facilitated by the fact that the error term $\epsilon_i = y_i - \mu_i$ can be represented by a scale mixture of normal distributions \citep{kotz2001asymmetric, kozumi2011invgam}, $\epsilon_i = \sigma(\theta_1\nu_i + \theta_2 z_i \sqrt{\nu_i})$, where $\theta_1 = (1 - 2\tau)/\{\tau(1 - \tau)\}$, $\theta_2^2 = 2/\{\tau(1 - \tau)\}$, $z_i \sim N(0,1)$, $\nu_i$ follows the standard exponential distribution, and $z_i$ and $\nu_i$ are independent.

There is no motivation for the AL distribution as a data-generating process. In fact, one of the main reasons for adopting quantile regression is to leave the conditional distribution of $y_i$ unspecified. Additionally, the AL distribution cannot be a data-generating distribution because its skewness depends on the quantile $\tau$ that we happen to be interested in (right-skewness for $\tau < .5$ and left-skewness for $\tau > .5$). Koenker and Machado (1999)\nocite{koenker1999ALD} refer to the AL distribution as ``rather implausible,'' \citet{sriram2013ALDconsistent} refer to it as ``misspecified,'' \citet{sriram2015sandwich} and \citet{yang2016posterior} use the term ``working likelihood,'' and \citet{geraci2019} uses the term ``pseudo-likelihood.''

As mentioned, specifying a likelihood is necessary for Bayesian inference. In addition, a parametric likelihood is also extremely useful, in Bayesian and non-Bayesian settings alike, for extending the linear quantile regression model in different ways. For example, the AL likelihood has been used to accommodate censoring \citep{yu2007tobit}, include random effects or latent variables \citep{geraci2007quantile, geraci2014linear, wang2012nonlinME}, and jointly model longitudinal and time-to-event outcomes \citep{huang2016joint}.

AL-based Bayesian quantile regression is implemented in publicly available software such as the R packages \cn{brms} \citep{burkner2017brms} and \cn{bayesQR} \citep{benoit2017bayesqr}, and in the \cn{bayes:qreg} command in the Stata software \citep{stata}.

The scale parameter $\sigma$ appears to be arbitrary, and $\sigma = 1$ has been used by Yu and Moyeed (2001)\nocite{yu2001bayesian} without any discussion and by Yu and Stander (2007)\nocite{yu2007tobit}, who refer to $\sigma$ as a ``nuisance parameter.'' \citet{sriram2013ALDconsistent} proved posterior consistency of the point estimator of $\beta(\tau)$ both when $\sigma = 1$ and when $\sigma$ is given a prior. Yue and Rue (2011) \nocite{yue2011scale} estimate the scale parameter, arguing that it ``makes asymmetric Laplace distributions more flexible to capture the unknown error distributions.''

The most common prior for $\sigma$ seems to be an inverse Gamma prior \citep[e.g.,][]{kozumi2011invgam, yue2011scale, wang2012nonlinME, sriram2015sandwich}, the conjugate prior under prior independence between $\sigma$ and $\beta(\tau)$ \citep{kozumi2011invgam}. This prior is used by the R package \cn{BayesQR} and Stata's \cn{bayes:qreg} command, although the latter also allows $\sigma$ to be set to a constant. \citet{reich2010flexible} use a uniform prior on $(0,10)$ for $\sigma$, and \cn{brms} uses a half-\emph{t} distribution with 3 degrees of freedom, which we will denote half-\emph{t}(3).

\subsection{Standard error adjustment}

While many simulation studies evaluating AL-based Bayesian quantile regression focus exclusively on point estimation, \citet{reich2010flexible}, \citet{sriram2015sandwich}, \citet{yang2016posterior}, \citet{lee2020dis}, and \citet{ji2022dis} show that model-based credible intervals have poor coverage.

Let $\beta^0(\tau)$ be the true parameters of interest and $\widehat{\beta}(\tau)$ their maximum likelihood estimates for a given dataset $D$. \citet{yang2016posterior} show that, for flat (improper) priors for the coefficients $\beta(\tau)$ and fixed $\sigma$, the posterior distribution of $\beta(\tau)$ is asymptotically multivariate normal,
\begin{equation}
p_n({\beta}(\tau)\mid D) \xrightarrow[]{d} N\left(\widehat{\beta}(\tau), \frac{\sigma D_1^{-1}}{n}\right), \label{eq:asympost}
\end{equation}
where
\[
D_1 = \lim_{n \to \infty} \frac{1}{n} \sum_{i=1}^{n} f_{\text{ALD}}(y_i | \mu_i^{0}, \sigma, \tau) x_i x_i^{\prime},
\]
with $\mu_i^0 = x_i^{\prime}\beta^0(\tau)$. See also \citet{sriram2015sandwich} for the same result when $\sigma = 1$ and $\beta(\tau)$ has an arbitrary proper prior, with Remark 4 generalizing the result to arbitrary fixed $\sigma$.

As shown in~(\ref{eq:asympost}), the asymptotic posterior covariance matrix is proportional to the arbitrary scale parameter $\sigma$. Therefore, not surprisingly, it has been demonstrated empirically that credible intervals based on the estimated posterior covariance matrix $\widehat{\Sigma}(\sigma)$, for arbitrarily fixed $\sigma$, do not have correct coverage \citep{sriram2015sandwich, yang2016posterior, lee2020dis, ji2022dis}.

\citet{yang2016posterior} proposed a simple adjustment to the posterior covariance matrix to improve the coverage of credible intervals. The adjustment is based on ~(\ref{eq:asympost}) and the asymptotic sampling distribution of the maximum likelihood estimator \citep[][p.~74]{koenker2005book}
\begin{equation}
\sqrt{n}\widehat{\beta}(\tau) \xrightarrow[]{d} N\left(\beta^0, \tau(1 - \tau) D_1^{-1} D_0 D_1^{-1}\right), \label{eq:asymmle}
\end{equation}
where
\[
D_0 = \lim_{n \to \infty} \frac{1}{n} \sum_{i=1}^{n} x_i x_i^{\prime}.
\]

The adjusted standard errors by \citet{yang2016posterior} are obtained by substituting $\frac{n}{\sigma} \widehat{\Sigma}(\sigma)$ for $D_1^{-1}$, based on~(\ref{eq:asympost}), and estimating $D_0$ by its finite-sample counterpart $n^{-1} \sum_{i=1}^{n} x_i x_i^{\prime}$ in the asymptotic ``sandwich'' covariance matrix $\frac{\tau(1 - \tau)}{n} D_1^{-1} D_0 D_1^{-1}$ from~(\ref{eq:asymmle}):
\begin{equation}
\widehat{\Sigma}^{\text{adj}}(\sigma) = \frac{\tau(1 - \tau)}{\sigma^2} \widehat{\Sigma}(\sigma) \left(\sum_{i=1}^{n} x_i x_i^{\prime}\right) \widehat{\Sigma}(\sigma).
\label{eq:adjse}
\end{equation}

Using similar reasoning, \citet{sriram2015sandwich} proposed a two-step approach. In the first step, estimate the posterior means $\tilde{\beta}(\tau)$ and posterior covariance matrix $\widehat{\Sigma}(\sigma)$ for $\beta(\tau)$ based on the AL likelihood with fixed $\sigma$. In the second step, construct a ``sandwich likelihood'' that is multivariate normal with means $\tilde{\beta}(\tau)$ and covariance matrix given in (\ref{eq:adjse}) and re-estimate the posterior means and covariance matrix for $\beta(\tau)$ with the sandwich likelihood replacing the AL likelihood. \citet{sriram2015sandwich} proves that the credible sets from the posterior from Step 2 merge asymptotically with the frequentist confidence sets under mild assumptions.

\citet{yang2016posterior} argue that $\widehat{\Sigma}^{\text{adj}}(\sigma)$ is asymptotically invariant to the value of $\sigma$ and yields asymptotically valid posterior inferences. However, in finite samples, the posterior distribution may not be well approximated by a multivariate normal distribution with covariance matrix $\frac{\sigma D_1^{-1}}{n}$, depending on the sample size $n$, the quantile level $\tau$, the value of the scale parameter $\sigma$, and data-generating process. When the approximation does not work, the choice of $\sigma$ may influence the adjusted covariance matrix.

\citet{yang2016posterior} recommend setting $\sigma$ equal to its maximum likelihood estimate under the AL likelihood for $\tau = 0.5$, the median. No proof or arguments are provided for this recommendation, and it is not emphasized, presumably due to the assumed approximate invariance of $\widehat{\Sigma}^{\text{adj}}(\sigma)$ to $\sigma$.

For the broader class of approaches that choose a likelihood because its maximum minimizes a loss function, Syring and Martin (2019) \nocite{syring2019scale} suggest selecting the scale parameter so that the credible intervals have approximately correct empirical coverage. Such an approach could be used in AL-based Bayesian quantile regression but is not further pursued here.

\section{Infinitesimal Jackknife (IJ) standard errors}
\label{bayes_loo_methods}

Infinitesimal jackknife standard errors were originally proposed for frequentist estimators \citep{jaeckel1972} and applied in covariance structure analysis by \citet{jennrich2008nonparametric}. A Bayesian version, which we will simply denote IJ for brevity, was introduced by Giordano and Broderick (2024)\nocite{giordano2023bayesian}, based on their earlier work \citep{giordano2018covariances, giordano2020, giordano2020jacknife}. \citet{giordano2020} motivates the IJ variance as an approximation to a resampling variance, such as bootstrap or jackknife. Since it would be computationally infeasible to perform MCMC estimation in a large number of resampled datasets, \citet{giordano2020jacknife} propose using a linear approximation for the posterior means of the parameters in resampled data. Let $w$ be an $n$-dimensional weight vector, with element $w_i$ equal to the number of times unit $i$ is represented in the resampled data. The part of the posterior distribution that changes in resampled data is the likelihood function, and we can express the log-likelihood for a given weight vector as
\[
\sum_{i=1}^n w_i \ell_i(D|\theta),
\]
where $\ell_i(D|\theta)$ is the log-likelihood contribution from unit $i$ for parameter vector $\theta$.

The linear approximation to the posterior expectation for the resampled data then is
\begin{align}
E(\theta \mid D, w )  &\approx E(\theta \mid D, w = 1_n) + \left.\frac{d E(\theta \mid D, w)}{d w^{\prime}}\right|_{w = 1_n} (w - 1_n) \label{eq:NIJ} \\
&= E(\theta \mid D, w = 1_n) + \operatorname{cov}_{\theta \mid D}[\theta, \ell(D \mid \theta)] (w - 1_n), \label{eq:BIJ}
\end{align}
where $\ell(D \mid \theta)$ is the $n$-dimensional vector of log-likelihood contributions from the units, and $1_n$ is an $n$-dimensional vector with all elements equal to 1, so that the first term is the posterior expectation for the original dataset. The result that the matrix of derivatives in (\ref{eq:NIJ}) is replaced by the posterior covariance matrix of the parameters and log-likelihood contributions in (\ref{eq:BIJ}) is given in Theorem 1 of \citet{giordano2018covariances} and derived in Section 2 of \citet{efron2015frequentist}, as outlined in online Appendix~A. This covariance matrix can be estimated straightforwardly by the corresponding empirical covariance matrix of the MCMC samples.

The empirical influence function for the posterior expectation is then given by
\[
I_i := n \left.\frac{d E(\theta \mid D, w)}{d w_i}\right|_{w = 1_n} = n \operatorname{cov}_{\theta \mid D}[\theta, \ell_i(D \mid \theta)]
\]
and the corresponding infinitesimal jackknife variance estimator is
\[
\hat{V}^{\text{IJ}} := \frac{1}{n(n - 1)} \sum_{i=1}^n (I_i - \overline{I})(I_i - \overline{I})^{\prime},
\]
where $\overline{I}$ is the sample mean of the empirical influence function.

\subsection{Clustered data}
\label{sec:IJclus}

Cross-sectional clustered data (e.g., patients nested in hospitals, residents nested in street blocks, students nested in classes) and longitudinal data violate the independence assumption of the AL likelihood and of IJ standard errors.

There have been several approaches to explicitly modeling between-cluster heterogeneity and/or within-cluster dependence in quantile regression. In a non-Bayesian setting, Koenker and Hallock (2004)\nocite{koenker2004longit} propose a model for longitudinal data that includes subject-specific fixed effects $\alpha_i(\tau)$ in the linear conditional quantile model. They also include an $\ell_1$ (Lasso) penalty for these fixed effects to shrink them towards a common intercept. Lamarche and Parker (2023)\nocite{lamarche2023} propose a bootstrapping procedure for the standard errors of this model. Also in a non-Bayesian setting, Geraci and Bottai (2007, 2014) \nocite{geraci2007quantile}\nocite{geraci2014linear} define linear quantile mixed models that resemble linear mixed models by including normally distributed cluster-specific random effects in the linear conditional quantile function, with an AL distribution for the (level-1) errors. Bayesian quantile regression models with cluster-specific random effects have also been proposed, both with an AL likelihood \citep[e.g.,][]{wang2012nonlinME, luo2012reff, yu2015hierar} and with more flexible densities for the level-1 errors such as an infinite mixture of normal densities \citep{reich2010flexible}.

In quantile regression models with cluster-specific fixed or random effects, the regression coefficients represent the relationship between cluster-specific quantiles and the covariates and do not have any meaningful marginal (over the clusters) interpretation \citep[see, e.g.,][]{reich2010flexible}. Here we focus on models for the marginal quantiles. Such a marginal approach was also adopted, for instance, by \citet{lipsitz1997gee} in Biostatistics, and Parente \& Silva (2016)\nocite{parente2016quantile} as well as \citet{hagemann2017cluster} in Econometrics. For example, \citet{hagemann2017cluster} ignores the clustering for point estimation and develops a bootstrap procedure for obtaining cluster-robust standard errors. Here we propose adopting such an approach but with a clustered-data version of the IJ standard errors instead of bootstrapping.

Specifically, letting $J$ be the number of clusters and $n$ the total sample size, we define the empirical influence function for cluster $j$ as
\[
I_j^{(cl)} := J \sum_{\substack{i \\ \text{in cluster } j}} \frac{I_i}{n}
\]
and estimate the variance as
\[
\hat{V}^{\text{IJ}}_{(cl)} := \frac{1}{J(J - 1)} \sum_{j=1}^J (I_j^{(cl)} - \overline{I^{(cl)}})(I_j^{(cl)} - \overline{I^{(cl)}})^{\prime}.
\]
This approach can be justified from a resampling perspective where entire clusters are resampled. Equivalently, we can define
\[
I_j^{(cl)} := J \operatorname{cov}_{\theta \mid D}\left[\theta, \ell_j^{(cl)}(D \mid \theta)\right],
\]
where
\[
\ell_j^{(cl)} := \sum_{\substack{i \\ \text{in cluster } j}} \ell_i(D \mid \theta).
\]

\section{Evaluation of standard errors for independent data}
\label{sec:sim1}

\subsection{Methods}

We consider two data-generating models: (1) location-shifted normal, and (2) location-shifted and scaled normal.

\paragraph{Model 1: Location-shifted normal model}

The data-generating model is
\[
y_i = \alpha + \beta x_i + \epsilon_i, \quad \epsilon_i \sim N(0, 1).
\]
The corresponding function for the $\tau^{\text{th}}$ conditional quantile is
\[
Q_{\tau}(y_i \mid x_i) = [\alpha + \Phi^{-1}(\tau)] + \beta x_i,
\]
so that the slope of $x_i$ is constant across quantile levels and the location (or intercept) at quantile level $\tau \neq 0.5$ is shifted by $\Phi^{-1}(\tau)$ relative to the intercept at the median, where $\Phi$ is the standard normal CDF.

\paragraph{Model 2: Location-shifted and scaled normal model}

The model is
\[
y_i = \alpha + \beta x_i + (1 + \gamma x_i) \epsilon_i, \quad \epsilon_i \sim N(0, 1),
\]
with conditional quantile function
\[
Q_{\tau}(y_i \mid x_i) = [\alpha + \Phi^{-1}(\tau)] + [\beta + \gamma \Phi^{-1}(\tau)] x_i.
\]
Now the slopes of $x_i$ at different quantile levels are no longer constant but are shifted by $\gamma \Phi^{-1}(\tau)$ relative to the slope at the median.

For the models above, we consider two choices of sample size ($n = 200$ and $n = 1000$), simulate $x_i$ from a standard Gaussian distribution, and fix $\alpha = \beta = 2$ for both models and $\gamma = 0.3$ for Model 2. We compare various kinds of approximate standard errors and corresponding  credible intervals (based on a normal approximation) for five different quantiles, $\tau \in \{0.1, 0.3, 0.5, 0.7, 0.9\}$.

First, we investigate sensitivity to the scale parameter of the following Bayesian methods with $\sigma$ set to $\{0.1, 0.2, 0.5, 1, 2, 5, 10\}$:
\begin{description}
\item [\lab{ALD}:] Unadjusted model-based posterior standard deviations.
    \item [\lab{IJf}:] Infinitesimal jackknife standard errors with fixed $\sigma$.
    \item [\lab{Yang}:] Adjusted standard errors proposed by \citet{yang2016posterior}.
\end{description}
Point estimation for these methods is performed using \cn{Stan}.

Second, we compare IJf, with $\sigma = 1$, with the following methods that do not rely on setting the scale parameter to an arbitrary constant:
\begin{description}
    \item \lab{boot}: Frequentist point estimates and bootstrap standard errors (\cn{quantreg}, R package version 5.95, by \citet{quantreg}, with the options \cn{se=boot} and \cn{bsmethod="xy"}).
    \item \lab{sandwich}: Frequentist point estimates and standard errors based on the sandwich estimator (\cn{quantreg}, R package version 5.95, by \citet{quantreg}, with option \cn{se=nid}).
    \item \lab{IJ}: Bayesian point estimates (posterior means) and IJ standard errors with \emph{estimated} $\sigma$ (\cn{brms} with AL likelihood and half-\emph{t}(3) prior for $\sigma$ and our \cn{IJSE} package).
    \item \lab{AdjBQR}: Posterior means and adjusted standard errors proposed by \citet{yang2016posterior} with $\sigma$ fixed at MLE for $\tau = 0.5$ (\cn{AdjBQR}, R package version 1.0, by \citet{wang2016BCQR}).
    \item \lab{brms}: Posterior means and model-based posterior standard deviations when $\sigma$ has a half-\emph{t}(3) distribution (\cn{brms}, R package version 2.20.1, by \citet{burkner2017brms}).
    \item \lab{BayesQR}: Posterior means and model-based posterior standard deviations when $\sigma$ has an inverse Gamma(0.01, 0.01) prior with scale parameter and shape parameter both equal to 0.01 (\cn{BayesQR}, R package version 2.4, by \citet{benoit2017bayesqr}).
\end{description}

In both simulation studies, all methods are applied to each of 100 replications. We assess how well the different versions of standard errors approximate the sampling standard deviations of the point estimates of the intercept and slope parameters by reporting the relative error defined as

\[
R_e = \sqrt{\frac{\overline{\text{se}^2}}{\operatorname{var}(\widehat{\beta})}} - 1,
\]

where $\overline{\text{se}^2}$ is the average squared standard error estimate across replications and $\operatorname{var}(\widehat{\beta})$ is the variance of the point estimates across replications. The Monte Carlo error (MCe) of this relative error is approximately \citep{white2010relerror}
\[
\text{MCe} = (R_e + 1) \sqrt{\frac{\operatorname{var}(\text{se}^2)}{\operatorname{var}(\widehat{\beta})^2} + \frac{1}{(2m - 1)}},
\]
where $m = 100$ is the number of replications. We also present approximate 95\% confidence intervals for the relative error based on the MCe.

Finally, we report empirical coverage of 90\% credible intervals (using a normality-based approximation centered at the posterior mean) along with exact binomial 95\% confidence intervals for the coverage.

\subsection{Results}

\begin{figure}[htbp]
    \centering
\includegraphics[scale=.20]{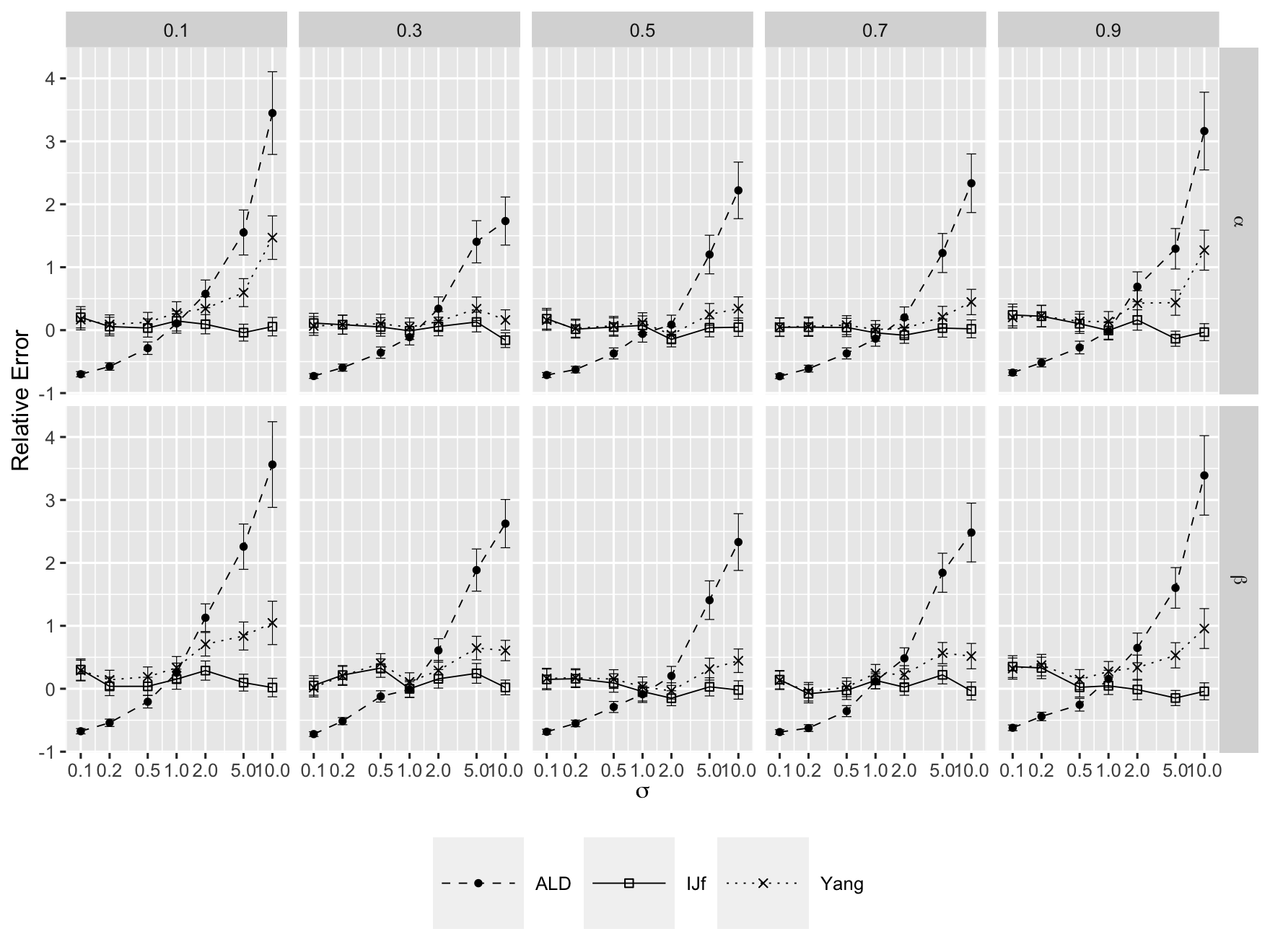} 
    \caption{\footnotesize{}Relative error with approximate 95\% confidence intervals for different types of standard errors as a function of the scale parameter $\sigma$ for Model 2 with $n=200$.}
    \label{fig:dif_scale}
\end{figure}

Figure \ref{fig:dif_scale} shows the relative error for \lab{ALD} (solid line), \lab{IJf} (dotted) and \lab{Yang} (dashed) as a function of the scale parameter $\sigma$ for the intercept $\alpha$ (row 1) and slope $\beta$ (row 2) and different quantile levels $\tau$ (columns) for Model 2 with $n=200$. Analogous figures for Model 1 and $n=1000$ are given in online Appendix B. It is clear that the model-based posterior standard deviations (denoted \lab{ALD}) can either over or under-estimate the uncertainty depending on the value of the scale parameter, with standard errors being as much as 300\% too large for the range of scale parameters considered. As $\sigma$ increases, the data appear less informative, leading to larger model-based posterior standard deviations. While \lab{Yang} performs better than \lab{ALD}, it produces unsatisfactory results (such as being over 50\% too large) for some of the values of $\sigma$ considered, especially for quantile levels 0.1 and 0.9. In contrast, the relative error of \lab{IJf} is close to zero across the range of scale parameters considered.

\begin{figure}[htbp]
  \footnotesize
\centering
\includegraphics[scale=.20]{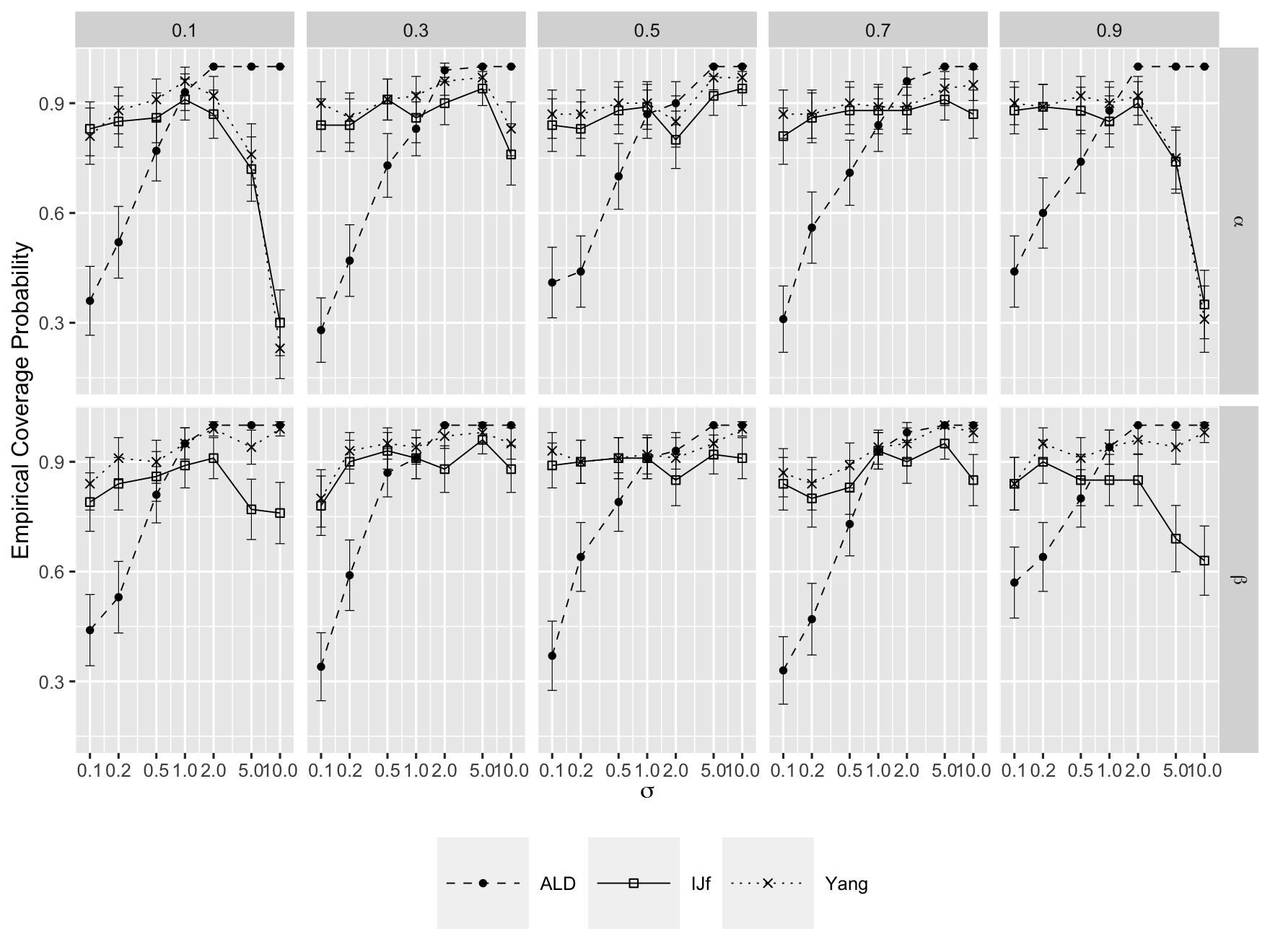}
\caption[Empirical coverage for Model 2 with $n=200$]{\label{fig:ecr_200}\footnotesize{}Empirical coverage of 90\% credible intervals for different types of standard errors as a function of the scale parameter $\sigma$ for different quantiles ($\tau$) for Model 2 with $n=200$.}
\end{figure}

Figure \ref{fig:ecr_200} shows the empirical coverage of 90\% credible intervals with exact binomial 95\% confidence intervals for for Model 2 with $n=200$, in the same format as Figure  \ref{fig:dif_scale}. Analogous graphs for Model 1 and $n=1000$ are given in online Appendix B.

As expected, \lab{ALD} performs poorly for both $\alpha$ and $\beta$, at all quantile levels, for most values of $\sigma$, with decreasing undercoverage as $\sigma$ increases to about 1 and then increasing overcoverage as $\sigma$ continues to increase. For non-extreme quantiles (i.e., $\tau = 0.3, 0.5 \text{ or } 0.7$), coverage of the proposed \lab{IJf} method is similar to that of the frequentist sandwich and bootstrap methods, all at around the nominal 0.9 level, across different choices of the scale parameter $\sigma$, sample size and data-generating model.

However, for the two extreme quantiles with large scale parameter (i.e., $\tau = 0.1 \text{ or } 0.9$ and $\sigma = 5 \text{ or } 10$), coverage of the \lab{IJf} intervals for both the intercept and the slope is too low, especially for the intercept. 
A reason for this undercoverage is that the Bayesian point estimator for the intercept is severely biased for extreme quantiles with large scale parameters, as also observed in previous simulation studies \citep[e.g.,][page 48-49]{lee2020dis}. For example, when $\tau=0.9$, $\sigma=10$, the bias for $\beta$ is estimated as 0.2 for Model 2 with $n=200$. In this simulation condition, \lab{IJf} has  under-coverage because its standard error reflects the sampling standard deviation, as shown in Figure \ref{fig:dif_scale}, whereas \lab{Yang} has over-coverage (0.99) because the adjusted standard error is too large. Specifically, the average \lab{Yang} standard error  is 0.2 compared with 0.09 for \lab{IJf}. Briefly, the reason for the bias seems to be that the posterior distribution of the intercept becomes increasingly skewed  as $\tau$ becomes more extreme (further from 0.5) and as $\sigma$ increases, so that the posterior mean differs increasingly from the posterior mode. However, the AL likelihood is chosen because its mode corresponds to the classical estimator. See also Section~\ref{sec:engel} for an illustration of this phenomenon. 

\begin{figure}[htbp]
    \centering
    \includegraphics[scale=0.19]{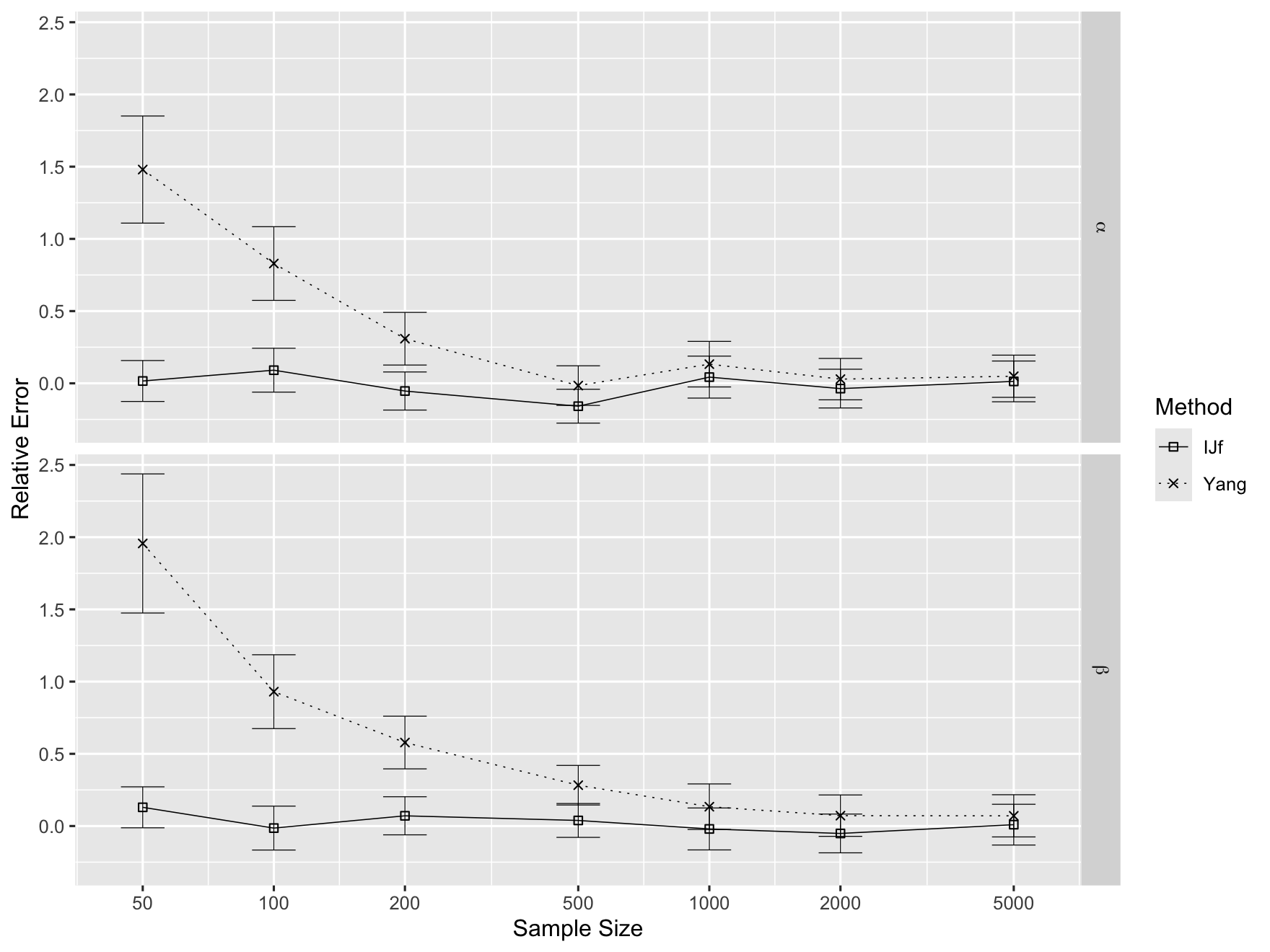}
    \caption{\footnotesize{}Relative error of \lab{IJf} and \lab{Yang} with approximate 95\% confidence intervals as a function of sample size for $\tau=0.7$, $\sigma=10$, and Model 2.}
    \label{fig:dif_sample_size}
\end{figure}

As \lab{Yang} relies on asymptotic results, we further investigated its performance, compared with \lab{IJf}, for $\sigma=10$ and $\tau=0.7$ as the sample size increases from 50 to 5000.  Figure \ref{fig:dif_sample_size} shows that the relative error for \lab{Yang} decreases as the sample size increases, as expected. However, the advantage of using \lab{IJf}  is clear here when the sample size is smaller than $500$.

\begin{figure}[htbp]
    \centering
    \includegraphics[scale=0.20]{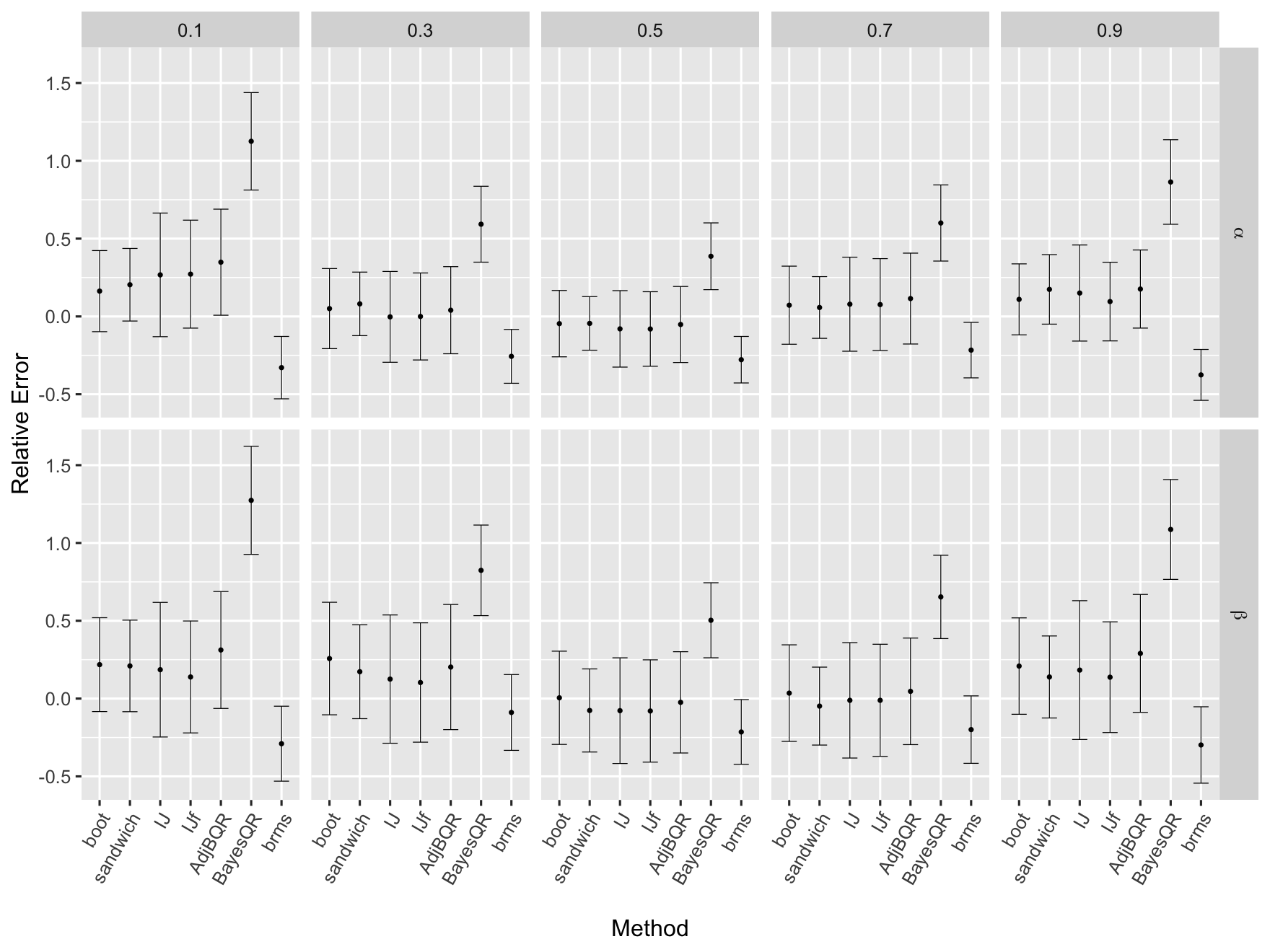}
    \caption{\footnotesize{}Relative error with approximate 95\% confidence intervals for different methods with $n=200$ and Model 2.}
    \label{fig:dif_methods}
\end{figure}

Figure~\ref{fig:dif_methods} shows the relative error for \lab{IJf} with $\sigma=1$ and for the methods that do not rely on fixing the scale parameter for Model 2 with $n=200$. The results for Model 1 and $n=1000$ are given in online Appendix B. \lab{IJ}, \lab{IJf}, and \lab{AdjBQR} (the adjusted standard errors) perform well, with relative errors similar to the frequentist methods (\lab{boot} and \lab{sandwich}), whereas the unadjusted Bayesian estimates that treat $\sigma$ as a free parameter (\lab{BayesQR} and \lab{brms}) perform worse. \lab{BayesQR} consistently performs rather poorly compared to all other methods, and hence we do not include it in the remainder of the paper. \lab{brms}, which is commonly used in practice, underestimates the standard errors for $\alpha$ across quantile levels and for $\beta$ at the extreme quantile levels ($\tau=0.1$ and $\tau=0.9$).

\clearpage

\section{Evaluation of standard errors for clustered data}
\label{bayes_ALD_cluster}

To make our simulation results comparable to the results presented by \citet{hagemann2017cluster} for cluster-robust inference for quantile regression, we follow his simulation setup. The data-generating model is
\[
y_{ij} = \frac{1}{10} u_{ij} + x_{ij} + x_{ij}^2 u_{ij},
\]
where $x_{ij}$ is the sum of $\sqrt{\rho} z_{j}$ and $\sqrt{1 - \rho} \epsilon_{ij}$ with $\rho \in [0,1)$, $z_{j} \sim N(0,1)$ and, independently, $\epsilon_{ij} \sim N(0,1)$, so that $x_{ij}$ has an intraclass correlation equal to $\rho$ and a variance of 1. Finally, $u_{ij} \sim N(0, \frac{1}{3})$. This outcome model implies that the conditional quantile functions are
\[
Q_\tau(y_{ij} \mid x_{ij}) = \alpha(\tau) + \beta_1(\tau) x_{ij} + \beta_2(\tau) x_{ij}^2,
\]
with $\alpha(\tau) = \frac{\Phi^{-1}(\tau)}{\sqrt{300}}$, $\beta_1(\tau) = 1$, and $\beta_2(\tau) = \frac{\Phi^{-1}(\tau)}{\sqrt{3}}$.

We consider two cluster sizes, $I \in \{10, 30\}$, three choices of number of clusters, $J \in \{10, 50, 100\}$, two choices of intraclass correlation, $\text{ICC} \in \{0.3, 0.8\}$, and five quantile levels, $\tau \in \{0.1, 0.3, 0.5, 0.7, 0.9\}$. We compare \lab{IJf} (with $\sigma=0.1$), with \lab{IJ} and  the current state-of-the-art clustered gradient bootstrap (denoted \lab{boot}) defined in \citet{hagemann2017cluster} for frequentist quantile regression.

The relative error of the standard errors and empirical coverage of the corresponding 90\% intervals are estimated and both are reported with 95\% confidence intervals. 

\subsection{Results}
Results for only $J=10$ clusters are not shown here because none of the methods consistently performed well in this case. As shown in online Appendix C, with $J=10$, the relative error for bootstrap standard errors was large (sometimes off the scale of the figures) for the extreme quantiles ($\tau=0.1$ and $\tau=0.9$), the reason being that the bootstrap standard error were extremely large for some simulated datasets in these conditions. While the relative error for \lab{IJ} and \lab{IJf} is close to zero, We observed undercoverage of \lab{IJ} and \lab{IJf} intervals  
when $J=10$, particularly for $\beta_2$ and for the extreme quantiles, possibly due to poor point estimation with the smaller total sample sizes ($n=100$ or $300$ for $J=10$).  In contrast, bootstrap intervals sometimes show overcoverage for extreme quantiles. We therefore recommend against using any of these methods for the extreme quantiles when $J=10$.

\begin{figure}[htbp]
\centering
\includegraphics[scale=.18]{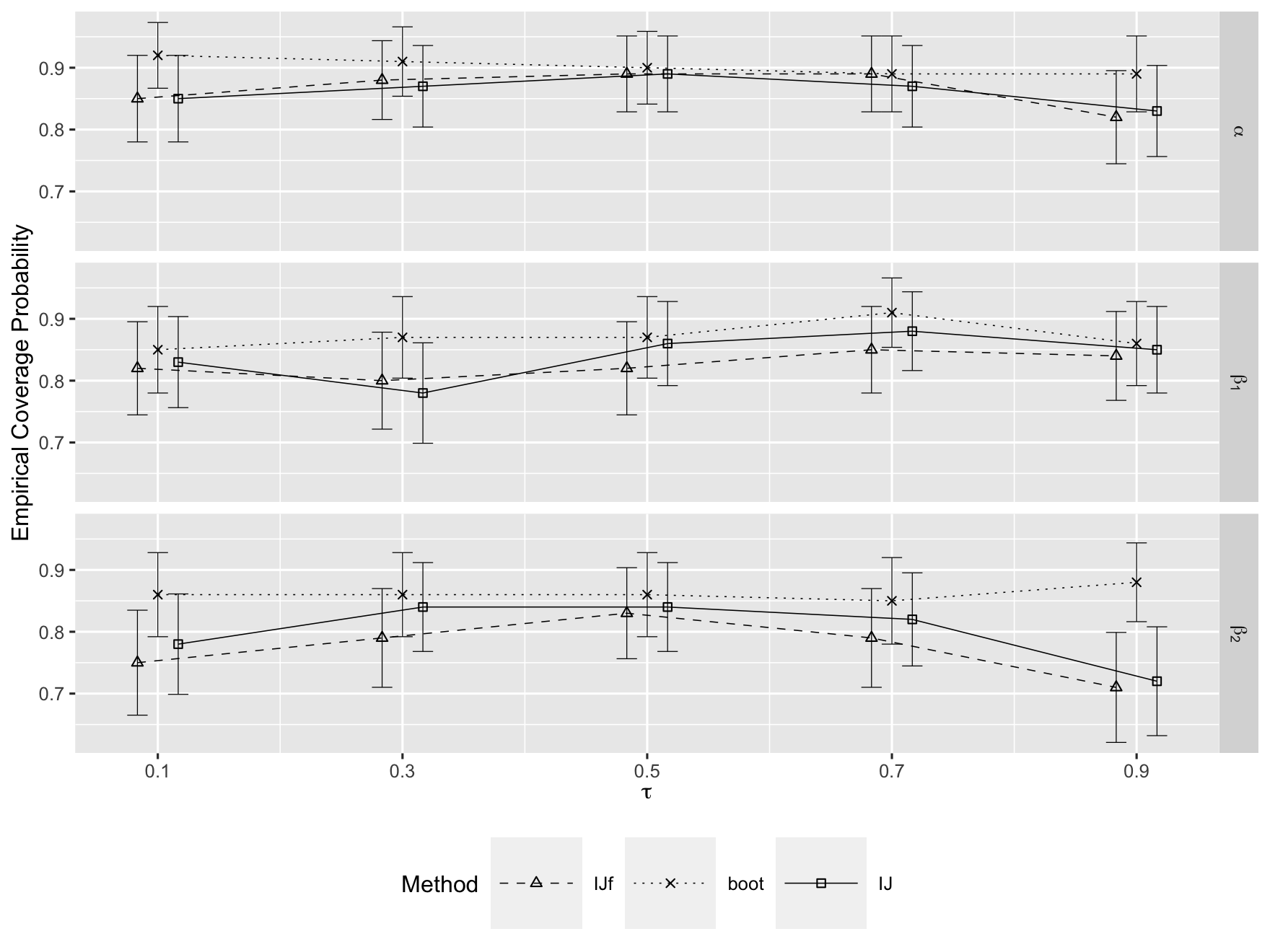}
\caption[Coverage for $I=10$, $J=50$]{\label{fig:I10J50cv}\footnotesize Empirical coverage probability as a function of $\tau$ for $I=10$, $J=50$, $\rho = 0.8$.}
\end{figure}

\begin{figure}[htbp]
\centering
\includegraphics[scale=.18]{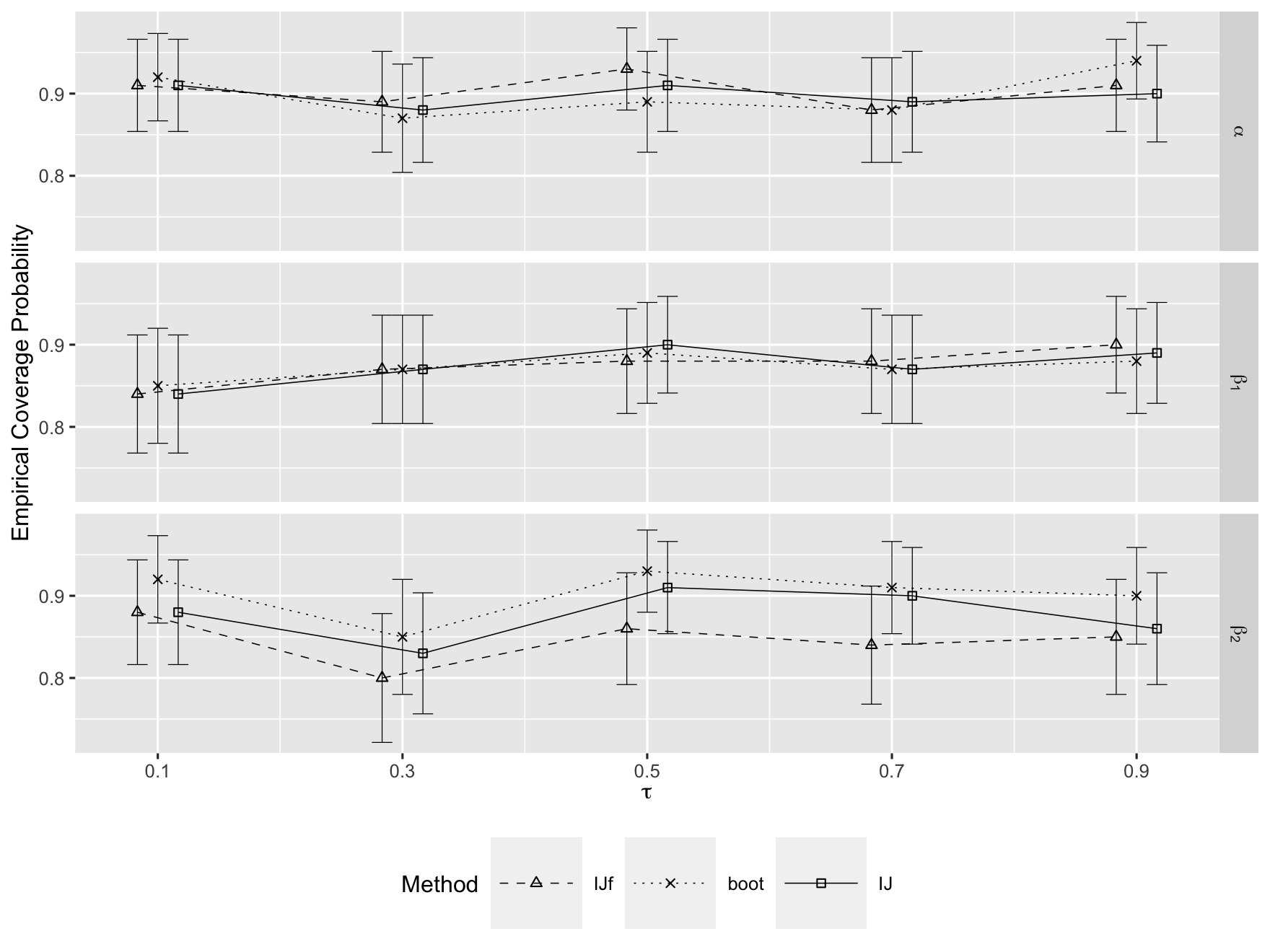}
\caption[Coverage for $I=30$, $J=100$, $\tau = .3$]{\label{fig:I30J100cv}\footnotesize Empirical coverage probability as a function of $\tau$ for $I=30$, $J=100$, $\rho = 0.8$.}
\end{figure}

Here we present the empirical coverage for the smallest sample size considered with $J>10$, i.e., $I=10$, $J=50$ in Figure \ref{fig:I10J50cv}, and for the largest sample size considered, i.e., $I=30$, $J=100$  in Figure \ref{fig:I30J100cv}, both for $\rho=0.8$. In both conditions, coverage tends to be closer to the nominal level for \lab{IJ} than for \lab{IJf} and the confidence intervals mostly span 0.9, except for $\beta_2$, at the extreme quantiles, with the smaller total sample size. 
The results for the other conditions are presented in online Appendix C
\clearpage

\section{Applications}
For the applications we will estimate the scale parameter (use \lab{IJ} rather than \lab{IJf}) as this approach does not rely on finding a suitable value of the scale parameter and tended to perform better than \lab{IJf} in the simulations. We will compare \lab{IJ}-based intervals with intervals based on alternative standard errors for  two real datasets.
\label{sec:realdata}
\subsection{Income and food expenditure}\label{sec:engel}

Here we follow Koenker and Bassett (1982)\nocite{koenker1982food} in analyzing data from \cite{engel1857} on household income and food expenditure. This dataset is also used in the Stata manual \citep{statamanual} for demonstrating their AL-based Bayesian quantile regression. As it is in the public domain, this is a useful benchmark dataset for comparing standard error estimates.

Engel used the data to show that ``The poorer a family, the greater the part of total expenditures must be spent on food,''  later referred to as Engel's law \citep{perthel1975Engel}. The data provided as \cn{Engel1867.dta} with the Stata software consists of annual household income and annual food expenditure, in thousands of Belgian Franks, for 235 European working-class households in the 19th century.

\begin{figure}[htbp]
    \includegraphics[scale=.7]{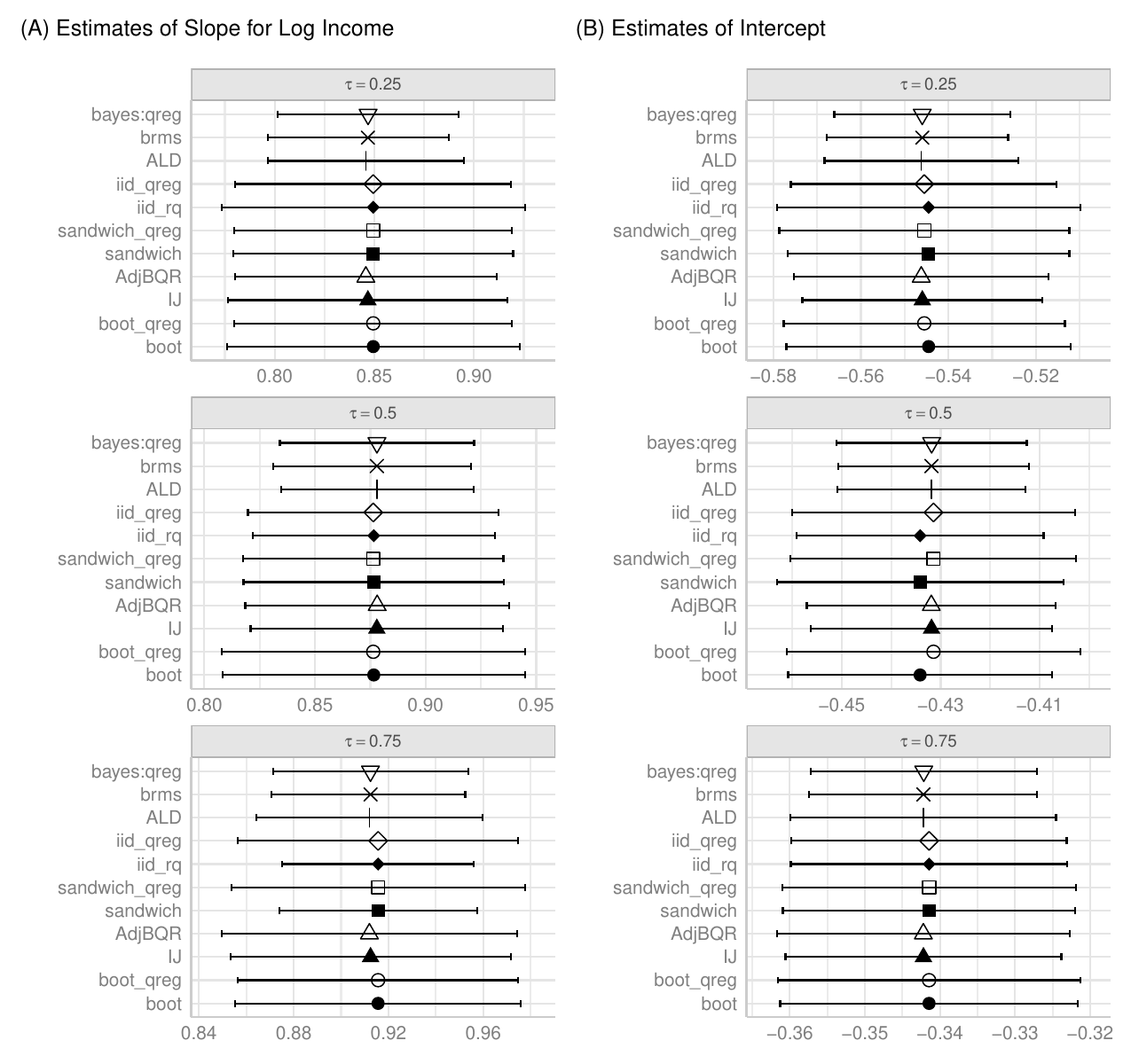}
    \caption{\footnotesize{}Comparison of slope and intercept estimates for log income using various quantile regression methods with approximate 95\% credible intervals/confidence intervals}
    \label{fig:engel_QR_estimates}
\end{figure}

Performing quantile regression of log-expenditure on log-income, Koenker and Bassett (1982)\nocite{koenker1982food} report slope estimates (that they refer to as Engel elasticities) of 0.85, 0.88, and 0.92 for $\tau=0.25$ $\tau=0.50$, and $\tau=0.75$, respectively. We fitted the quantile regression models to Engel’s data using various methods, and Figure~\ref{fig:engel_QR_estimates} shows the point estimates with 95\% credible intervals/confidence intervals  for the slope (panel A) and intercept (panel B) for $\tau = 0.25$, $\tau = 0.50$, and $\tau = 0.75$. The methods used were the same as used in the simulations (\lab{ALD}, here with $\sigma$ set to its MLE at the median, \lab{IJ}, \lab{boot}, \lab{sandwich}, \lab{AdjBQR}, \lab{brms}) plus the following:

\begin{description}
    \item \lab{bayes:qreg}: AL-based Bayesian quantile regression using the  \cn{bayes:qreg} command in Stata \citep{statamanual}. This method uses an inverse Gamma(0.01, 0.01) prior for $\sigma$.
    \item \lab{iid\_qreg}: Frequentist quantile regression using the Stata \cn{qreg} command with the default method for computing standard errors, assuming independent and identically distributed (i.i.d.) units.
    \item \lab{iid\_rq}: Frequentist quantile regression using the \cn{rq()} function in the \cn{quantreg} R package, with the option \cn{se = iid}. The asymptotic covariance matrix for i.i.d. units is computed as described by \cite{koenker1978regression}.
    \item \lab{sandwich\_qreg}: Frequentist quantile regression using the Stata \cn{qreg} command with the \cn{vce(robust)} option to compute the sandwich estimator for non-i.i.d. units by \cite{hendricks1992hierarchical}.
    \item \lab{boot\_qreg}: Frequentist bootstrapped quantile regression using the \cn{bsqreg} command in Stata. This method estimates the variance-covariance matrix by randomly resampling the data \citep{angelis1993analytical}.
\end{description}

The point estimates and standard error estimates for \cn{bayesQR} \citep{benoit2017bayesqr} were so divergent from the other methods that we decided not to include it in Figure~\ref{fig:engel_QR_estimates}. Readers interested in the estimates obtained from \cn{bayesQR} can refer to online Appendix D.

The first notable observation from Figure~\ref{fig:engel_QR_estimates} is that the point estimates for both the slope and intercept are generally similar across methods, whereas the standard error estimates vary significantly. Most importantly, the standard errors produced by the \cn{brms} package in R and the  \cn{bayes:qreg} command in Stata, both of which are publicly available and likely to be chosen by applied researchers, are considerably smaller than for other methods. As shown by the relative error results in Figure~\ref{fig:dif_methods}, this underestimation appears to be a general pattern. The \lab{ALD} method shows similar levels of underestimation of standard errors to \lab{brms} and \lab{bayes:qreg}.

Second, among the AL-based Bayesian quantile regression options, the two methods that apply some form of standard error correction, i.e., \lab{AdjBQR} and \lab{IJ}, generally produce standard errors comparable in magnitude to the frequentist methods using sandwich estimators. The bootstrap standard errors tend to be larger, and the \lab{sandwich} and \lab{iid\_rq} methods underestimate the standard errors for the slope to a degree similar to \lab{brms} when $\tau = 0.75$. This underestimation seems to be related to the local estimate of the sparsity calculation in the \cn{rq()} function of the \cn{quantreg} package and suggests caution when using \cn{quantreg} for extreme quantiles, as the Stata commands do not show this problem.

\begin{figure}[htbp]
\begin{tabular}{c@{\hspace{-1mm}}c}
\includegraphics[scale=.4]{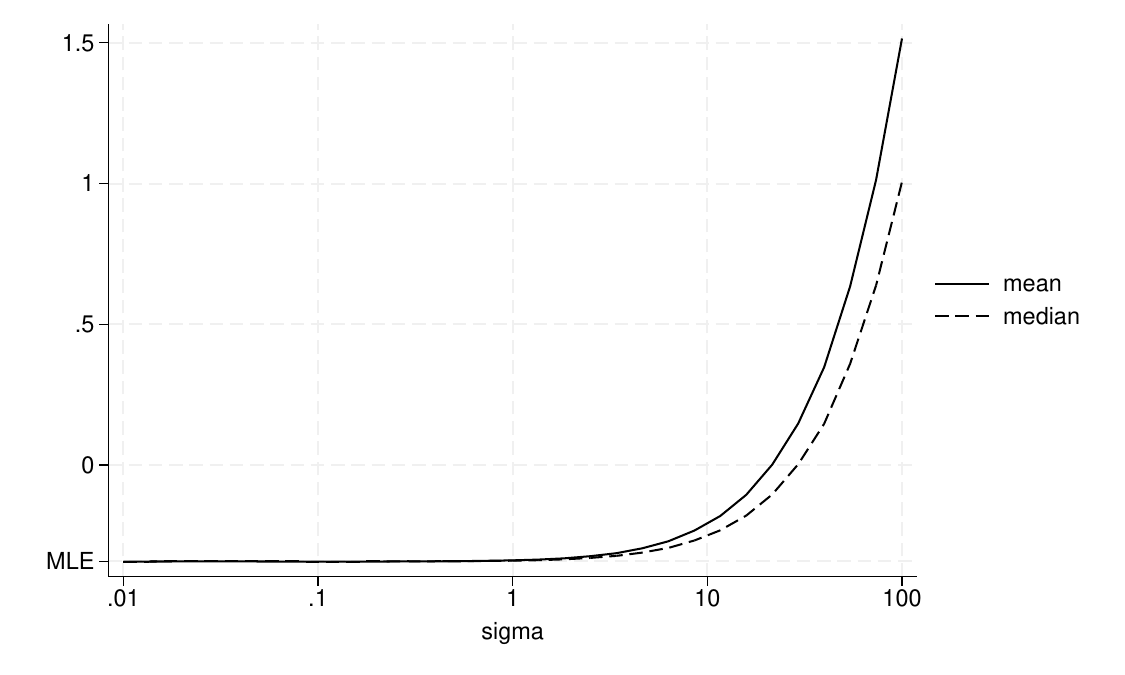}&\includegraphics[scale=.4]{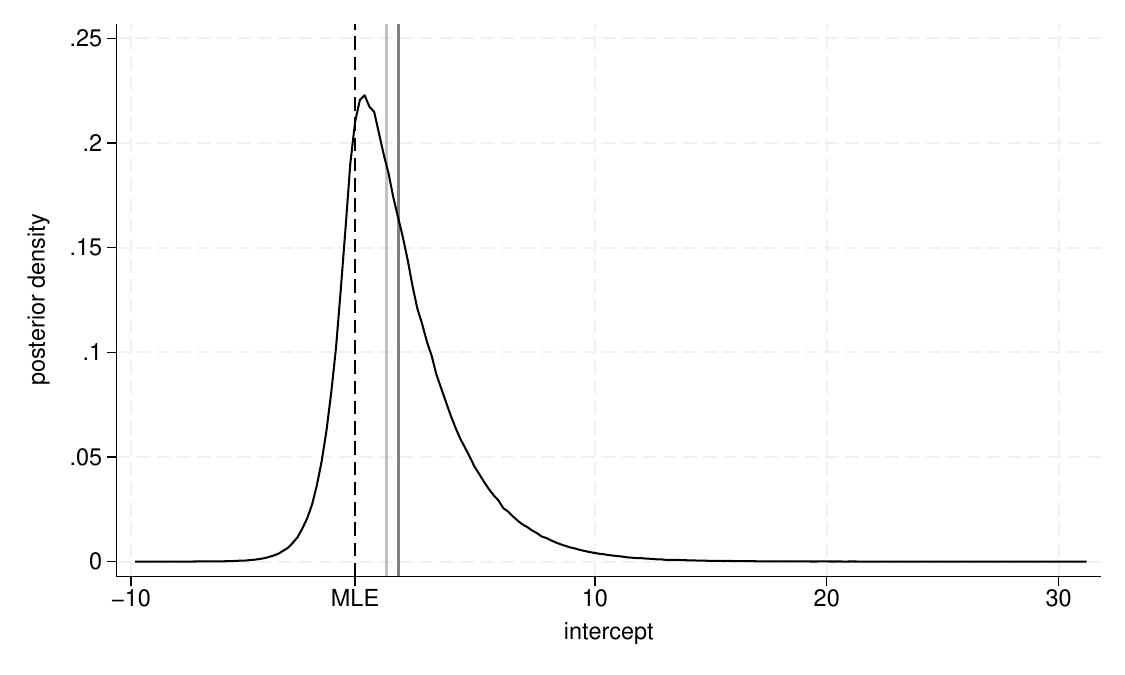}
\end{tabular}
\caption{\footnotesize{}Comparison of posterior mean, median and MLE for $\sigma$ ranging from 0.01 to 100 (left panel) and for $\sigma=100$, with kernel density estimate of posterior density (right panel) for the Engel data with $\tau=0.75$\label{fig:engelcurves}}
\end{figure}

We can also use this dataset to demonstrate that, for an AL likelihood with fixed $\sigma$ and for $\tau=0.75$, the posterior density for the intercept becomes increasingly right-skewed as $\sigma$ increases.
The left panel of Figure~\ref{fig:engelcurves} shows that the difference between posterior mean and median of the intercept increases as $\sigma$ increases and that both point estimates increasingly differ from the MLE or posterior mode. The right panel shows the kernel density estimate of the posterior density for $\sigma=100$, for which the considerable right-skew of the posterior leads to the posterior mean being larger than the posterior median which in turn is larger than the posterior mode and MLE. (Note that $\sigma=100$ seems large, but the same results would be obtained for $\sigma=1$ by changing the scale of the response variable correspondingly.)

\subsection{Project STAR}\label{sec:star}

In this section, we analyze data from the Tennessee Student/Teacher Achievement Ratio (STAR) experiment to compare our IJ standard errors for clustered data with estimates presented by \cite{hagemann2017cluster}, who utilized a bootstrap method to obtain cluster-robust standard errors.

In the 1985--1986 school year, Project STAR randomly assigned incoming kindergarten students in 79 Tennessee schools to one of three class types: small classes (13-17 students), regular-size classes (22-25 students), and regular-size classes with a full-time teacher’s aide. Teachers were also randomly assigned to these class types. The study included 6,325 students across 325 classrooms, with 5,727 students having complete data for analysis. \cite{hagemann2017cluster} aimed to estimate the effect of classroom type on student achievement, with standard errors clustered at the classroom level to account for any unobserved between-classroom heterogeneity due to peer effects and unobserved teacher characteristics.

To replicate Hagemann's (2017)\nocite{hagemann2017cluster} results, we fit a quantile regression model to math and reading test scores from the end of the 1985-1986 school year. In this model, the main explanatory variables of interest are the two treatment dummies:  \cn{small}, an indicator for the student being assigned to a small class, and  \cn{regaide}, an indicator for the student being assigned to a regular size class with an aide. Additionally, we control for a set of pre-treatment covariates specified in \cite{hagemann2017cluster}, along with school fixed effects.

\begin{figure}[htbp]
\centering
\includegraphics[scale=.5]{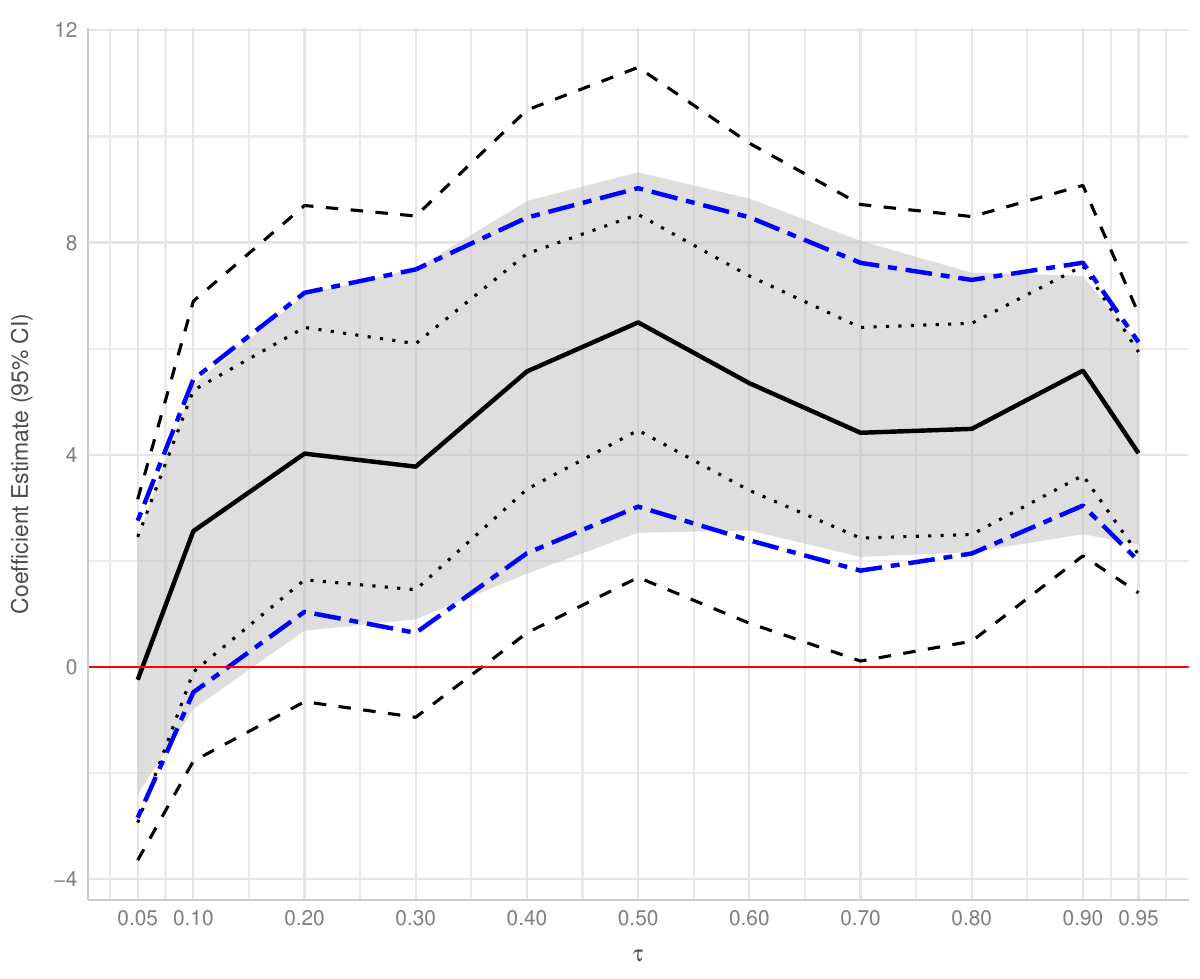}
\caption{\footnotesize{}Quantile regression coefficient estimates and 95\% confidence intervals for the treatment dummy \cn{small} as a function of quantile level $\tau$. \label{fig:star_replication}}
\end{figure}

Figure~\ref{fig:star_replication} replicates Figure 3 in \cite{hagemann2017cluster} and additionally presents our IJ standard error results. The solid black line represents the quantile regression coefficient estimates for the treatment dummy \cn{small} across eleven different quantile levels $\tau$, presented as ticks on the x-axis. The dotted lines show the 95\% pointwise confidence intervals that do not account for clustering in classrooms. These intervals were obtained by specifying the \cn{se = "ker"} option for the \cn{rq()} function in the \cn{quantreg} package, which uses a kernel estimate of the sandwich as proposed by \cite{powell1991estimation}. The gray band shows the 95\% pointwise cluster-robust  wild gradient bootstrap confidence intervals proposed by \cite{hagemann2017cluster}, computed by  the \cn{boot.rq()} function in the \cn{quantreg} package with the \cn{cluster} option. We used 999 bootstrap simulations to calculate these intervals.

The two-dashed blue lines in Figure~\ref{fig:star_replication} represent  95\% pointwise confidence intervals based on our IJ standard errors for clustered data. Here the \cn{brms} package was used for point estimation, with an AL likelihood and a half-\emph{t}(3) prior for $\sigma$, and our  \cn{IJSE} package was used to compute IJ standard errors for clustered data as described in Section \ref{sec:IJclus}. These IJ confidence intervals  closely match those based on the cluster-robust bootstrap. (The outermost black dashed lines, included to replicate Figure 3 in \cite{hagemann2017cluster}, represent a 95\% bootstrap confidence band for the entire quantile regression coefficient function).

\section{Concluding Remarks}
\label{bayes_loo_conclusion}

We have argued and shown empirically that model-based posterior standard deviations are not valid when the scale parameter $\sigma$ is set to an arbitrary constant, a fact that is not widely known even though it has previously been shown by \cite{sriram2015sandwich}, \cite{yang2016posterior} and others. Selecting $\sigma$ to achieve good coverage \citep{syring2019scale} is one option and using the expression for the asymptotic covariance matrix to adjust the standard errors \citep{sriram2015sandwich,yang2016posterior} is another option. However, when the sample size is not very large and a quantile other than the median is of interest, the adjustment by \cite{yang2016posterior} can perform poorly unless $\sigma$ is set to a ``good'' value, one option being its MLE at the median. 

Estimating $\sigma$ along with the model parameters instead of setting it to an arbitrary constant produces better-performing model-based posterior standard deviations, but they were substantially too low  across most of our simulation conditions and worse than the adjusted standard errors by \cite{yang2016posterior} with $\sigma$ set to its MLE at the median.

We propose using Bayesian Infinitesimal Jackknife (IJ) standard errors \citep{giordano2023bayesian} because they have low relative errors regardless of the value that $\sigma$ has been set to. Coverage is also good except at extreme quantiles when $\sigma$ is such that the point estimates are biased. We have seen little discussion in the literature of bias due to a poor choice of $\sigma$. We showed that the bias tends to increase with $\sigma$, as the posterior distributions of the regression parameters become increasingly skewed, particularly for the intercept parameter. 

If $\sigma$ is set to a constant, we therefore recommend comparing the posterior mean and median of the intercept or using other ways of assessing the skewness of the posterior. When skewness is severe, estimation should be repeated with a smaller value of $\sigma$. Alternatively,  $\sigma$ can be estimated along with the model parameters, and our empirical results suggest that this approach works quite well. 

Besides its good performance compared with alternative methods, the IJ approach has several other advantages. One advantage is that IJ standard errors for any function $f(\beta(\tau))$ of the model parameters can be obtained by estimating the posterior covariance matrix between $f(\beta(\tau))$ and the log-likelihood contributions \citep{giordano2023bayesian}. Another advantage, demonstrated in this paper, is that the IJ approach can be used for clustered data, with similar performance as frequentist estimation with bootstrap standard errors for clustered data. Finally, IJ standard errors are applicable for other models besides quantile regression, whenever frequentist standard errors are required or model-based posterior standard deviations are not valid due to model misspecification. Because of  the ease of computation, we believe that IJ standard errors will become a standard component of the Bayesian toolkit. We hope that our \cn{IJSE} package in R will facilitate that development. For any model estimated with \cn{brms}, for which log-likelihood contributions have been computed, our package will compute IJ standard errors for clustered or independent data. At the very least, they can be used to evaluate whether the Bayesian model is well calibrated.

\small{}

\normalsize{}
\appendix
\clearpage

\section*{Online appendices}
\setcounter{page}{1}
\setcounter{figure}{0}
\section*{Appendix A: Derivation of approximation for Bayesian estimates in resampled data}

Here we outline the derivation of the result in equation~(\ref{eq:BIJ}), following section 2 of \citet{efron2015frequentist}.

To start, the posterior expectation for the reweighted sample can be written as
\begin{equation}
 E(\theta | D, w)
=
\frac{\int \theta \exp (L(D | \theta, w)) p(\theta) d\theta}
{\int \exp (L(D | \theta, w)) p(\theta)d\theta},
\label{eq:posexp}
\end{equation}
where $L(D|\theta,w) = \sum_i \ell_i(D|\theta)$ is the log-likelihood for the reweighted data. To simplify notation, we will use $\ell_i(\theta):= \ell(D|\theta)$ and $L(\theta,w):= L(D|\theta,w)$.

We can derive the required derivatives by
defining $M(w)$ to be the numerator of (\ref{eq:posexp}) and
$N(w)$ to be the corresponding denominator. 
The partial derivatives of $M(w)$  and $N(w)$ with respect to $w$ are
\begin{align}
& M^{\prime}(w) = \int L^{\prime}(\theta, w)\theta \exp(L(\theta, w))p(\theta) \mathrm{d} \theta = N(w) E [L^{\prime}(\theta, w)\theta]\\
& N^{\prime}(w) = \int L^{\prime}(\theta, w) \exp(L(\theta, w))p(\theta) \mathrm{d} \theta = N(w) E [L^{\prime}(\theta, w)],
\end{align}
where the expectations are the posterior expectations (over the parameters given the data) and  $L^{\prime}(\theta,w)$  is
equal to the vector of log-likelihood contributions from the units, $L^{\prime}(\theta,w)=\ell(\theta):=(\ell_1(\theta),\ldots,\ell_n(\theta))^\intercal$  since
$$
\left. \frac{d \sum_i w_i \ell_i( \theta)}{d w_{i}}\right|_{w=1} = \ell_i(\theta).
$$

We now use the fact that $(\frac{M}{N})^{\prime}=(\frac{M} {N})\left(\frac{M^{\prime}}{M}  -\frac{N^{\prime}}{N}\right)$ to obtain
$$
\begin{aligned}
\left.\frac{d E(\theta | D, w)}{d w^{\intercal}}\right |_{w=1_{n}} &=\frac{M(w)}{N(w)}\left\{\frac{ N(w) E [L^{\prime}(\theta, w)\theta]}{M(w)}-\frac{N(w) E [L^{\prime}(\theta, w)]}{N(w)}\right\} \\
&=E(\theta|D,w)\left\{\frac{ E [\ell(\theta)\theta]}{E(\theta| D, w)}-E [\ell(\theta)]\right\} \\
&= E (\ell(\theta)\theta)-E(\theta|D,w)]E [\ell(\theta)] \\
&=\operatorname{cov}_{\theta|D}[\theta, \ell(\theta) ].
\end{aligned}
$$

\section*{Appendix B: Further simulation results for independent data}
In this section, we provide all the results that are not shown in the main text for simulation results for independent data.

\begin{figure}[htbp]
    \centering
    \includegraphics[scale=.19]{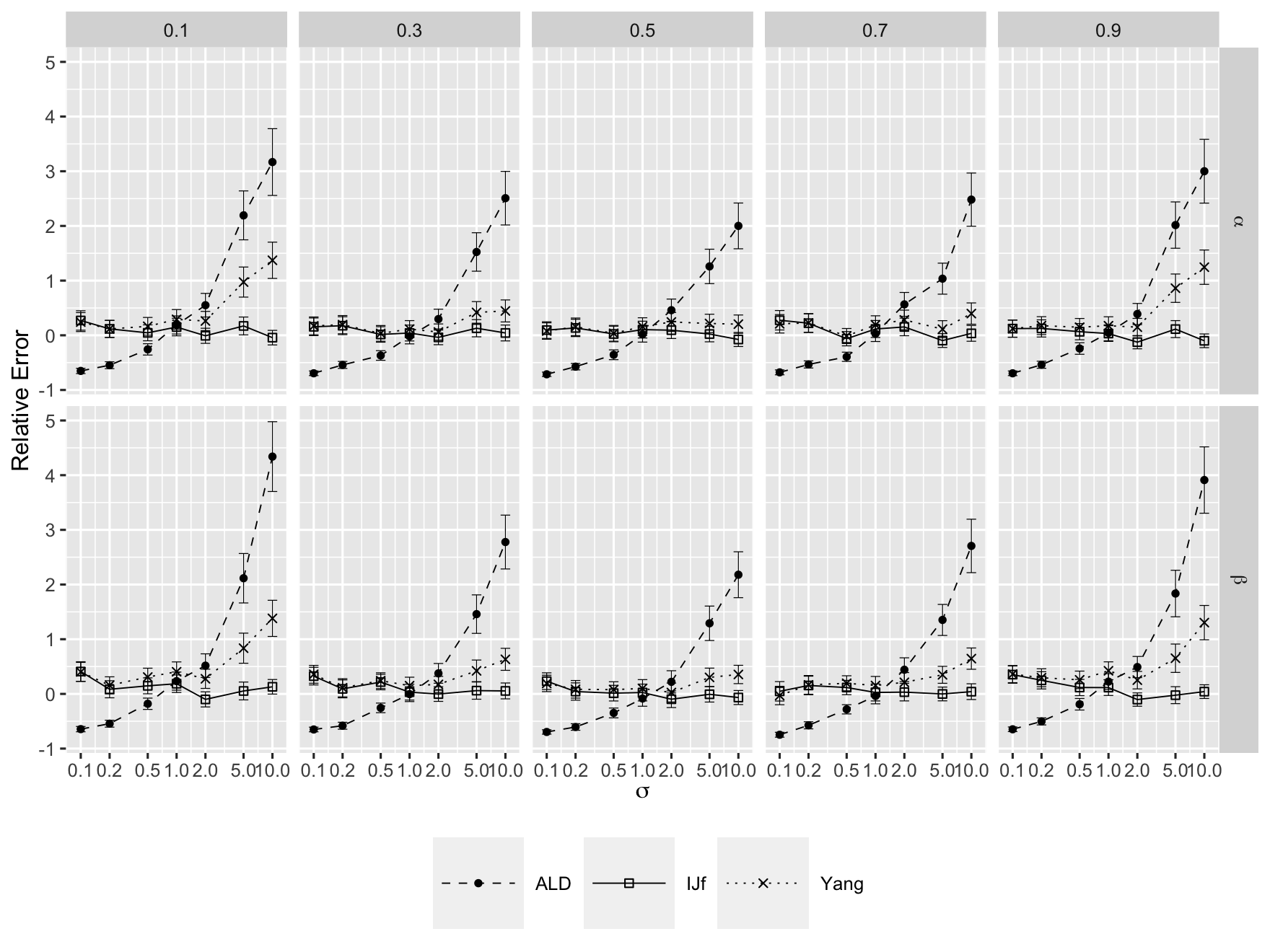}
    \caption{Relative error of different types of standard errors as a function of scale parameters for Model 1 with $n=200$.}
    \label{fig:re_200m1}
\end{figure}

\begin{figure}[htbp]
    \centering
    \includegraphics[scale=.19]{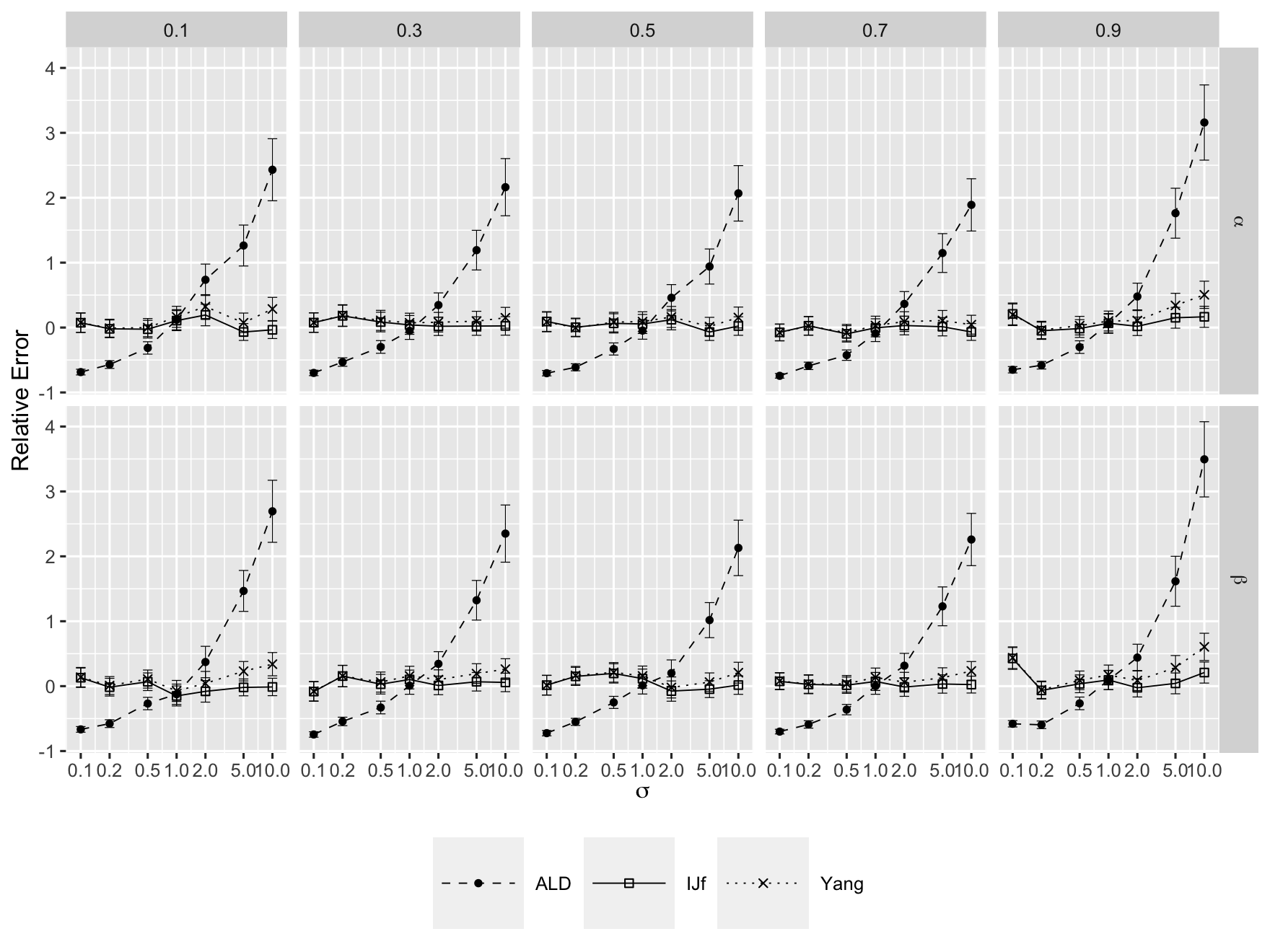}
    \caption{\footnotesize{}Relative error of different types of standard errors as a function of scale parameters for Model 1 with $n=1000$.}
    \label{fig:re_1000m1}
\end{figure}

\begin{figure}[htbp]
    \centering
    \includegraphics[scale=.19]{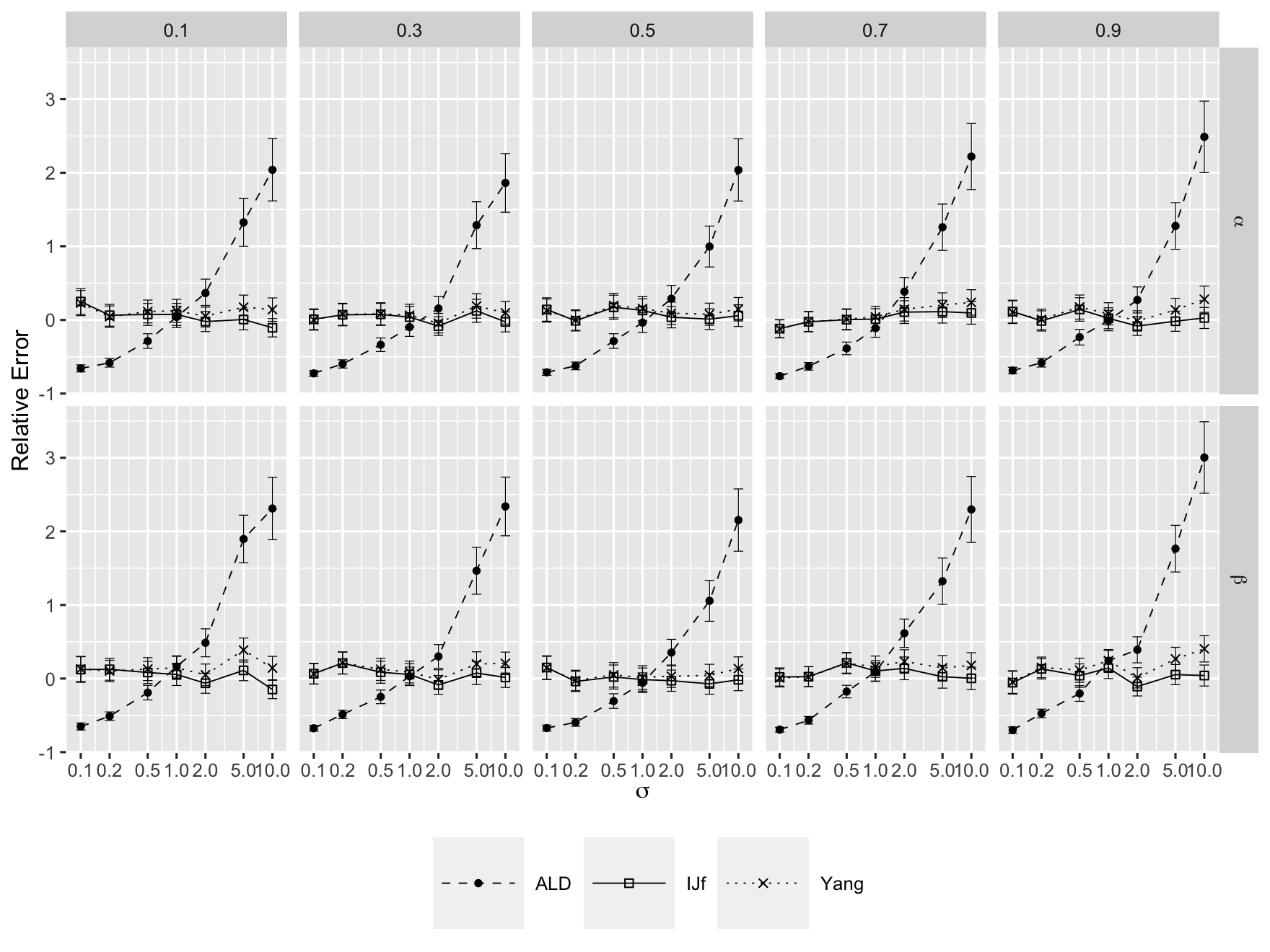}
    \caption{\footnotesize{}Relative error of different types of standard errors as a function of scale parameters for Model 2 with $n=1000$.}
    \label{fig:re_1000m2}
\end{figure}

\begin{figure}[hptp]
  \footnotesize
 \vspace{-6mm}
\begin{center}
\begin{tabular}{c}
\includegraphics[scale=.19]{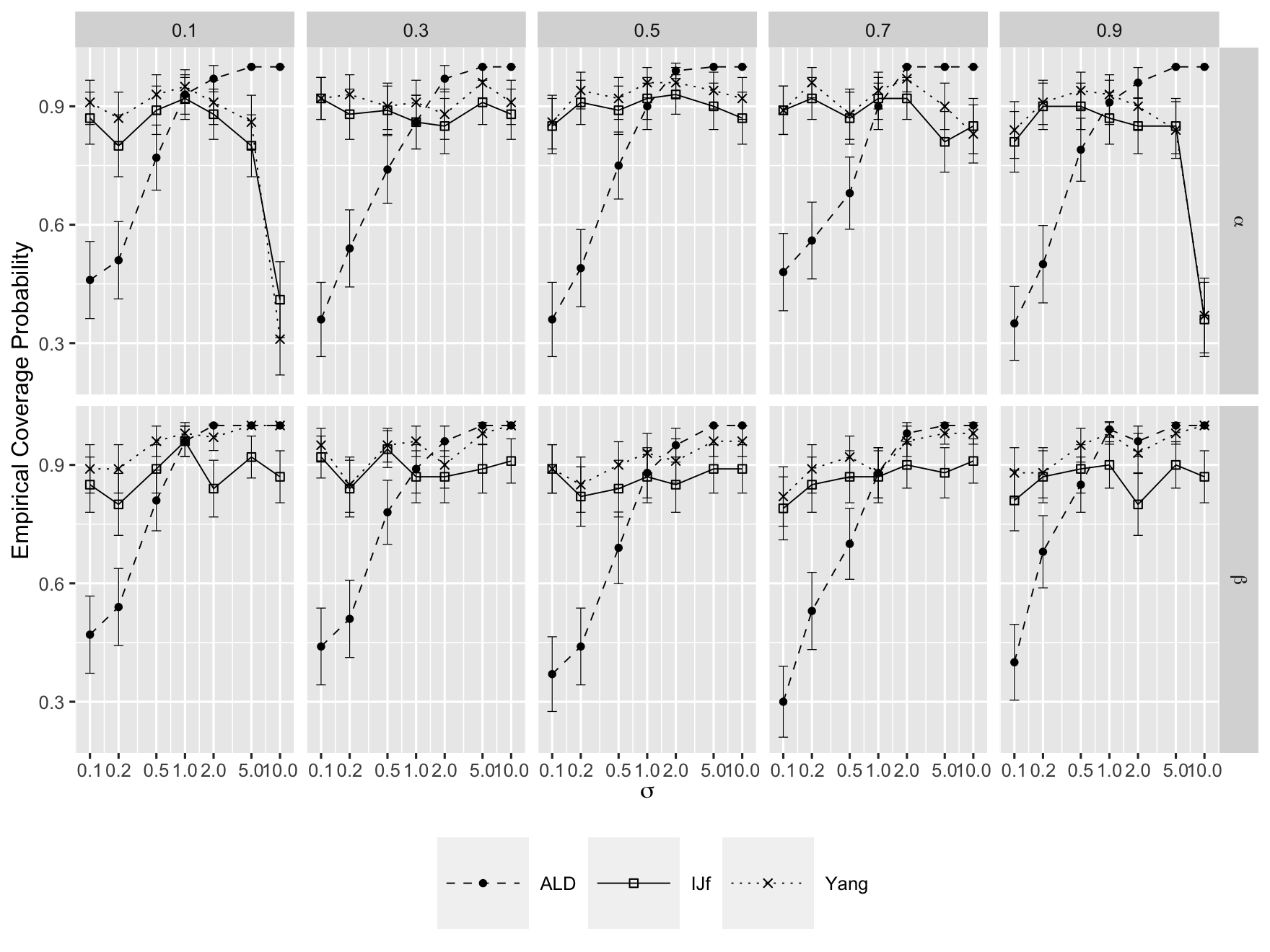}
\end{tabular}
\end{center}
\vspace{-10mm}
\caption[Coverage for different SE methods for $n=200$]{\footnotesize\label{fig:ecr_200m1} Empirical coverage for different SE methods as a function of the scale parameter $\sigma$ for different quantiles ($\tau$) for Model 1 with $n=200$.}
\end{figure}

\begin{figure}[hptp]
  \footnotesize
 \vspace{-6mm}
\begin{center}
\begin{tabular}{c}
\includegraphics[scale=.19]{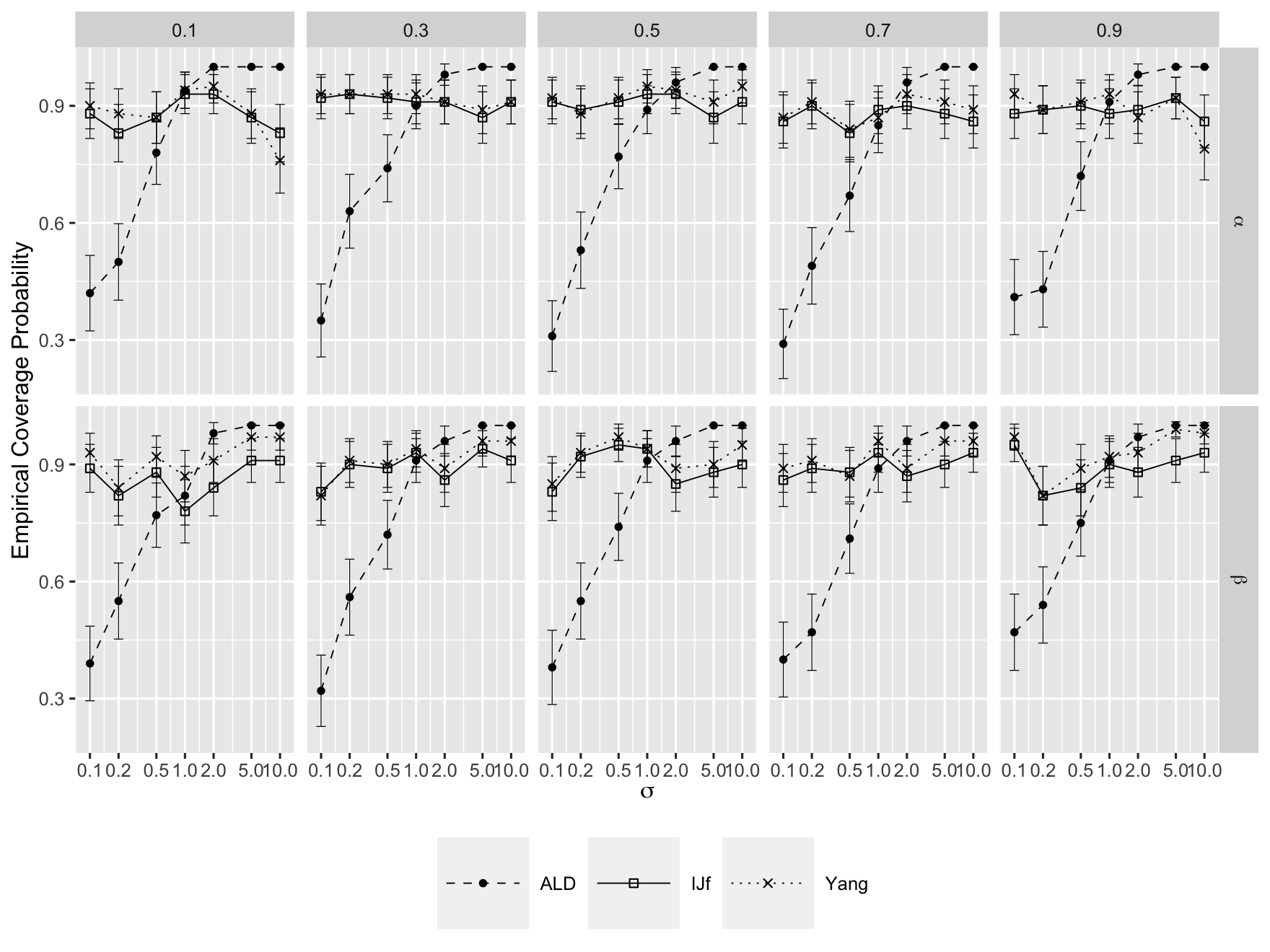}\\
\end{tabular}
\end{center}
\vspace{-10mm}
\caption[Coverage for different SE methods for $n=1000$]{\footnotesize\label{fig:ecr_1000m1} Empirical coverage for different SE methods as a function of the scale parameter $\sigma$ for different quantiles ($\tau$) for Model 1 with $n=1000$.}
\end{figure}

\begin{figure}[hptp]
  \footnotesize
 \vspace{-6mm}
\begin{center}
\begin{tabular}{c}
\includegraphics[scale=.19]{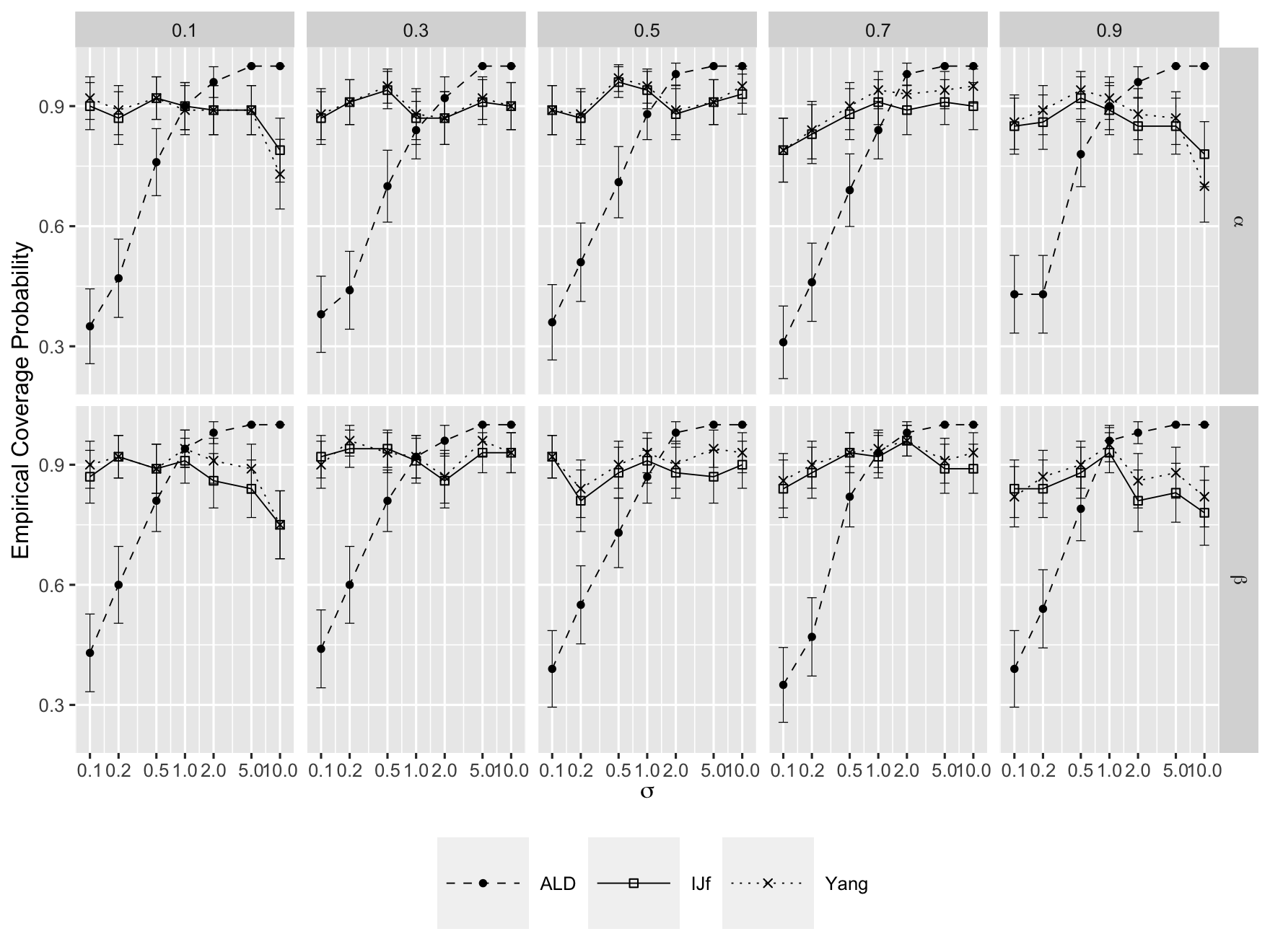}
\end{tabular}
\end{center}
\vspace{-10mm}
\caption[Coverage for different SE methods for $n=1000$]{\footnotesize\label{fig:ecr_1000m2} Empirical coverage for different SE methods as a function of the scale parameter $\sigma$ for different quantiles ($\tau$) for Model 2 with $n=1000$.}
\end{figure}

Relative error for methods that set $\sigma$ to a constant are shown for Model 1 with $n=200$  in Figure~\ref{fig:re_200m1}, and the analogous results for $n=1000$ for Model 1 and 2 are shown in Figures~\ref{fig:re_1000m1} and ~\ref{fig:re_1000m2}, respectively.
Empirical coverage for the same methods are shown for Model 1 with $n=200$ in Figure~\ref{fig:ecr_200m1}, and the analogous results for $n=1000$ for Model 1 and 2 are shown in Figures~\ref{fig:ecr_1000m1} and ~\ref{fig:ecr_1000m2}, respectively.

\begin{figure}
    \centering
    \includegraphics[scale=.19]
    {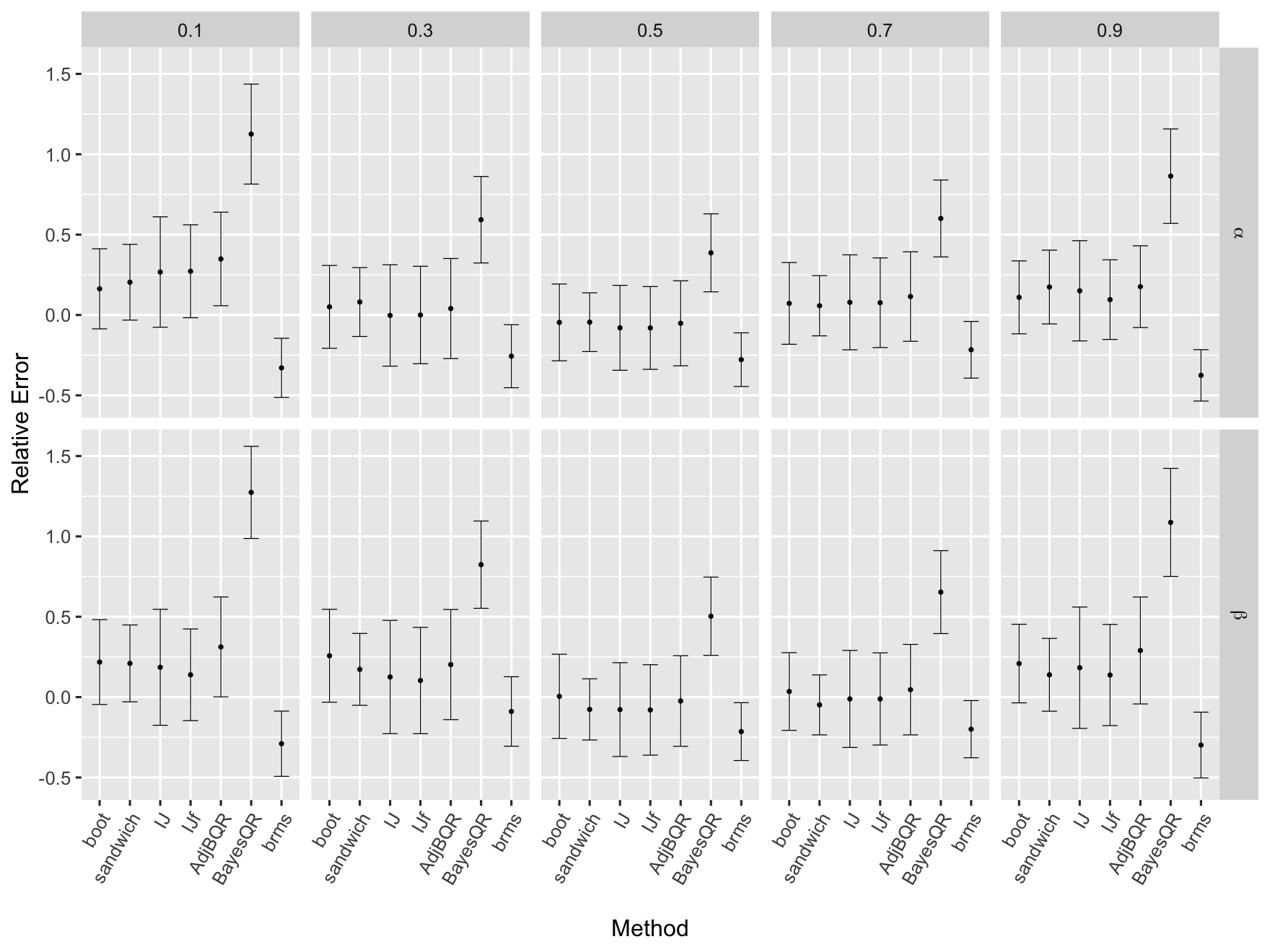}
    \caption{Relative error with approximate 95\% confidence intervals for different methods with $n=200$ and Model 1.}
    \label{fig:dif_methods_200m1}
\end{figure}

\begin{figure}
    \centering
    \includegraphics[scale=.19]
    {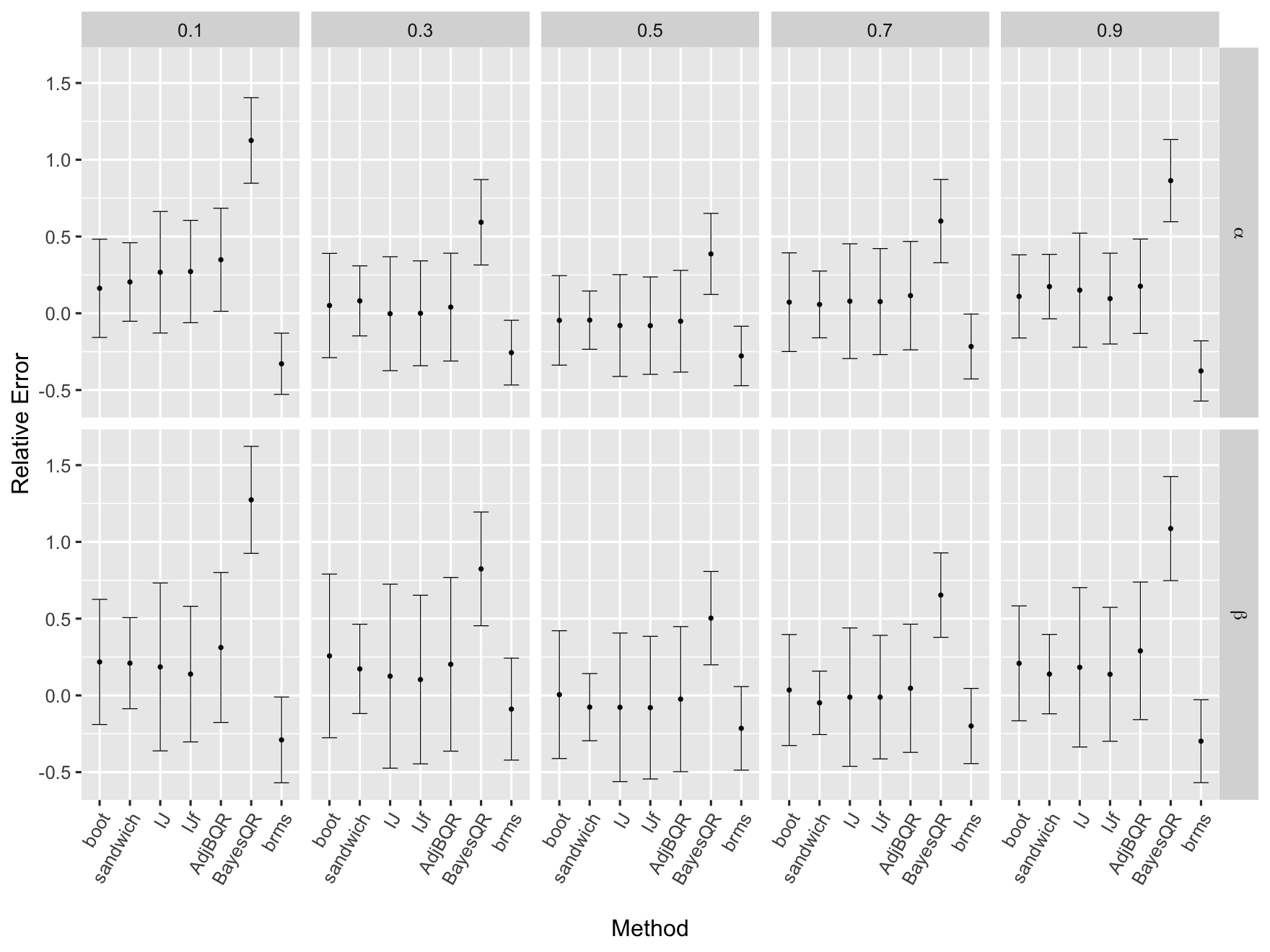}
    \caption{Relative error with approximate 95\% confidence intervals for different methods with $n=1000$ and Model 1.}
    \label{fig:dif_methods_1000m1}
\end{figure}

\begin{figure}
    \centering
    \includegraphics[scale=.19]
    {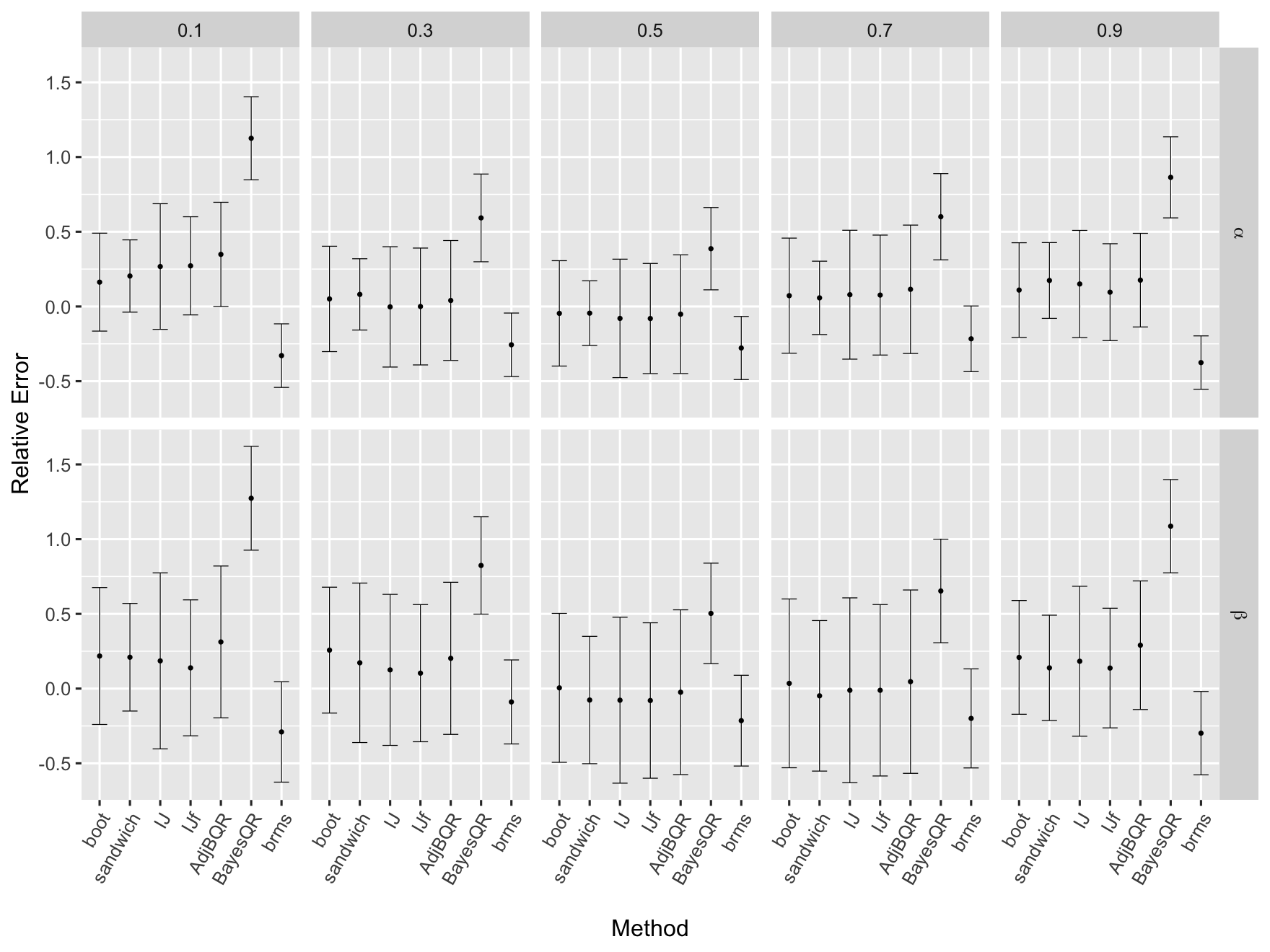}
    \caption{Relative error with approximate 95\% confidence intervals for different methods with $n=1000$ and Model 2.}
    \label{fig:dif_methods_1000m2}
\end{figure}

Relative error for the methods that do not fixe $\sigma$ are shown for Model 1 with $n=200$  in Figure~\ref{fig:dif_methods_200m1} and  for $n=1000$ for Model 1 and 2  in Figures~\ref{fig:dif_methods_1000m1} and ~\ref{fig:dif_methods_1000m2}, respectively.

\clearpage

\section*{Appendix C: Further simulation results for clustered data}
In this section, we provide all the results that are not shown in the main text for simulation results for clustered data.

\begin{figure}[htbp]
 \footnotesize
\begin{center}
\begin{tabular}{c}
\includegraphics[scale=.6]{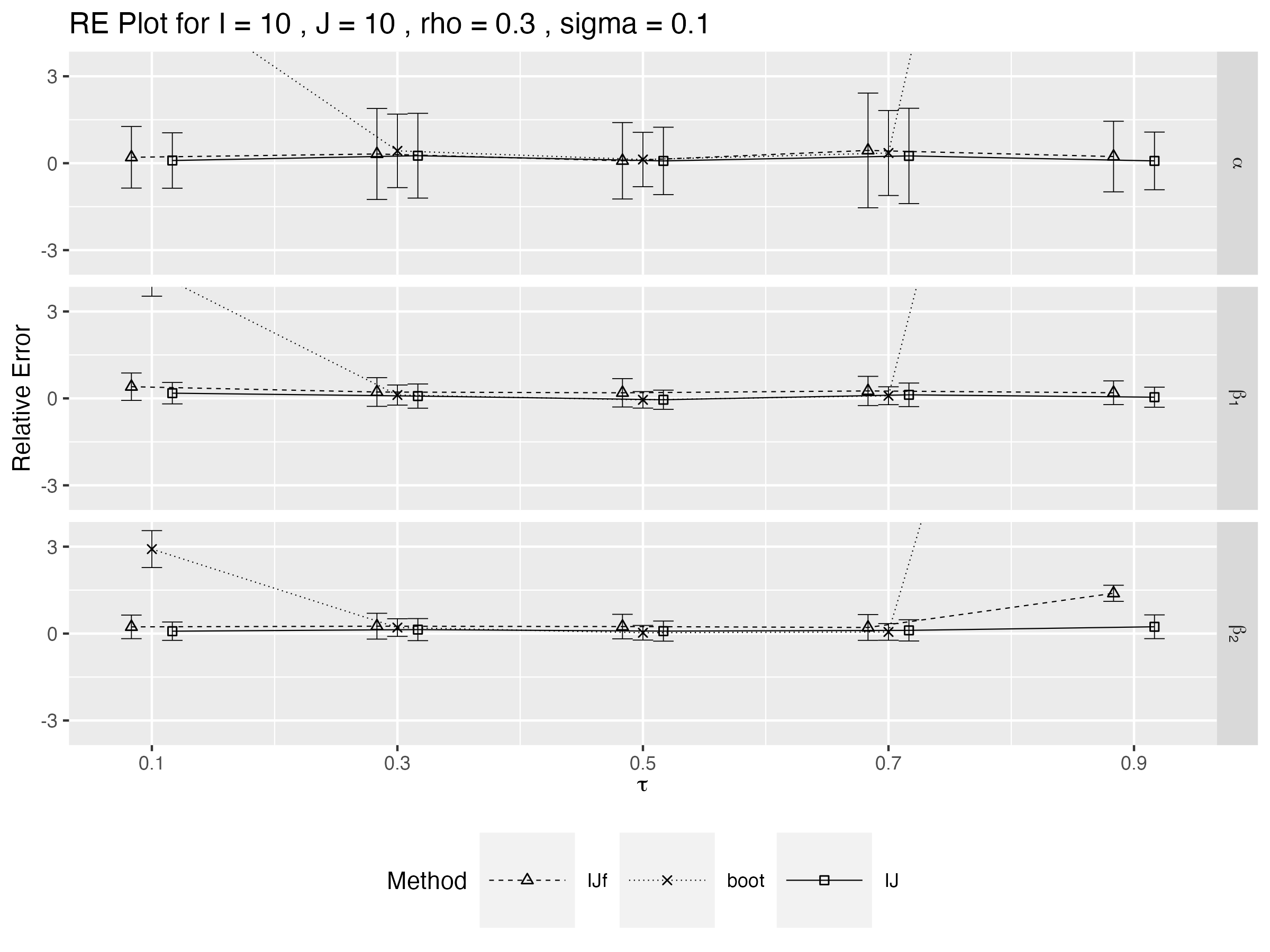}\\
\vspace{-1mm}
\end{tabular}
\end{center}
\vspace{-10mm}
\caption[Relative error estimates for $I=10$, $J=10$, $\rho = .3$]{\label{fig:I10J10re03}\footnotesize Relative error as a function of $\tau$ for $I=10$, $J=10$, $\rho = .3$, and $\sigma=0.1$.}
\end{figure}

\begin{figure}[htbp]
 \footnotesize
 \vspace{-6mm}
\begin{center}
\begin{tabular}{c}
\includegraphics[scale=.6]{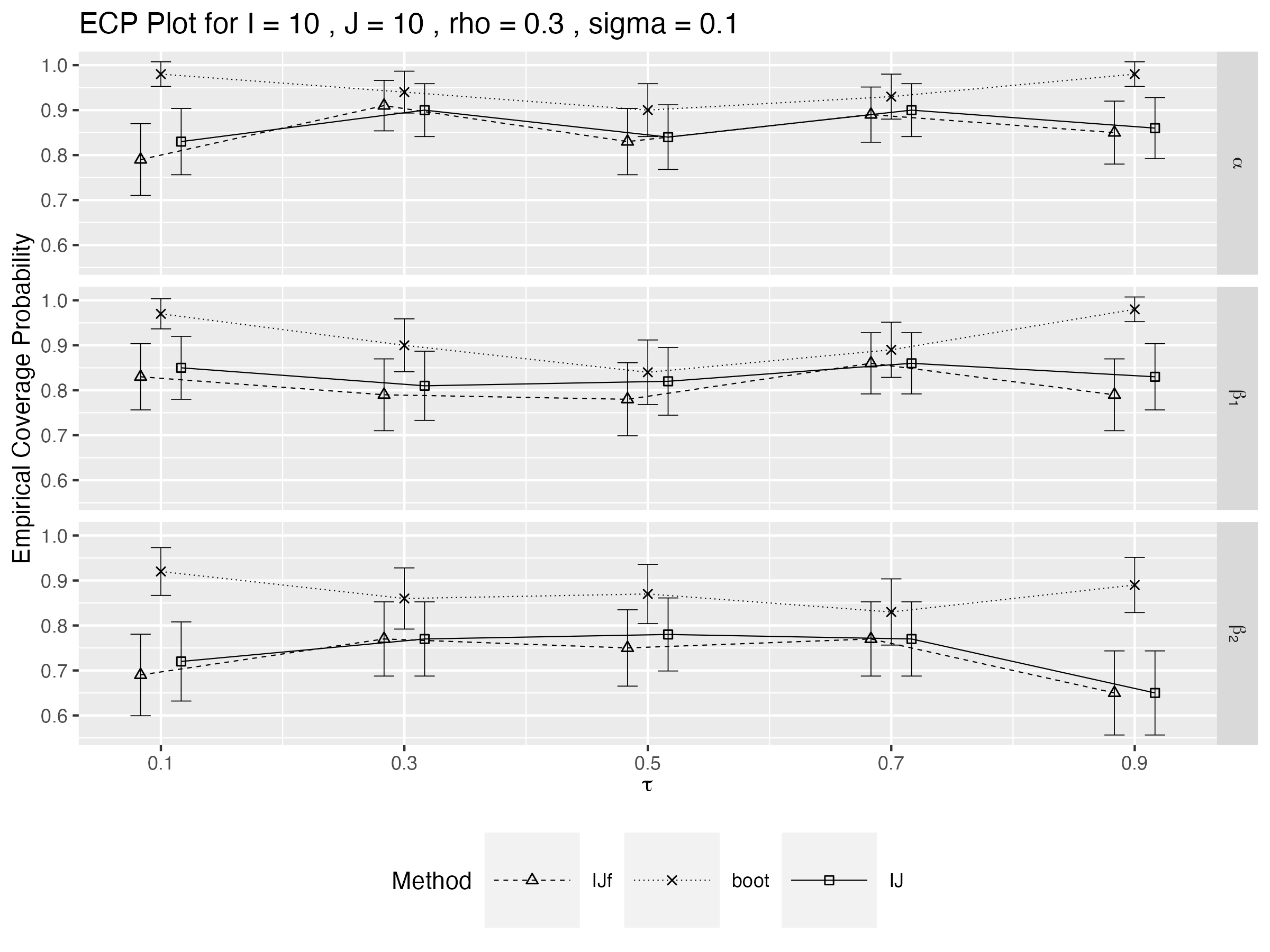}\\
\vspace{-1mm}
\end{tabular}
\end{center}
\vspace{-10mm}
\caption[Empirical coverage for $I=10$, $J=10$, $\rho = .3$]{\label{fig:I10J10cv03}\footnotesize Empirical coverage probability as a function of $\tau$ for $I=10$, $J=10$, $\rho = .3$, and $\sigma=0.1$.}
\end{figure}

\begin{figure}[htbp]
 \footnotesize
 \vspace{-6mm}
\begin{center}
\begin{tabular}{c}
\includegraphics[scale=.6]{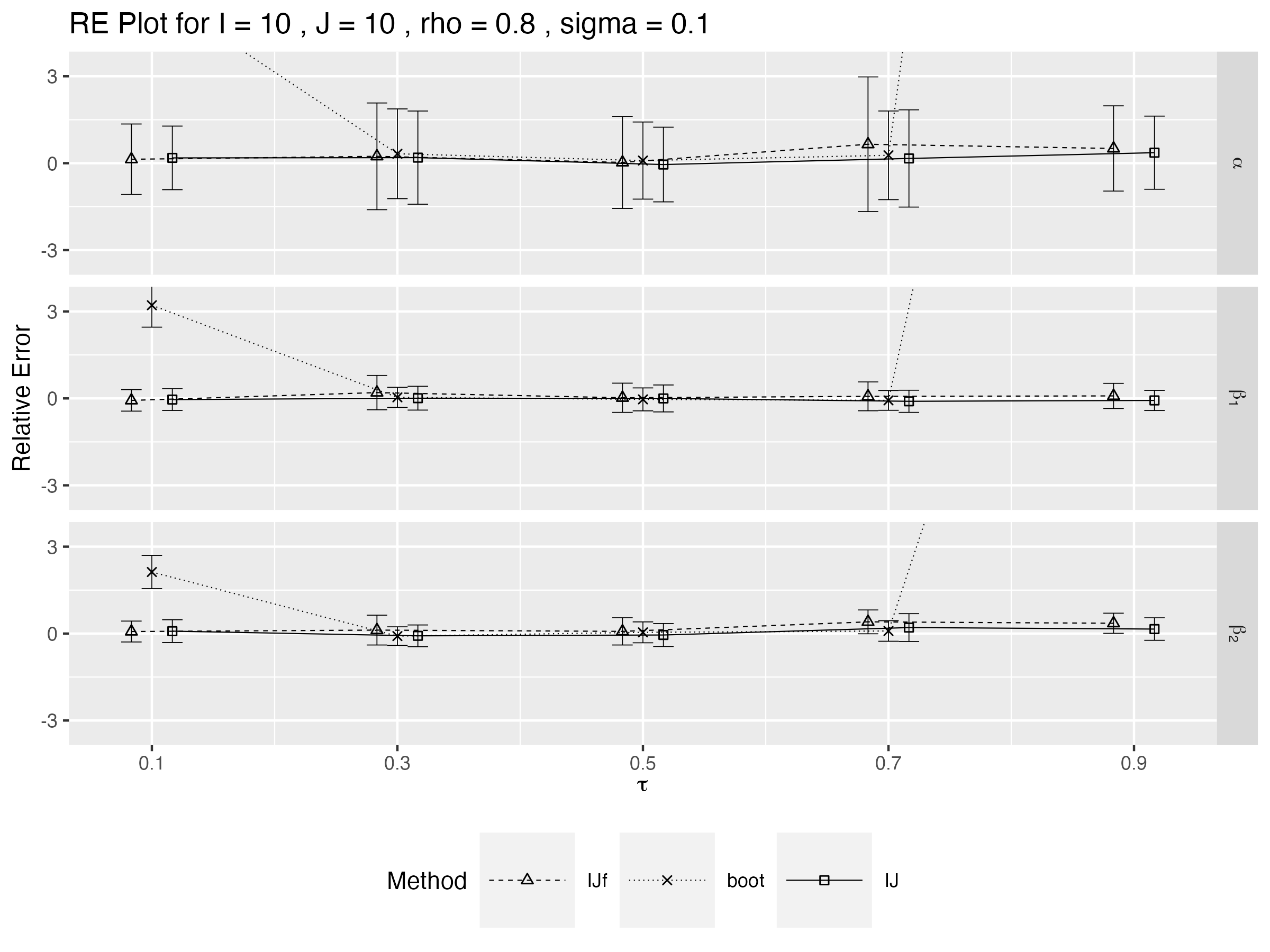}\\
\vspace{-1mm}
\end{tabular}
\end{center}
\vspace{-10mm}
\caption[Relative error estimates for $I=10$, $J=10$, $\rho = .8$]{\label{fig:I10J10re08}\footnotesize Relative error as a function of $\tau$ for $I=10$, $J=10$, $\rho = .8$, and $\sigma=0.1$.}
\end{figure}

\begin{figure}[htbp]
 \footnotesize
 \vspace{-6mm}
\begin{center}
\begin{tabular}{c}
\includegraphics[scale=.6]{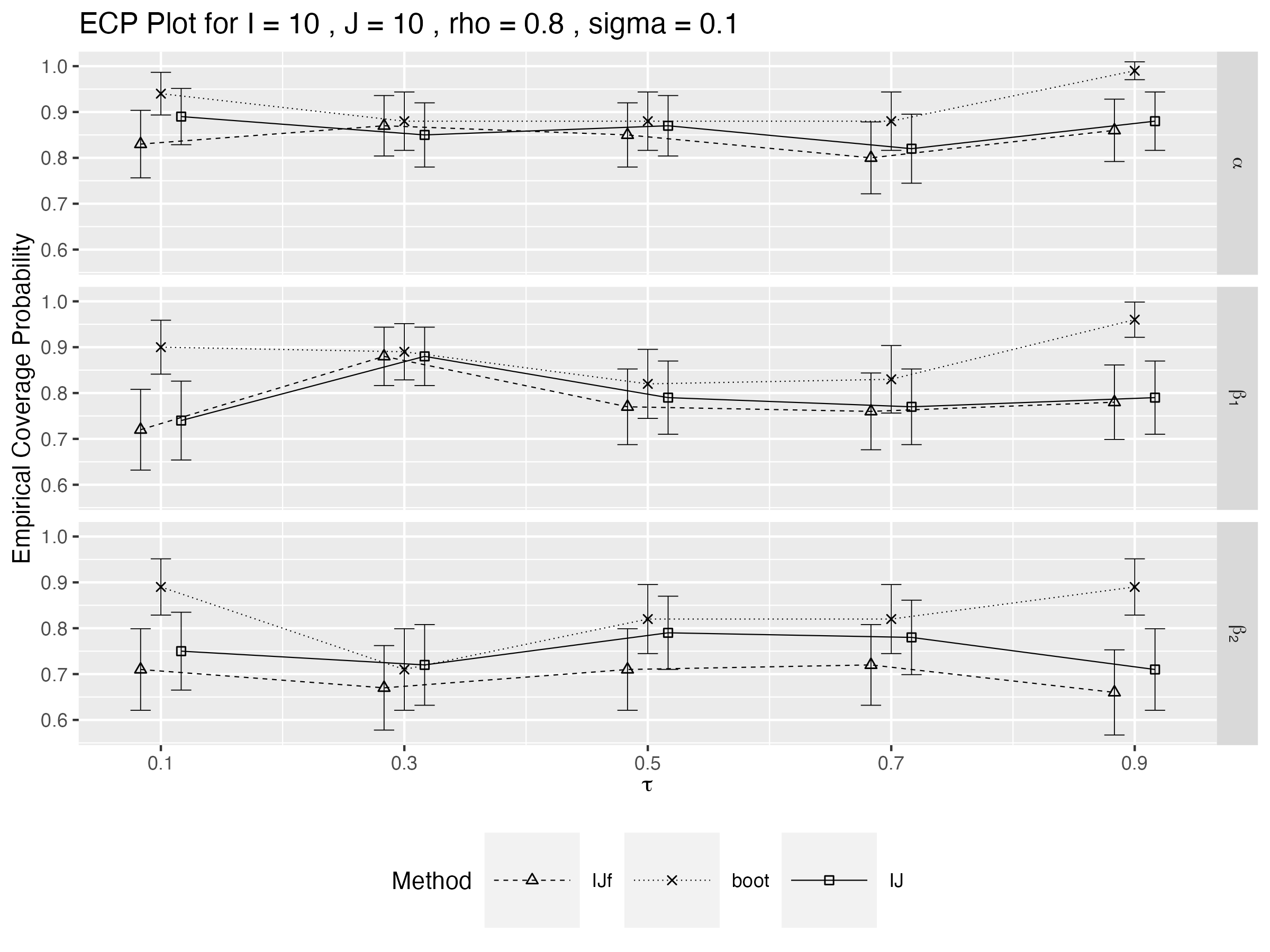}\\
\vspace{-1mm}
\end{tabular}
\end{center}
\vspace{-10mm}
\caption[Empirical coverage for $I=10$, $J=10$, $\rho = .8$]{\label{fig:I10J10cv08}\footnotesize Empirical coverage probability as a function of $\tau$ for $I=10$, $J=10$, $\rho = .8$, and $\sigma=0.1$.}
\end{figure}

\begin{figure}[htbp]
 \footnotesize
 \vspace{-6mm}
\begin{center}
\begin{tabular}{c}
\includegraphics[scale=.6]{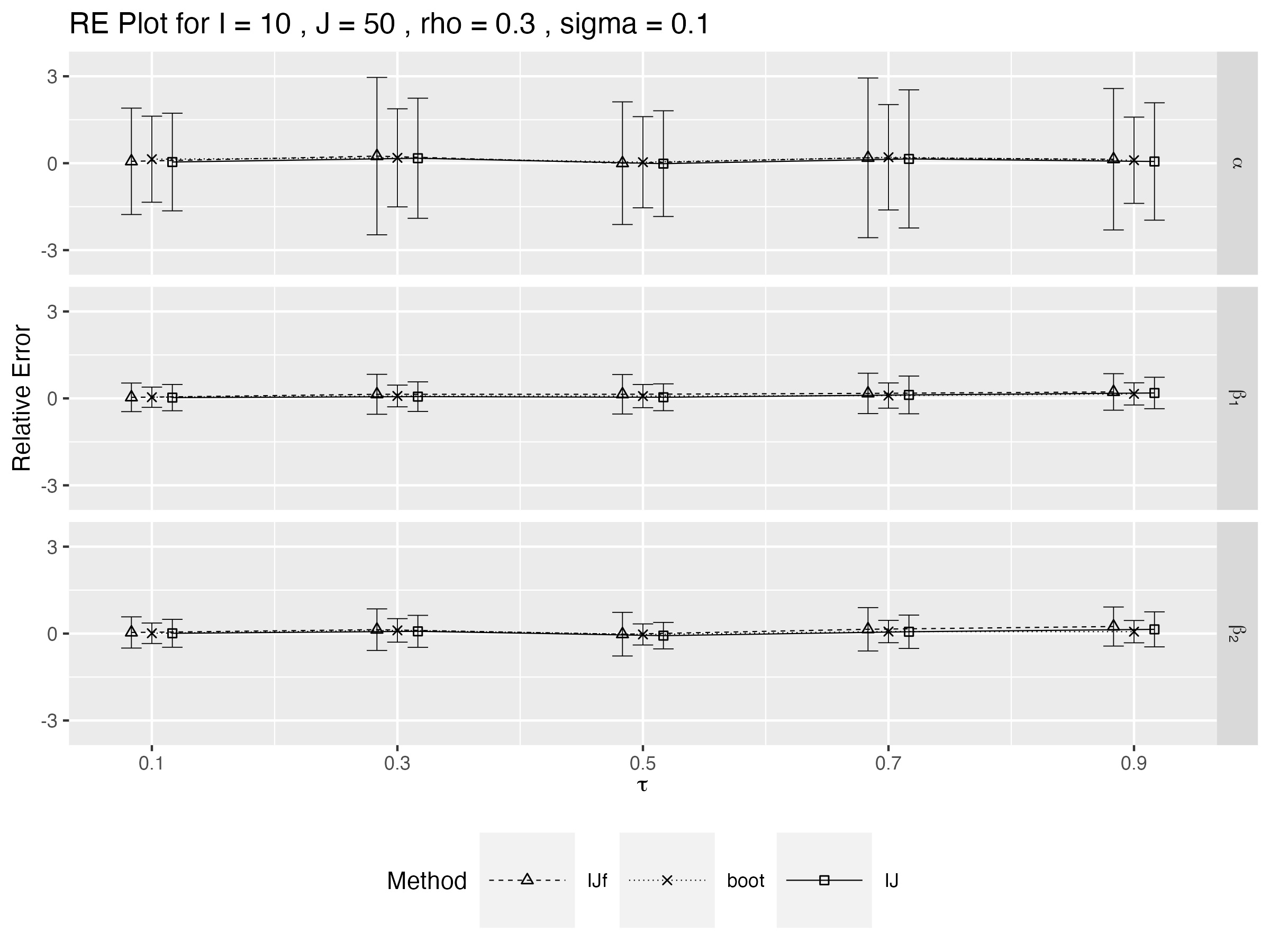}\\
\vspace{-1mm}
\end{tabular}
\end{center}
\vspace{-10mm}
\caption[Relative error estimates for $I=10$, $J=50$, $\rho = .3$]{\label{fig:I10J50re03}\footnotesize Relative error as a function of $\tau$ for $I=10$, $J=50$, $\rho = .3$, and $\sigma=0.1$.}
\end{figure}

\begin{figure}[htbp]
 \footnotesize
 \vspace{-6mm}
\begin{center}
\begin{tabular}{c}
\includegraphics[scale=.6]{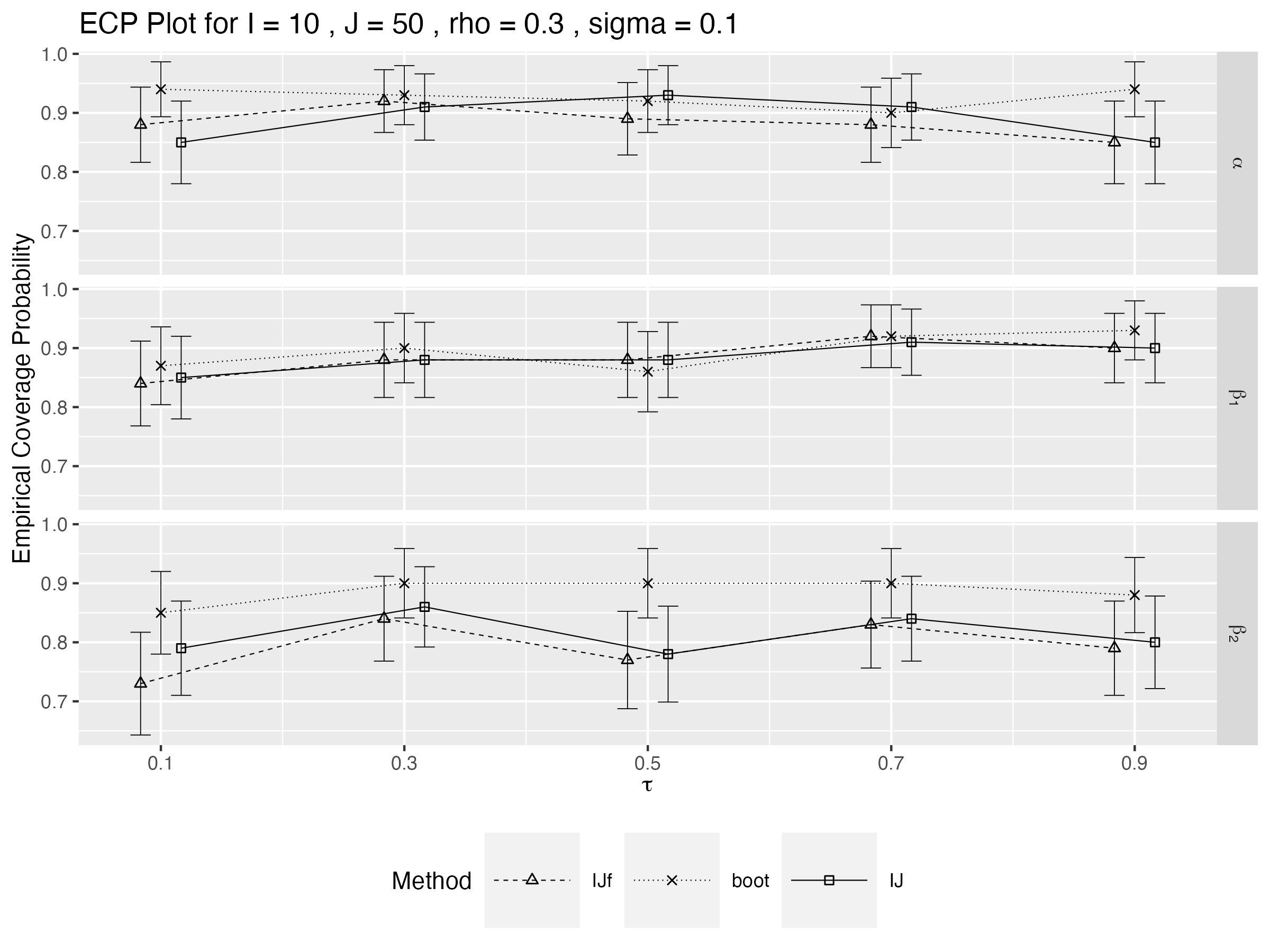}\\
\vspace{-1mm}
\end{tabular}
\end{center}
\vspace{-10mm}
\caption[Empirical coverage for $I=10$, $J=50$, $\rho = .3$]{\label{fig:I10J50cv03}\footnotesize Empirical coverage probability as a function of $\tau$ for $I=10$, $J=50$, $\rho = .3$, and $\sigma=0.1$.}
\end{figure}

 \begin{figure}[htbp]
 \footnotesize

\begin{center}
\begin{tabular}{c}
\includegraphics[scale=.6]{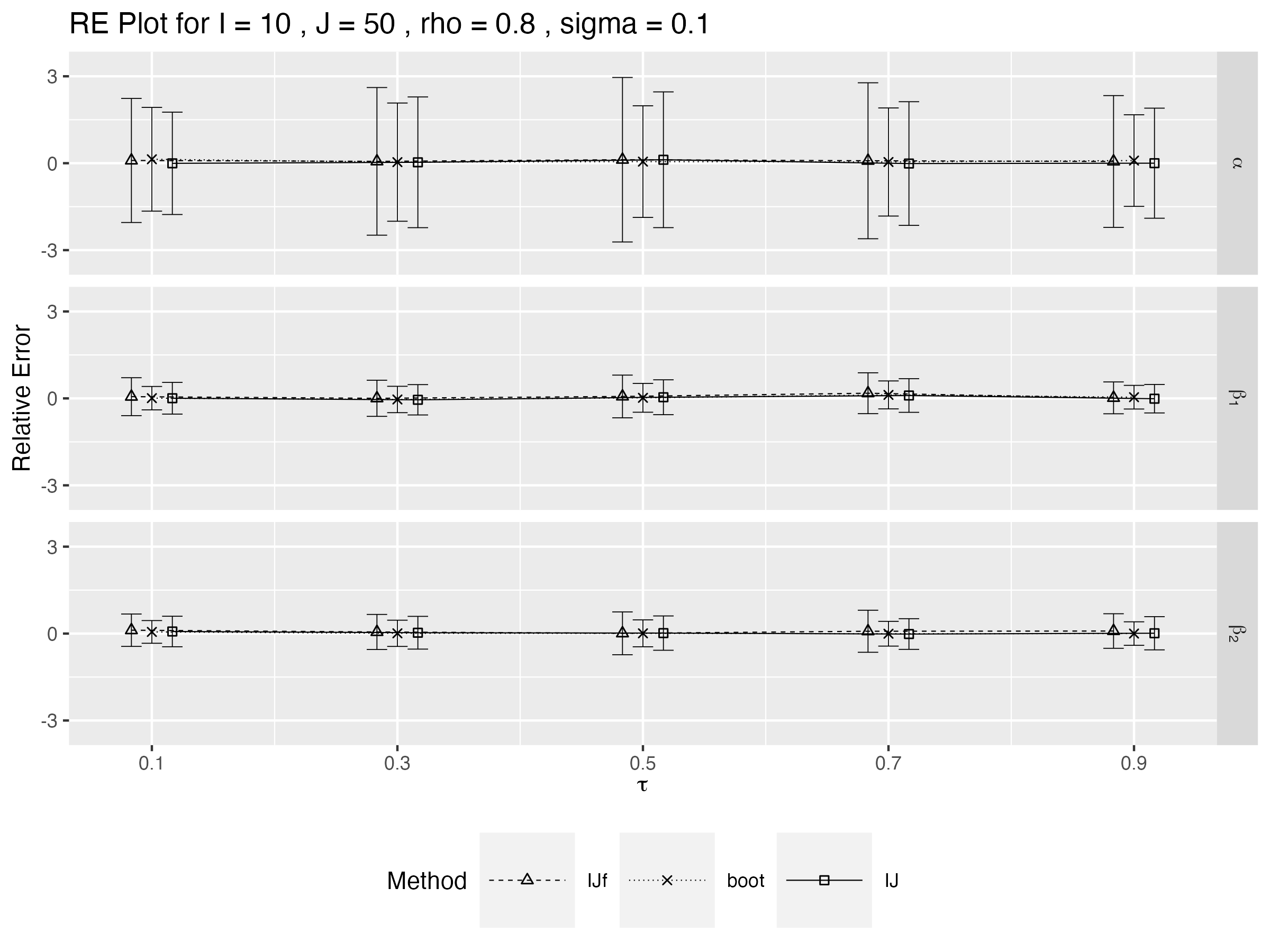}\\
\vspace{-1mm}

\end{tabular}
\end{center}
\vspace{-10mm}
\caption[Relative error estimates for $I=10 $, $J=50$]{\label{fig:I10J50re}\footnotesize Relative error as a function of $\tau$ for $I=10$, $J=50$, $\rho = .8$, and $\sigma=0.1$.}
\end{figure}

\begin{figure}[htbp]
 \footnotesize
 \vspace{-6mm}
\begin{center}
\begin{tabular}{c}
\includegraphics[scale=.6]{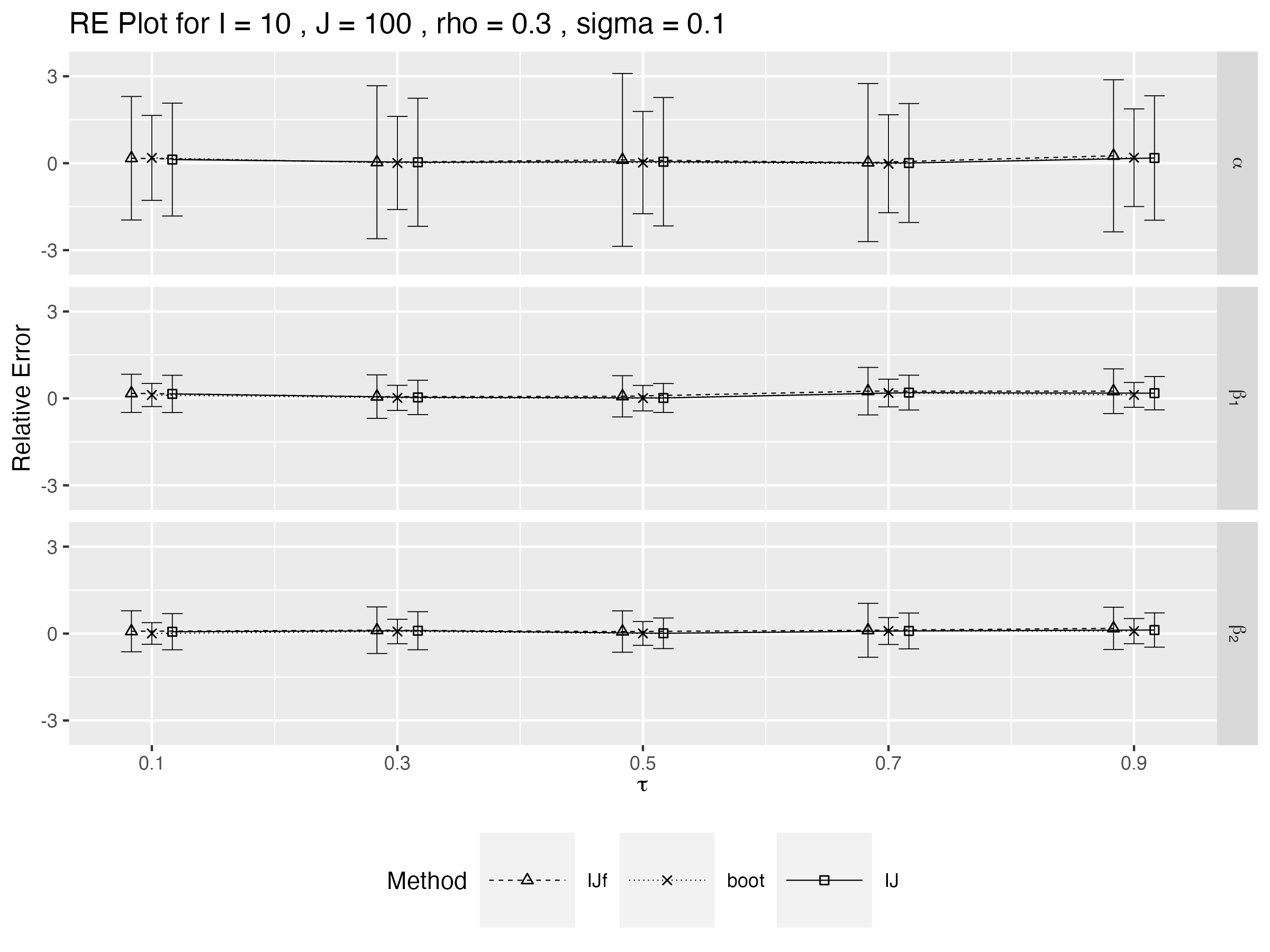}\\
\vspace{-1mm}
\end{tabular}
\end{center}
\vspace{-10mm}
\caption[Relative error estimates for $I=10$, $J=100$, $\rho = .3$]{\label{fig:I10J100re03}\footnotesize Relative error as a function of $\tau$ for $I=10$, $J=100$, $\rho = .3$, and $\sigma=0.1$.}
\end{figure}

\begin{figure}[htbp]
 \footnotesize
 \vspace{-6mm}
\begin{center}
\begin{tabular}{c}
\includegraphics[scale=.6]{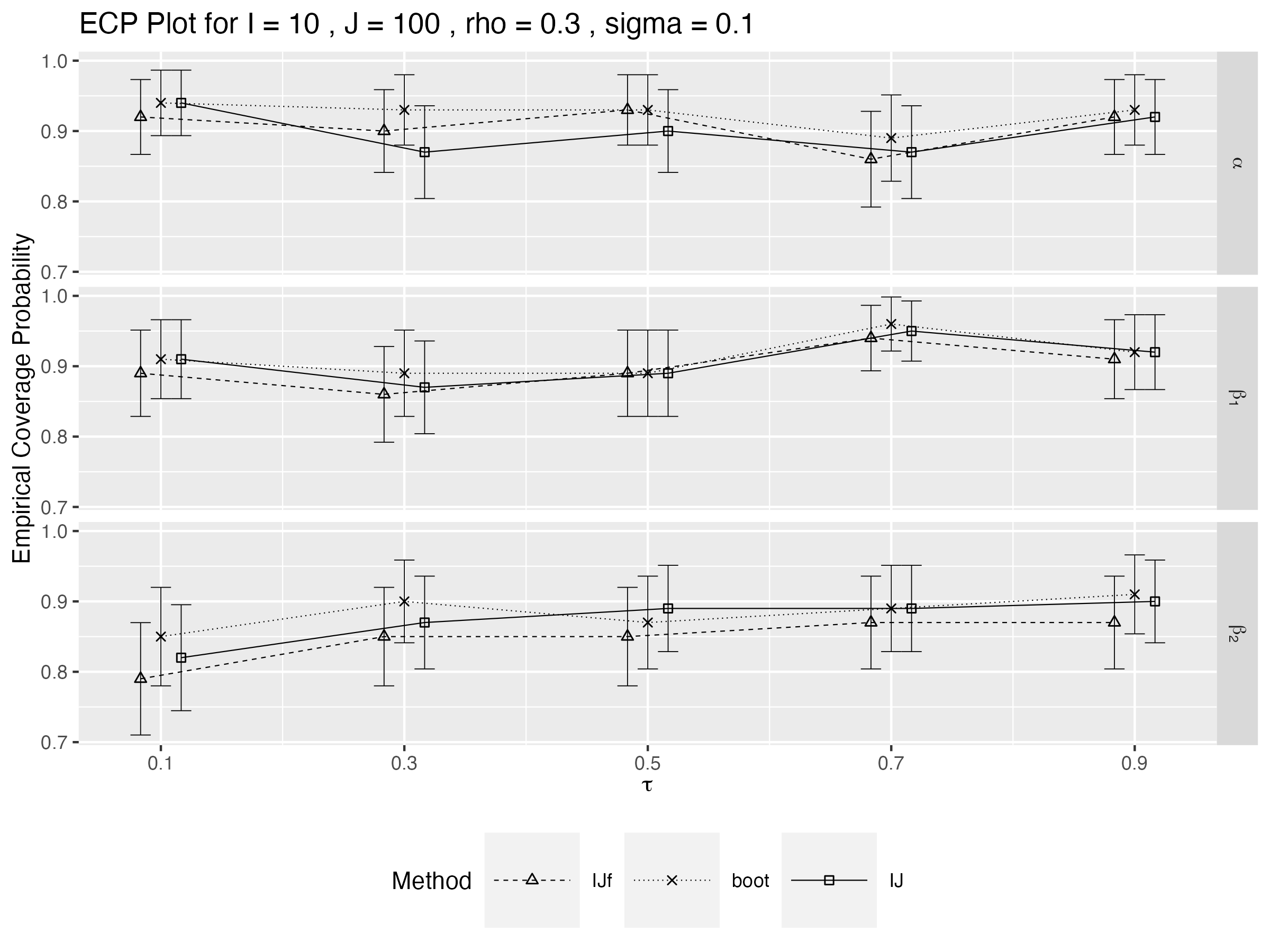}\\
\vspace{-1mm}
\end{tabular}
\end{center}
\vspace{-10mm}
\caption[Empirical coverage for $I=10$, $J=100$, $\rho = .3$]{\label{fig:I10J100cv03}\footnotesize Empirical coverage probability as a function of $\tau$ for $I=10$, $J=100$, $\rho = .3$, and $\sigma=0.1$.}
\end{figure}

\begin{figure}[htbp]
 \footnotesize
 \vspace{-6mm}
\begin{center}
\begin{tabular}{c}
\includegraphics[scale=.6]{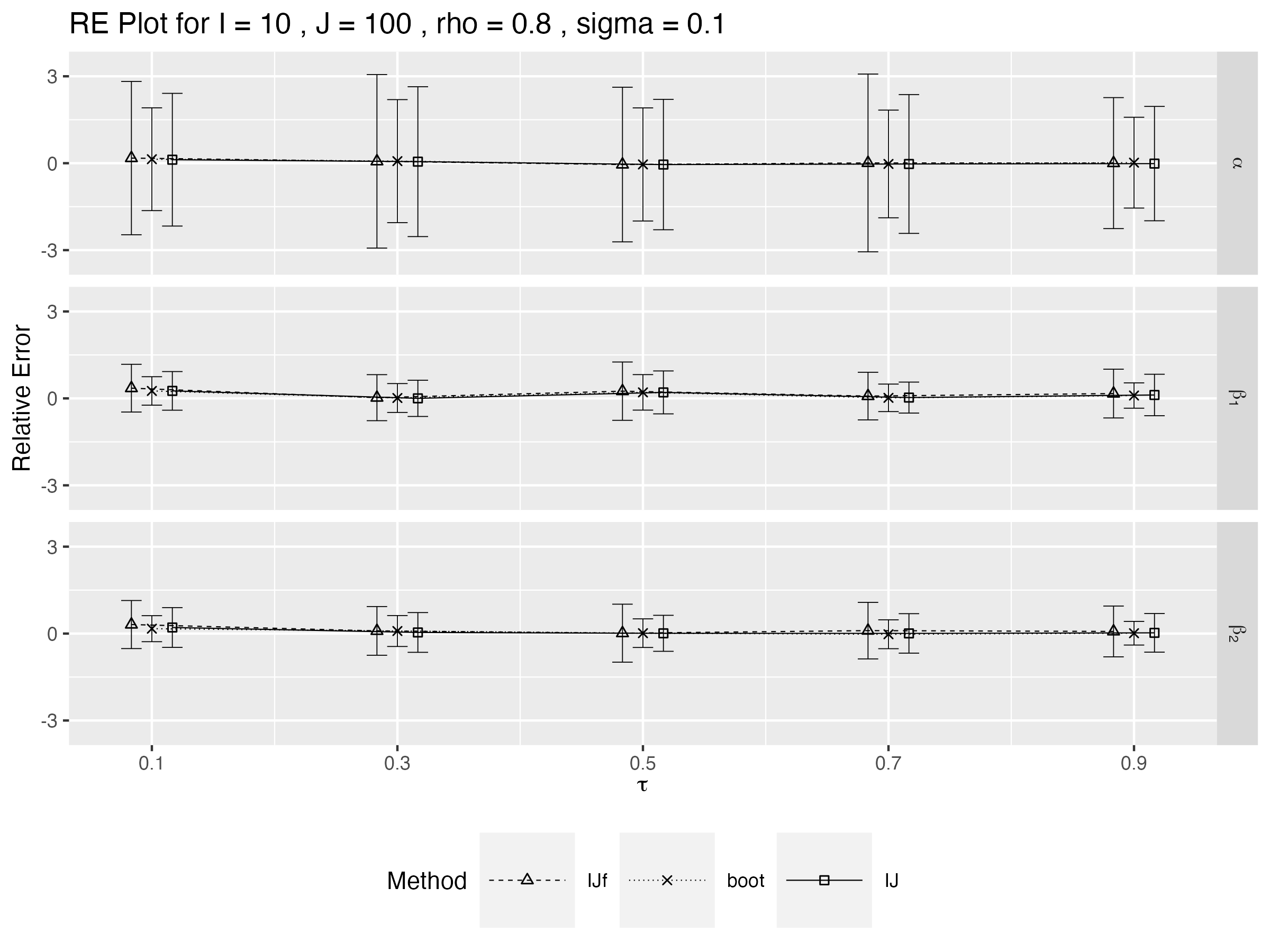}\\
\vspace{-1mm}
\end{tabular}
\end{center}
\vspace{-10mm}
\caption[Relative error estimates for $I=10$, $J=100$, $\rho = .8$]{\label{fig:I10J100re08}\footnotesize Relative error as a function of $\tau$ for $I=10$, $J=100$, $\rho = .8$, and $\sigma=0.1$.}
\end{figure}

\begin{figure}[htbp]
 \footnotesize
 \vspace{-6mm}
\begin{center}
\begin{tabular}{c}
\includegraphics[scale=.6]{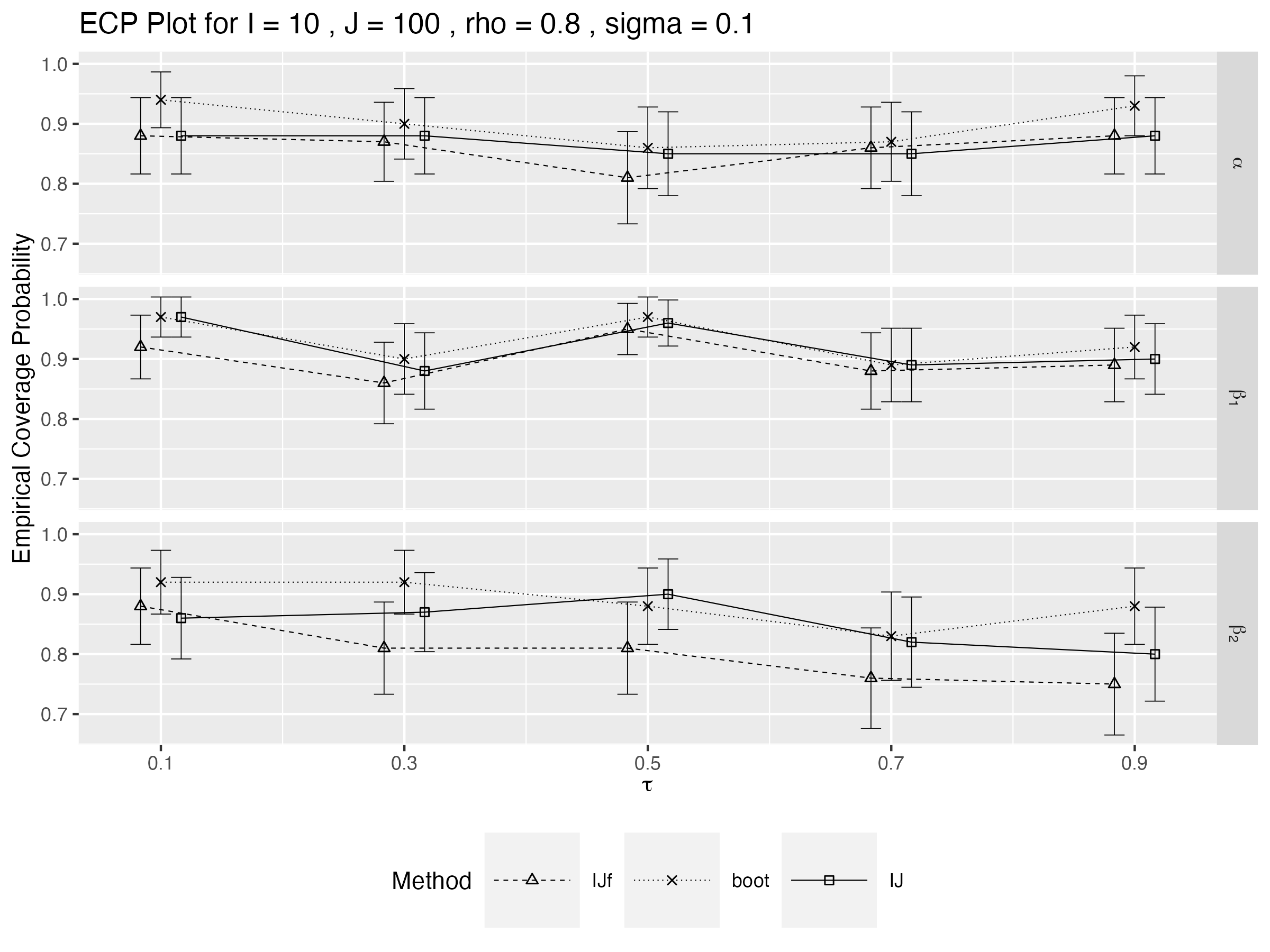}\\
\vspace{-1mm}
\end{tabular}
\end{center}
\vspace{-10mm}
\caption[Empirical coverage for $I=10$, $J=100$, $\rho = .8$]{\label{fig:I10J100cv08}\footnotesize Empirical coverage probability as a function of $\tau$ for $I=10$, $J=100$, $\rho = .8$, and $\sigma=0.1$.}
\end{figure}

\begin{figure}[htbp]
 \footnotesize
 \vspace{-6mm}
\begin{center}
\begin{tabular}{c}
\includegraphics[scale=.6]{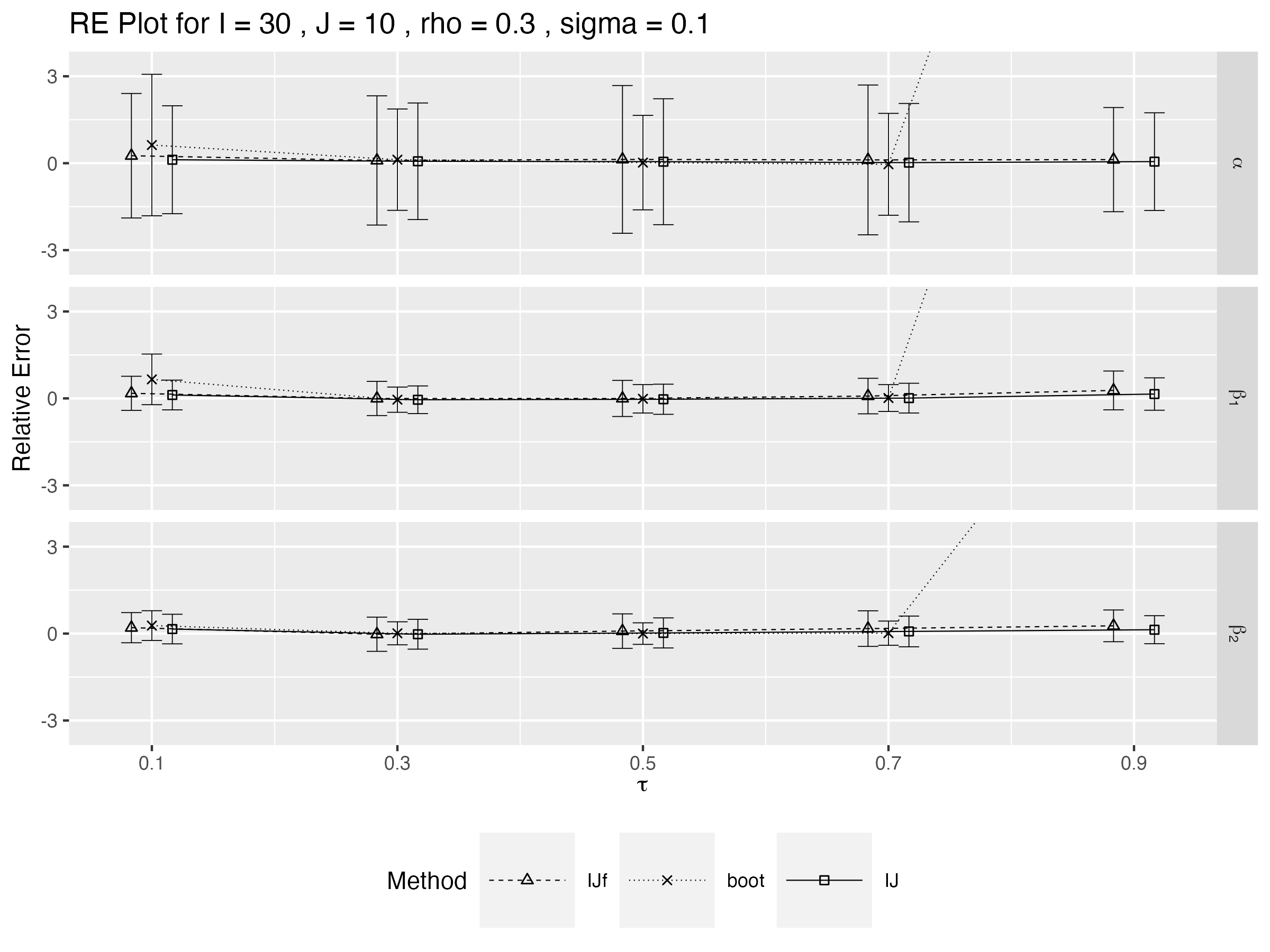}\\
\vspace{-1mm}
\end{tabular}
\end{center}
\vspace{-10mm}
\caption[Relative error estimates for $I=30$, $J=10$, $\rho = .3$]{\label{fig:I30J10re03}\footnotesize Relative error as a function of $\tau$ for $I=30$, $J=10$, $\rho = .3$, and $\sigma=0.1$.}
\end{figure}

\begin{figure}[htbp]
 \footnotesize
 \vspace{-6mm}
\begin{center}
\begin{tabular}{c}
\includegraphics[scale=.6]{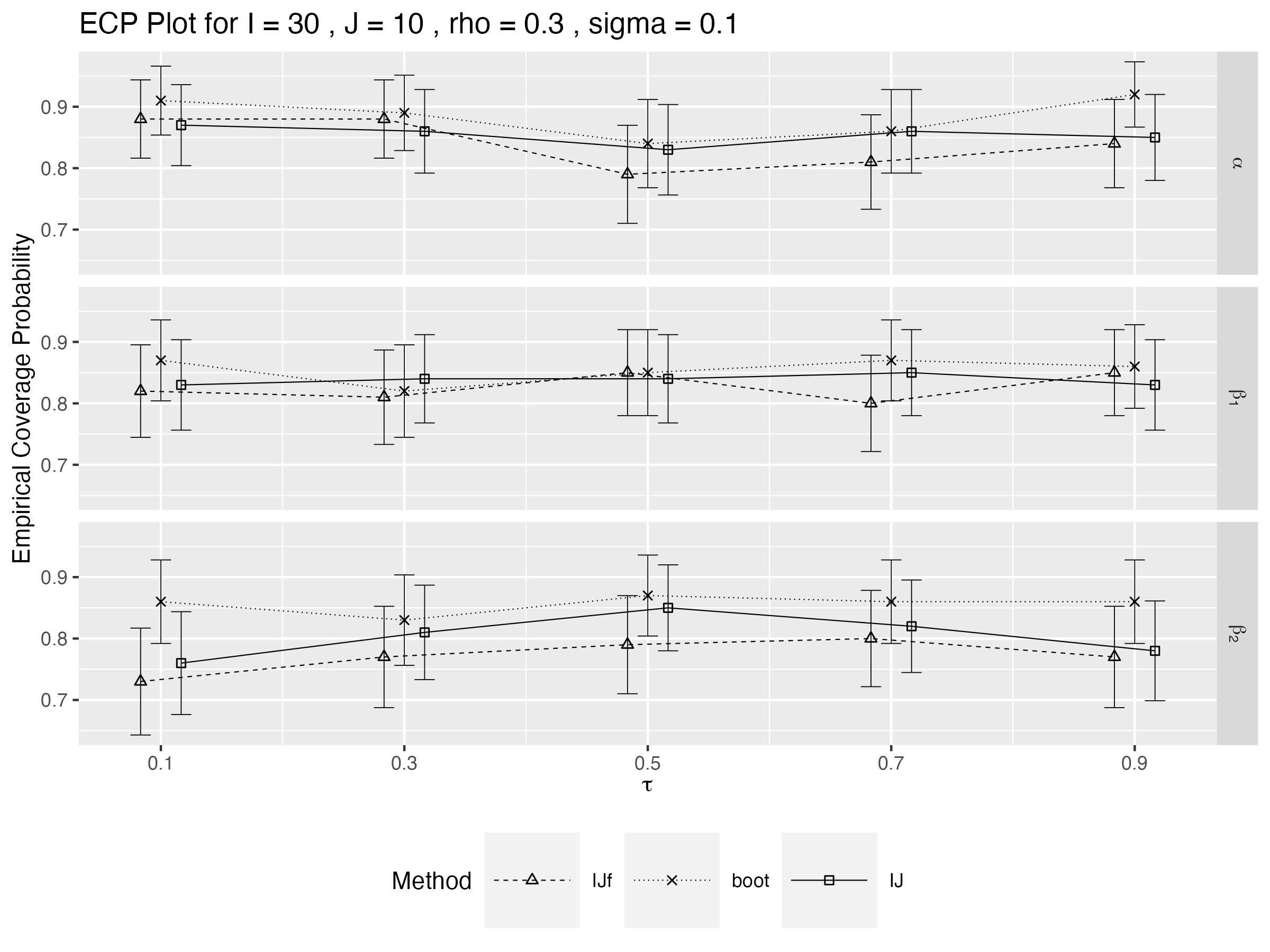}\\
\vspace{-1mm}
\end{tabular}
\end{center}
\vspace{-10mm}
\caption[Empirical coverage for $I=30$, $J=10$, $\rho = .3$]{\label{fig:I30J10cv03}\footnotesize Empirical coverage probability as a function of $\tau$ for $I=30$, $J=10$, $\rho = .3$, and $\sigma=0.1$.}
\end{figure}

\begin{figure}[htbp]
 \footnotesize
 \vspace{-6mm}
\begin{center}
\begin{tabular}{c}
\includegraphics[scale=.6]{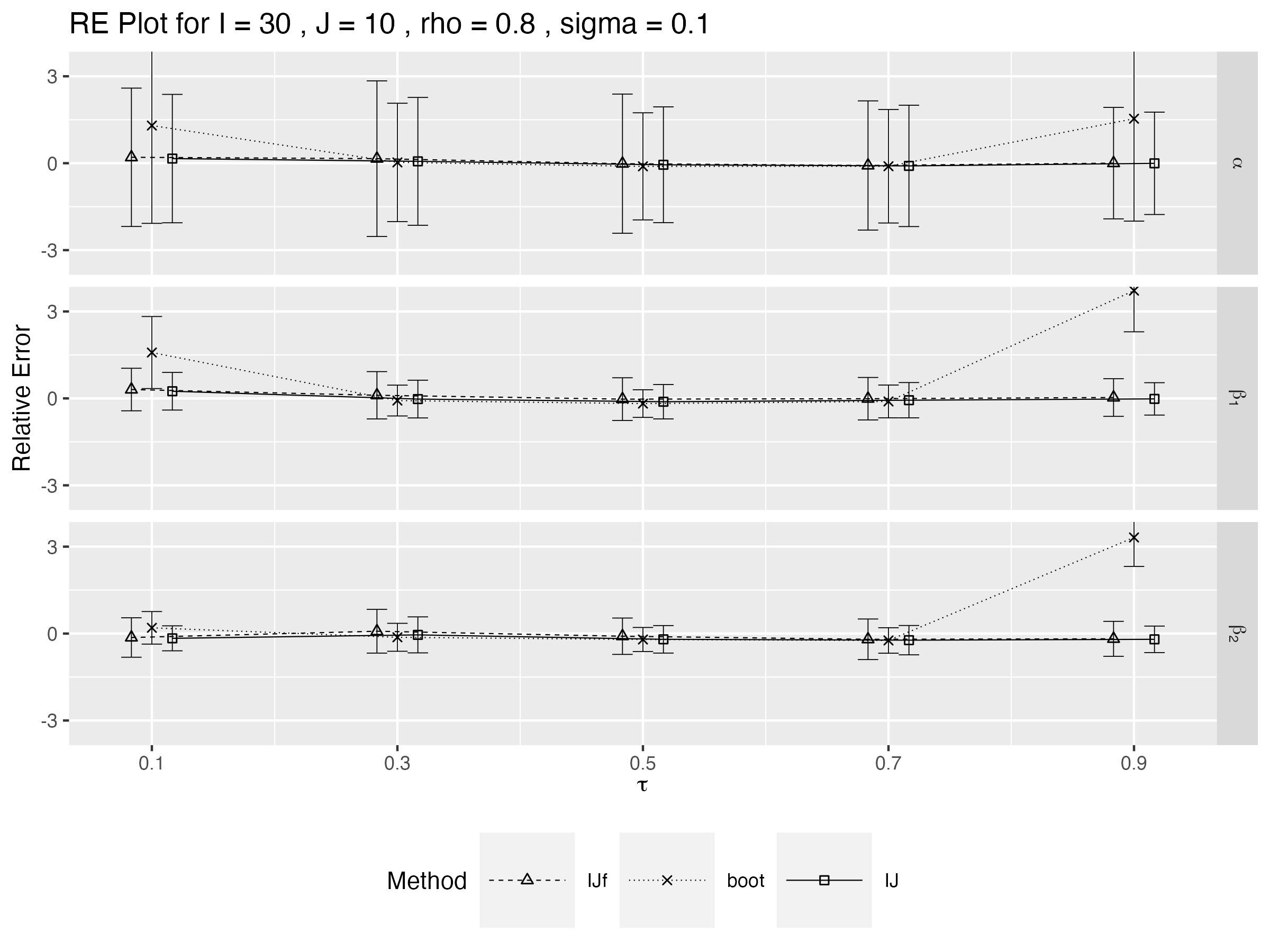}\\
\vspace{-1mm}
\end{tabular}
\end{center}
\vspace{-10mm}
\caption[Relative error estimates for $I=30$, $J=10$, $\rho = .8$]{\label{fig:I30J10re08}\footnotesize Relative error as a function of $\tau$ for $I=30$, $J=10$, $\rho = .8$, and $\sigma=0.1$.}
\end{figure}

\begin{figure}[htbp]
 \footnotesize
 \vspace{-6mm}
\begin{center}
\begin{tabular}{c}
\includegraphics[scale=.6]{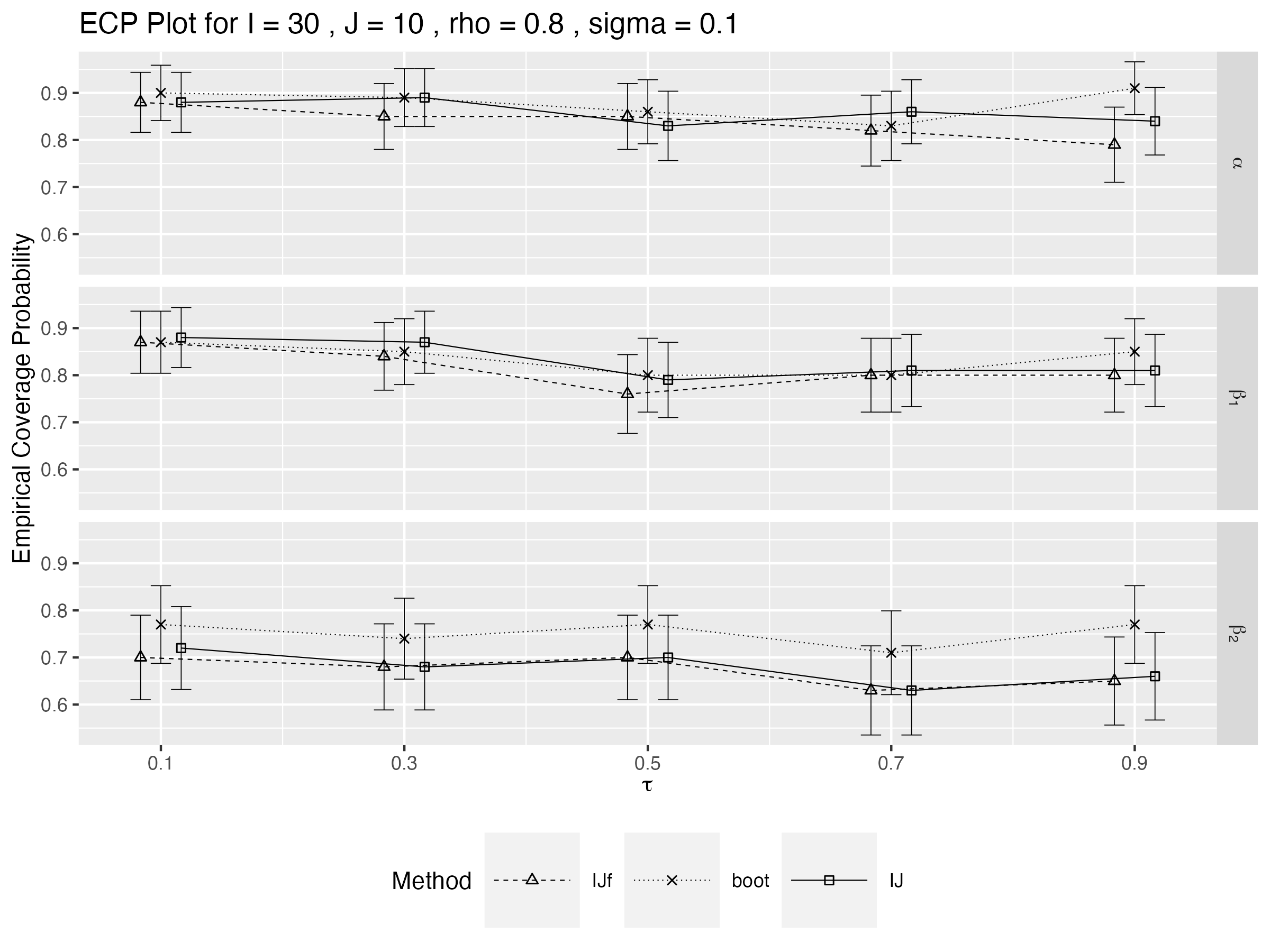}\\
\vspace{-1mm}
\end{tabular}
\end{center}
\vspace{-10mm}
\caption[Empirical coverage for $I=30$, $J=10$, $\rho = .8$]{\label{fig:I30J10cv08}\footnotesize Empirical coverage probability as a function of $\tau$ for $I=30$, $J=10$, $\rho = .8$, and $\sigma=0.1$.}
\end{figure}

\begin{figure}[htbp]
 \footnotesize
 \vspace{-6mm}
\begin{center}
\begin{tabular}{c}
\includegraphics[scale=.6]{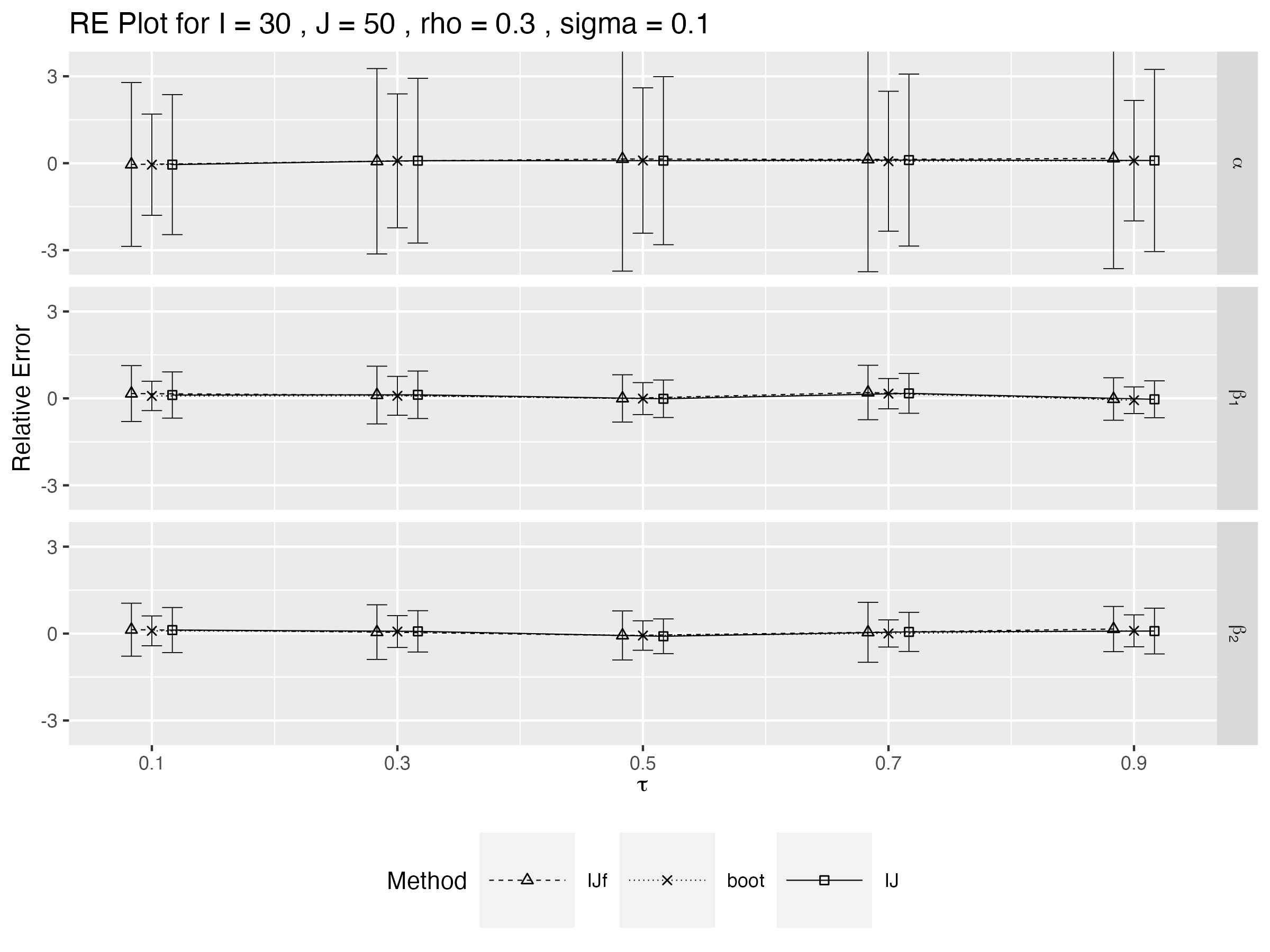}\\
\vspace{-1mm}
\end{tabular}
\end{center}
\vspace{-10mm}
\caption[Relative error estimates for $I=30$, $J=50$, $\rho = .3$]{\label{fig:I30J50re03}\footnotesize Relative error as a function of $\tau$ for $I=30$, $J=50$, $\rho = .3$, and $\sigma=0.1$.}
\end{figure}

\begin{figure}[htbp]
 \footnotesize
 \vspace{-6mm}
\begin{center}
\begin{tabular}{c}
\includegraphics[scale=.6]{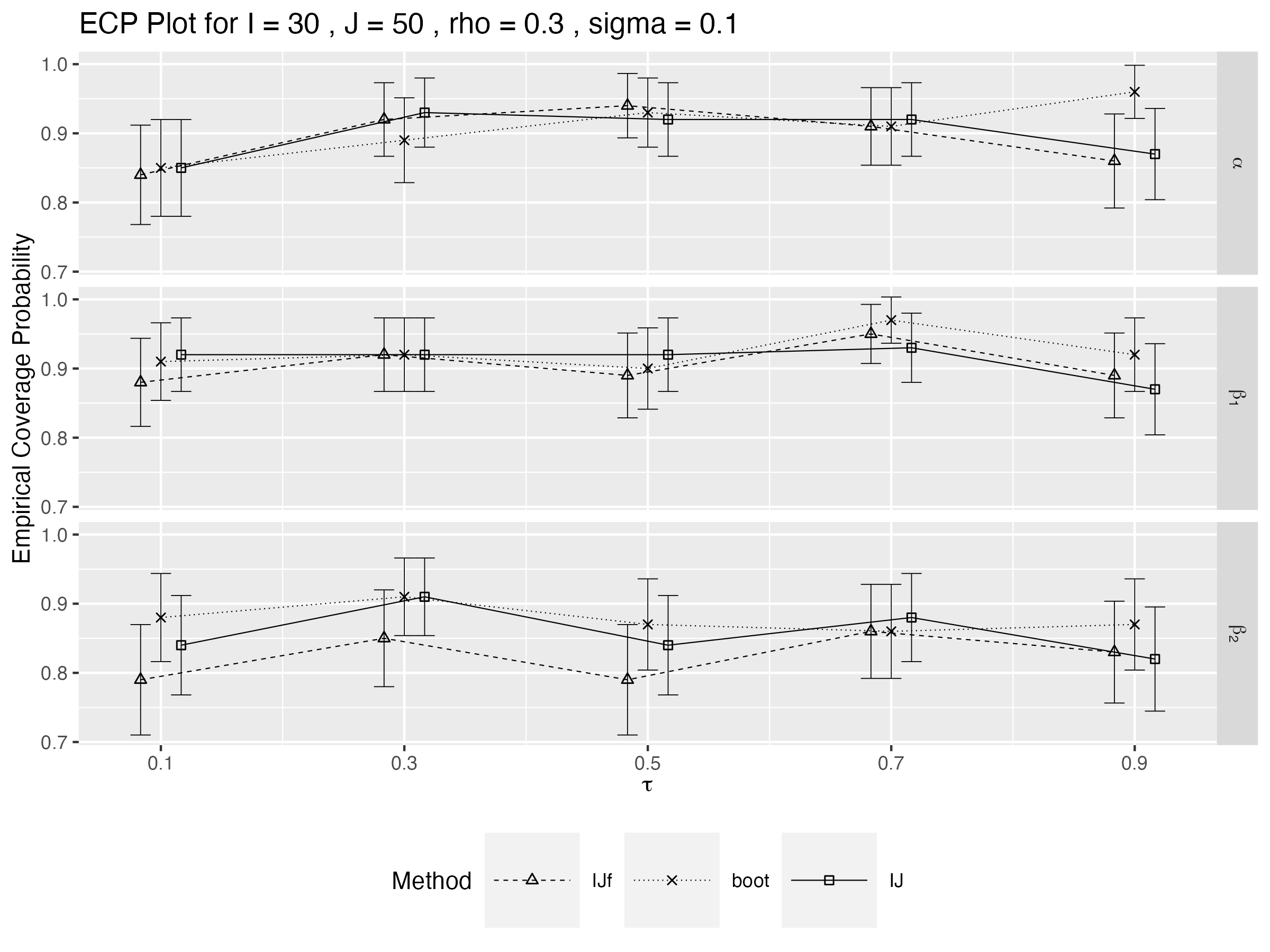}\\
\vspace{-1mm}
\end{tabular}
\end{center}
\vspace{-10mm}
\caption[Empirical coverage for $I=30$, $J=50$, $\rho = .3$]{\label{fig:I30J50cv03}\footnotesize Empirical coverage probability as a function of $\tau$ for $I=30$, $J=50$, $\rho = .3$, and $\sigma=0.1$.}
\end{figure}

\begin{figure}[htbp]
 \footnotesize
 \vspace{-6mm}
\begin{center}
\begin{tabular}{c}
\includegraphics[scale=.6]{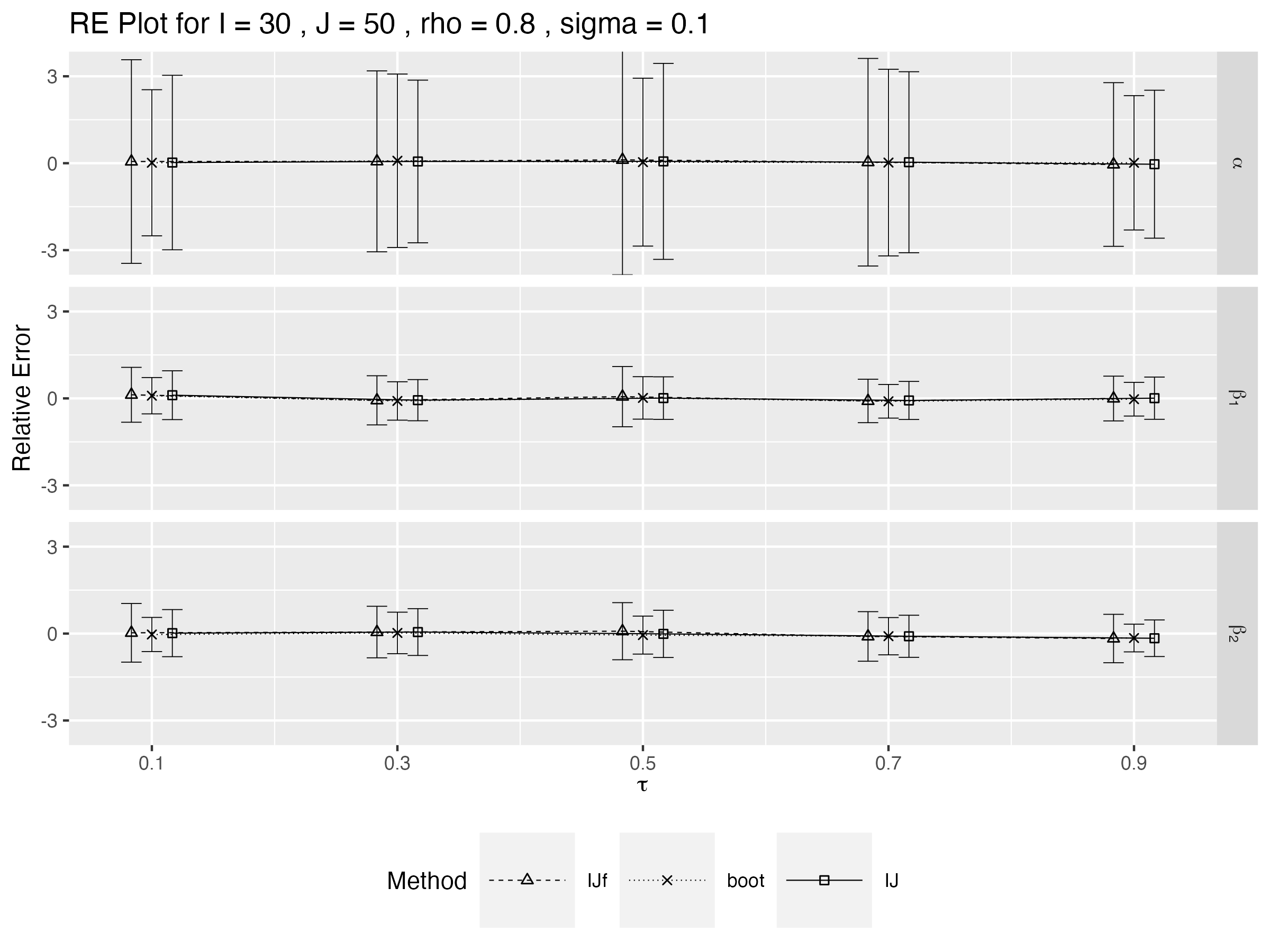}\\
\vspace{-1mm}
\end{tabular}
\end{center}
\vspace{-10mm}
\caption[Relative error estimates for $I=30$, $J=50$, $\rho = .8$]{\label{fig:I30J50re08}\footnotesize Relative error as a function of $\tau$ for $I=30$, $J=50$, $\rho = .8$, and $\sigma=0.1$.}
\end{figure}

\begin{figure}[htbp]
 \footnotesize
 \vspace{-6mm}
\begin{center}
\begin{tabular}{c}
\includegraphics[scale=.6]{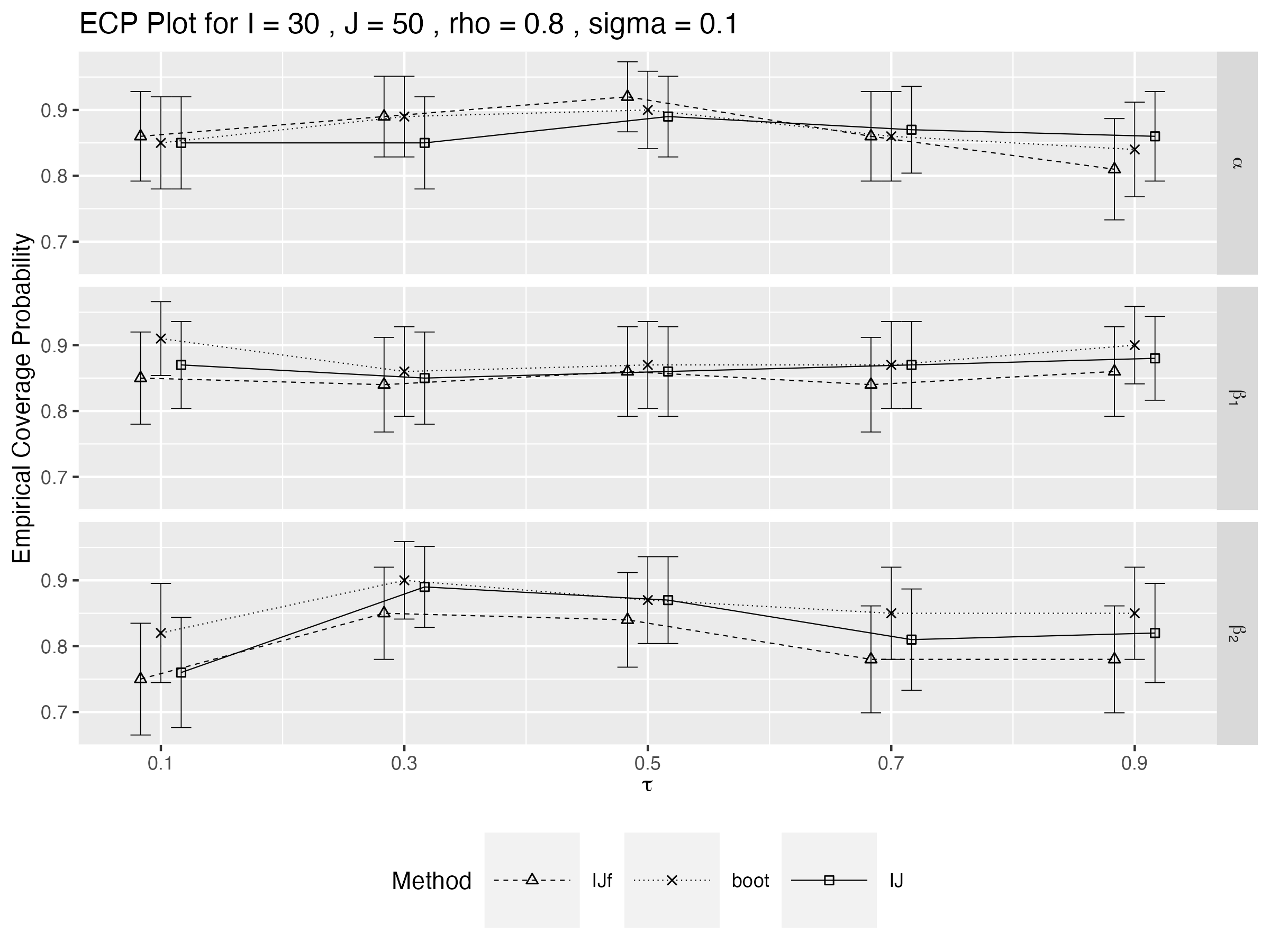}\\
\vspace{-1mm}
\end{tabular}
\end{center}
\vspace{-10mm}
\caption[Empirical coverage for $I=30$, $J=50$, $\rho = .8$]{\label{fig:I30J50cv08}\footnotesize Empirical coverage probability as a function of $\tau$ for $I=30$, $J=50$, $\rho = .8$, and $\sigma=0.1$.}
\end{figure}

\begin{figure}[htbp]
 \footnotesize
 \vspace{-6mm}
\begin{center}
\begin{tabular}{c}
\includegraphics[scale=.6]{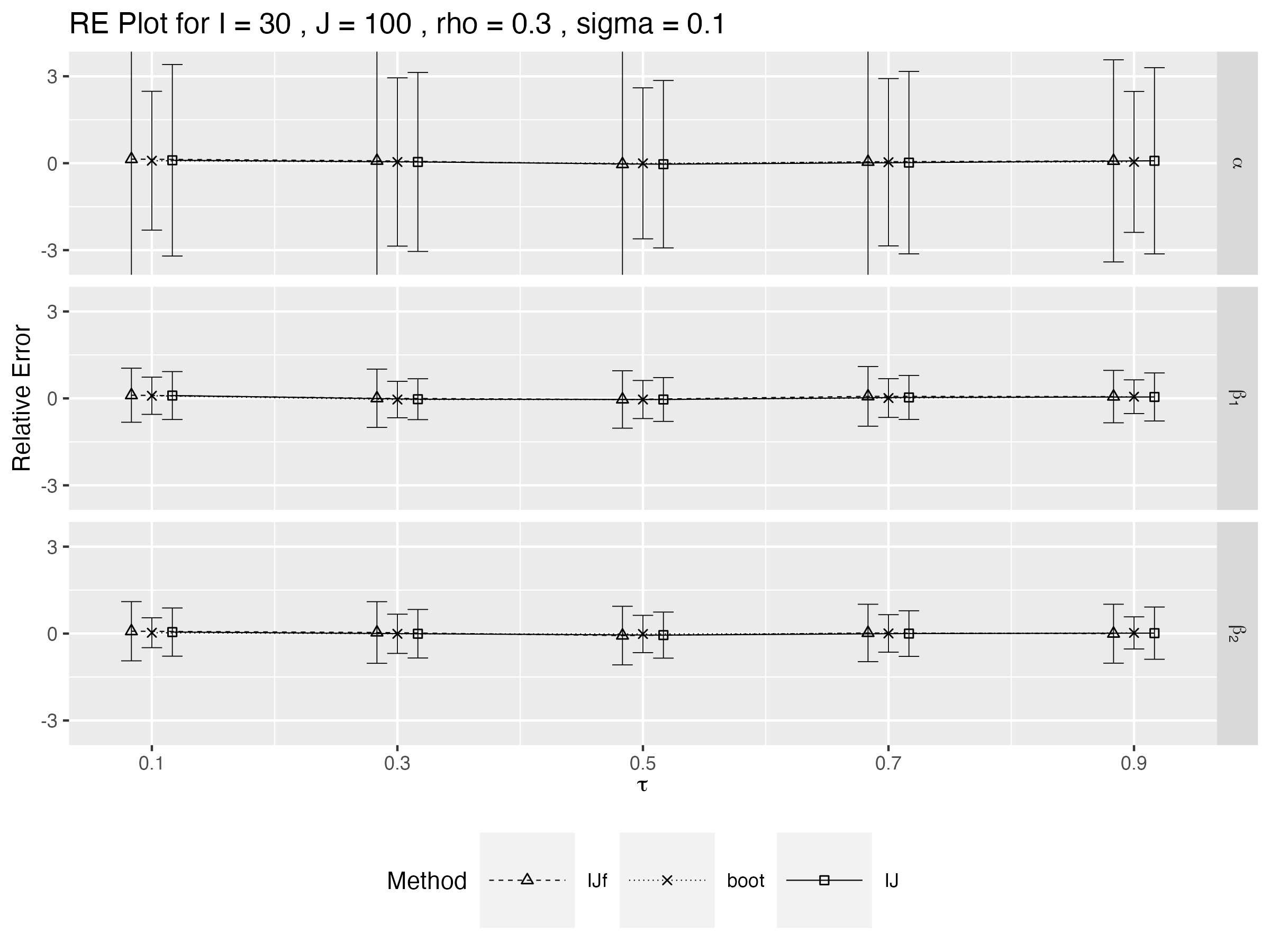}\\
\vspace{-1mm}
\end{tabular}
\end{center}
\vspace{-10mm}
\caption[Relative error estimates for $I=30$, $J=100$, $\rho = .3$]{\label{fig:I30J100re03}\footnotesize Relative error as a function of $\tau$ for $I=30$, $J=100$, $\rho = .3$, and $\sigma=0.1$.}
\end{figure}

\begin{figure}[htbp]
 \footnotesize
 \vspace{-6mm}
\begin{center}
\begin{tabular}{c}
\includegraphics[scale=.6]{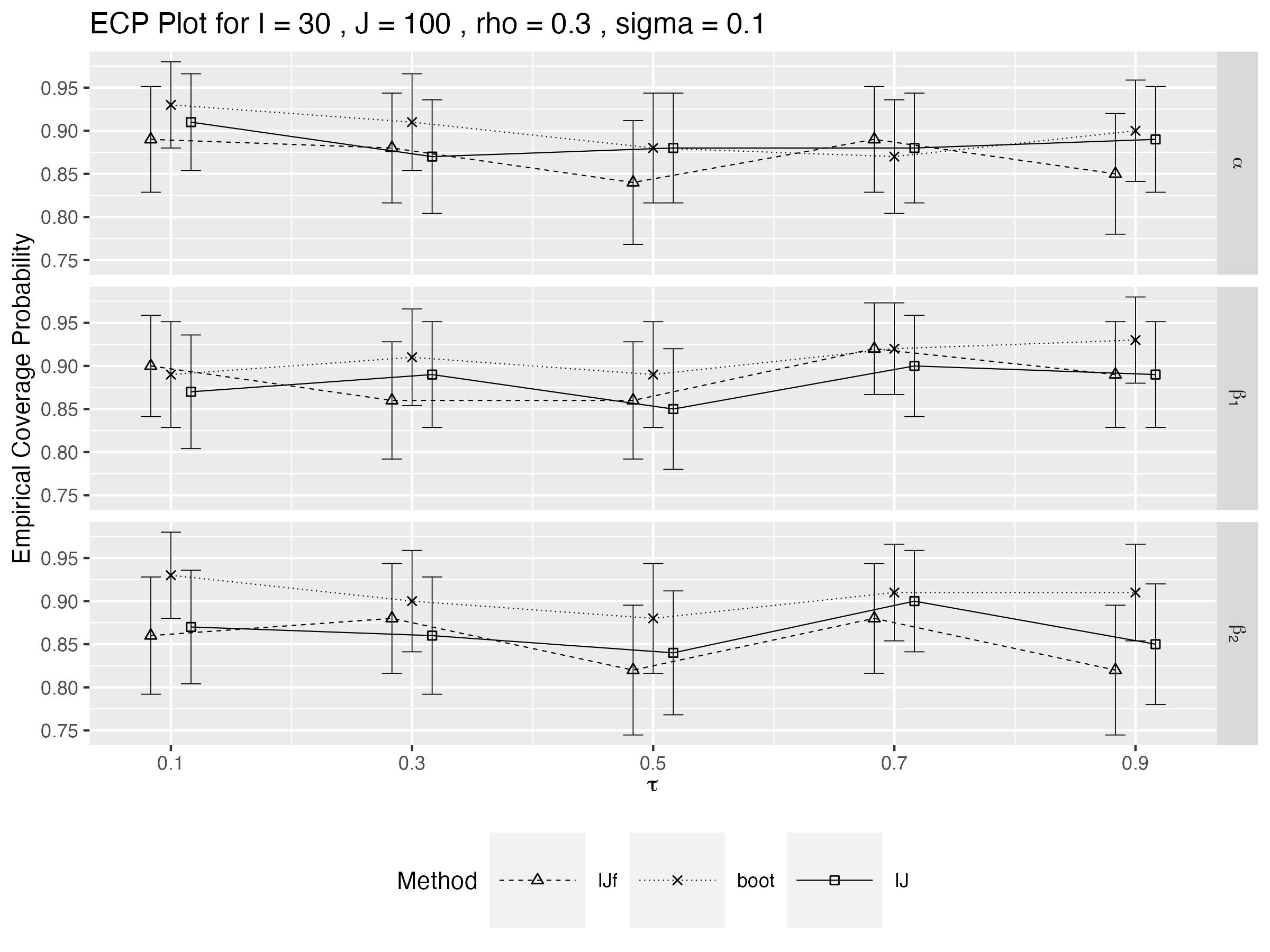}\\
\vspace{-1mm}
\end{tabular}
\end{center}
\vspace{-10mm}
\caption[Empirical coverage for $I=30$, $J=100$, $\rho = .3$]{\label{fig:I30J100cv03}\footnotesize Empirical coverage probability as a function of $\tau$ for $I=30$, $J=100$, $\rho = .3$, and $\sigma=0.1$.}
\end{figure}

\begin{figure}[htbp]
 \footnotesize
 \vspace{-6mm}
\begin{center}
\begin{tabular}{c}
\includegraphics[scale=.6]{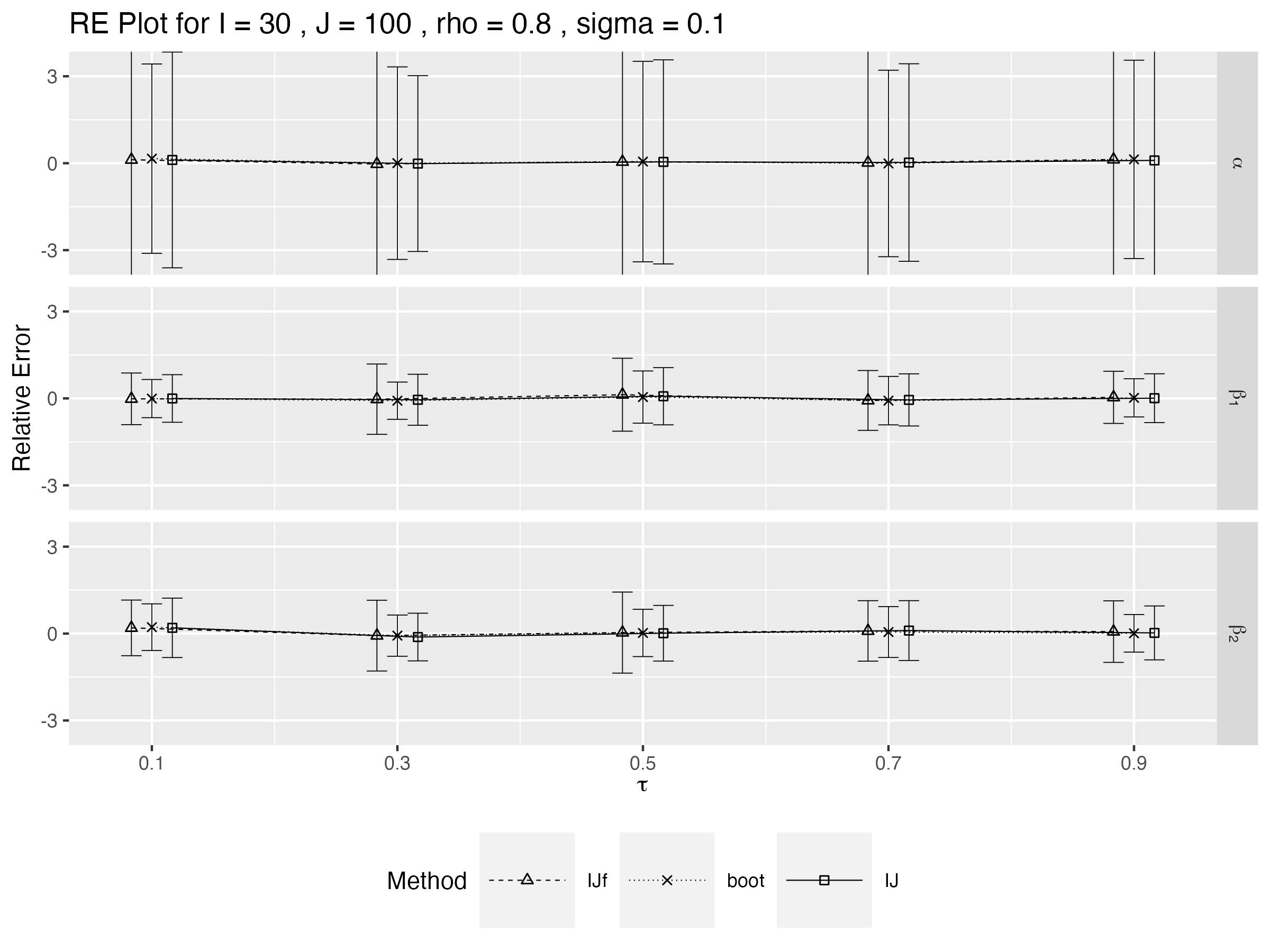}\\
\vspace{-1mm}
\end{tabular}
\end{center}
\vspace{-10mm}
\caption[Relative error estimates for $I=30$, $J=100$, $\rho = .8$]{\label{fig:I30J100re08}\footnotesize Relative error as a function of $\tau$ for $I=30$, $J=100$, $\rho = .8$, and $\sigma=0.1$.}
\end{figure}

\begin{figure}[htbp]
 \footnotesize
 \vspace{-6mm}
\begin{center}
\begin{tabular}{c}
\includegraphics[scale=.6]{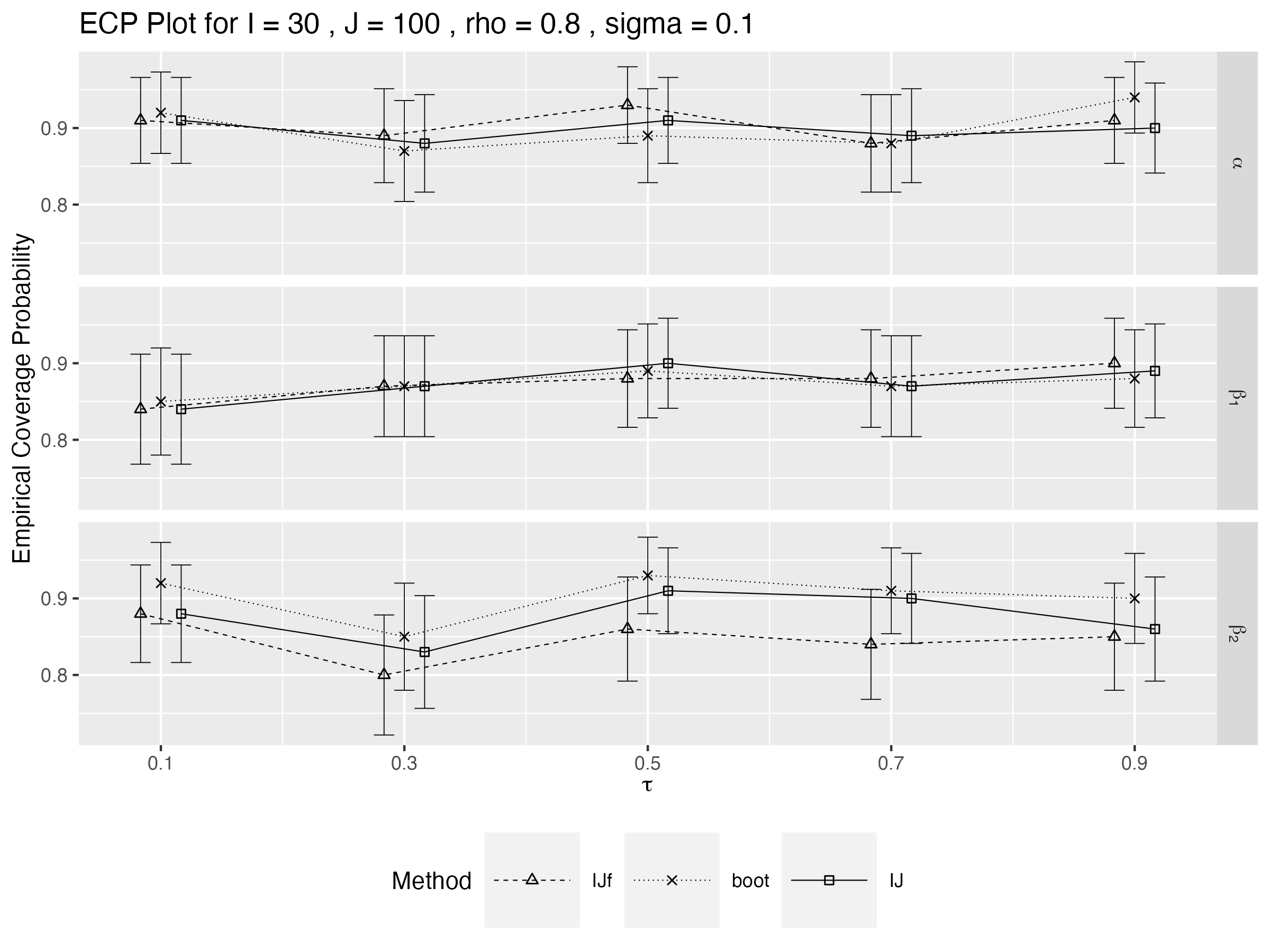}\\
\vspace{-1mm}
\end{tabular}
\end{center}
\vspace{-10mm}
\caption[Empirical coverage for $I=30$, $J=100$, $\rho = .8$]{\label{fig:I30J100cv08}\footnotesize Empirical coverage probability as a function of $\tau$ for $I=30$, $J=100$, $\rho = .8$, and $\sigma=0.1$.}
\end{figure}


\clearpage

\section*{Appendix D: Further application results}

\begin{minipage}{\textwidth}
\centering
\includegraphics[scale=.7]{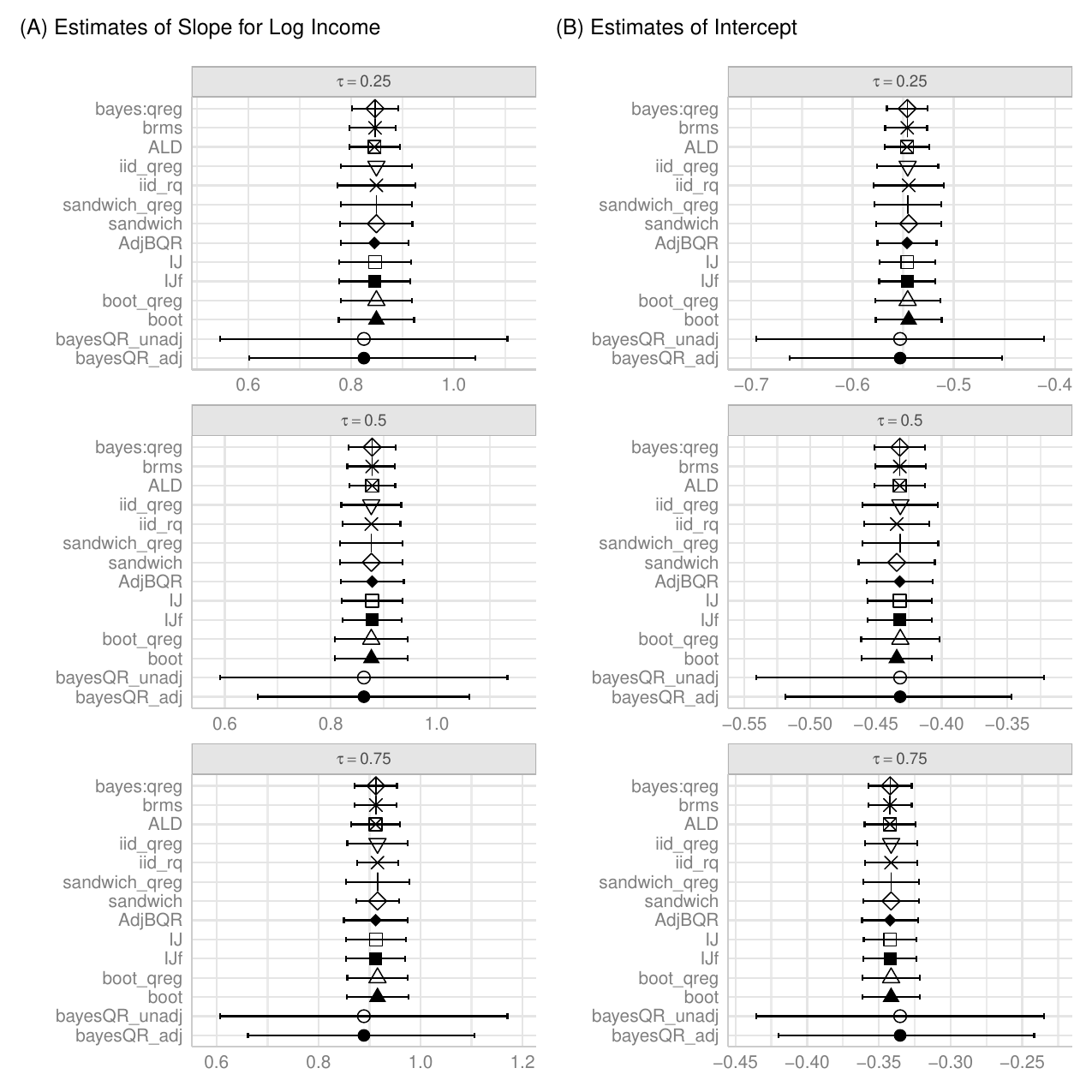}

\captionof{figure}{Comparison of slope and intercept estimates for log income using various quantile regression methods with approximate 95\% credible intervals/confidence intervals, including the results from \cn{bayesQR}.}
\label{fig:engel_QR_estimates_bayesQR}
\label{fig:engel_QR_estimates_bayesQR}
\end{minipage}

\end{document}